\documentclass[aps,reprint,prl,longbibliography,superscriptaddress]{revtex4-1}
\usepackage{}
\usepackage{amssymb}
\usepackage{amsmath}
\usepackage{dcolumn}
\usepackage{bm}
\usepackage{graphicx}
\usepackage{mathrsfs}
\usepackage[colorlinks,linkcolor=blue,anchorcolor=blue,citecolor=blue,urlcolor=blue]{hyperref}

\begin{document}

\title{Unconventional  quantum sound-matter interactions in spin-optomechanical-crystal hybrid systems }

\author{Xing-Liang Dong}
\affiliation {MOE Key Laboratory for Nonequilibrium Synthesis and Modulation of Condensed Matter,
Shaanxi Province Key Laboratory of Quantum Information and Quantum Optoelectronic Devices,
School of  Physics, Xi'an Jiaotong University, Xi'an 710049, China}

\author{Peng-Bo Li}
\email{lipengbo@mail.xjtu.edu.cn}
\affiliation {MOE Key Laboratory for Nonequilibrium Synthesis and Modulation of Condensed Matter,
Shaanxi Province Key Laboratory of Quantum Information and Quantum Optoelectronic Devices,
School of  Physics, Xi'an Jiaotong University, Xi'an 710049, China}
\affiliation {Theoretical Quantum Physics Laboratory, RIKEN Cluster for Pioneering Research, Wako-shi, Saitama 351-0198, Japan}

\author{Tao Liu}
\affiliation {Theoretical Quantum Physics Laboratory, RIKEN Cluster for Pioneering Research, Wako-shi, Saitama 351-0198, Japan}
\affiliation{School of Physics and Optoelectronics, South China University of Technology,  Guangzhou 510640, China}

\author{Franco Nori}
\affiliation {Theoretical Quantum Physics Laboratory, RIKEN Cluster for Pioneering Research, Wako-shi, Saitama 351-0198, Japan}
\affiliation {Department of Physics, The University of Michigan, Ann Arbor, Michigan 48109-1040, USA}

\date{\today}

\begin{abstract}
We predict a set of unusual quantum acoustic phenomena resulting from sound-matter interactions in a fully tunable solid-state platform,
in which an array of solid-state spins in diamond are coupled to quantized acoustic waves in a one-dimensional (1D) optomechanical crystal.
We find that, by a spatially varying laser drive that introduces a position-dependent phase in the optomechanical interaction, the mechanical
band structure can be tuned \emph{in situ}, consequently leading to unconventional quantum sound-matter interactions.
We show that quasi-chiral sound-matter interactions can occur, with tunable ranges from bidirectional to quasi-unidirectional, when the spins are resonant with the bands. When the solid-state spins' frequency  lies within the acoustic band-gap,
we demonstrate the emergence of an exotic polariton bound state,
which can mediate long-range tunable, odd-neighbor and complex spin-spin interactions.
This work expands the present exploration of quantum phononics and can have wide applications
in quantum simulation and quantum information processing.
\end{abstract}

\maketitle


Research of light-matter interactions in nanostructures injects new vitality to quantum optics
\cite{RevModPhys.90.031002}.
The confinement of electromagnetic waves to small dimensions and engineered structures
not only results in an enhanced light-matter coupling,
but also gives rise to new quantum phenomena, such as chiral light-matter interactions
\cite{Quantum2014Mitsch,PhysRevLett.113.237203,Bliokh1448,Spin2015Bliokh,Chiral2017Lodahl,triolo2017spin,PhysRevResearch.2.023003},
many-body physics in a band gap
\cite{Subwavelength2015AGT,Quantum2015Douglas,Hood10507,doi:10.1021/acsphotonics.8b01455,Belloeaaw0297},
and topological photonics
\cite{PhysRevLett.119.023603,Topological2019Bliokh,RevModPhys.91.015006,PhysRevLett.124.083603}.
These ideas are also explored in circuit quantum electrodynamics with superconducting qubits
\cite{Atomic2011You,vanLoo1494,GU20171,PhysRevLett.124.023603} and  optomechanical systems
\cite{Habraken_2012,RevModPhys.86.1391,PhysRevX.5.031011,Dynamically2017Kim,Optomechanical2017Verhagen,
Thermal2018Seif,Rakich_2018,Nonreciprocal2019Xu,PhysRevB.101.085108,PhysRevLett.124.083601}.

Phonons,  the quanta of mechanical waves,
are potential candidates for implementing on-chip quantum information processing and networks,
because the speed of acoustic waves is much slower than that of light
\cite{Gustafsson207,PhysRevX.5.031031,PhysRevX.8.041027,Bienfait368,PhysRevLett.124.053601}.
Moreover, mechanical systems can interface with various quantum emitters, ranging from superconducting circuits
\cite{RevModPhys.85.623,Chu199,Circuit2017Manenti,Quantum2018Satzinger},
and quantum dots
\cite{PhysRevLett.105.037401}, to solid-state defects
\cite{Kolkowitz1603,PhysRevLett.112.036405,PhysRevApplied.4.044003,
PhysRevLett.116.143602,PhysRevLett.117.015502,PhysRevX.6.041060,
Lee2017Topical,PhysRevLett.118.223603,PhysRevB.97.205444,PhysRevApplied.10.024011,Coherent2020Maity}.
In particular, hybrid systems composed of defect centers in diamond and phononic nanostructures (such as phononic waveguides or crystals)
provide a promising platform for quantum applications due to their long coherent times and scalability
\cite{Balasubramanian2009Ultralong,Rabl2010A,PhysRevLett.110.156402,Bar2013Solid,PhysRevLett.119.223602,
PhysRevLett.120.213603,PhysRevLett.121.123604,PhysRevLett.125.153602,BandPeng,Lemonde_2019,PhysRevResearch.2.013121,PhysRevA.101.042313}.
However, compared to light,
the realization of chiral sound-matter interactions remains an outstanding challenge
due to lack of polarization in phonons.
Moreover, rich physics resulting from quantum sound-matter interactions near acoustic band gaps remains largely unexplored.

In this work, we explore unconventional quantum interactions between sound and matter in a fully tunable solid-state device, with silicon-vacancy (SiV) centers in diamond coupled to the acoustic waves in a 1D optomechanical crystal. We show that, through utilizing optomechanical crystals \cite{Electromagnetically2011Safavi,PhysRevLett.112.153603,
PhysRevX.5.041051,Burek:16,Sipahigil847,PhysRevApplied.8.024026,Evans662,Cady_2019,PhysRevA.99.053852,PhysRevLett.123.183602}, the acoustic band structures of sound waves  can be tuned \emph{in situ} by a suitably designed laser drive that introduces a position-dependent phase in the optomechanical interaction. Moreover, we find a number of unprecedented phenomena resulting from the interactions between solid-state spins and acoustic waves with tunable bands. When the spins are resonant with the bands, we predict a quasi-chiral sound-matter interaction with a tunable range from bidirectional to quasi-unidirectional. When the spins' frequency lies within the middle bandgap, we analyze the emergence of an exotic bound state with alternating photon and phonon components. This polariton bound state can be exploited to mediate long-range tunable,
odd-neighbor and complex spin-spin interactions.
The exclusive advantage of this highly tunable solid-state system is that, the band structures of the acoustic waves and the resulting sound-matter interactions can be tuned \emph{in situ}; thus providing a promising platform for the exploration of unusual quantum acoustic phenomena. This work opens new routes for quantum acoustics and could have applications in quantum simulation and quantum information processing, including the simulation of spin models \cite{Belloeaaw0297}, quantum state transfer \cite{PhysRevLett.120.213603,Lemonde_2019} and entangled states preparation
\cite{PhysRevA.85.042306,Stannigel_2012} via unidirectional phonon channels, etc.

\begin{figure}[t]
\includegraphics[scale=0.25]{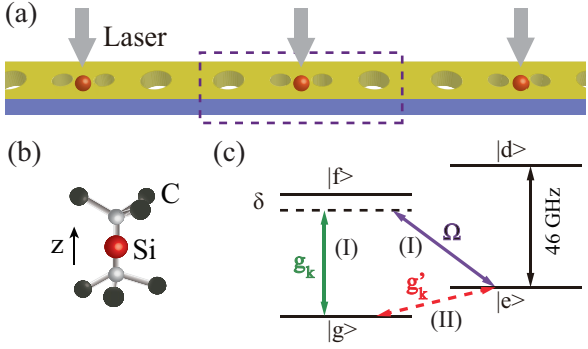}
\caption{\label{fig1}(Color online)
(a) Schematic of a 1D optomechanical crystal embedded with SiV centers arrays.
The photons and phonons are co-localized in a nanocavity and two adjacent nanocavities are coupled via the tunnelling effects
or a waveguide.
Physical structure (b) and  coupling scheme of the electronic ground state (c) of SiV center.}
\end{figure}
\emph{Model.--}
We consider a hybrid system of an array of SiV centers  integrated to an optomechanical crystal, as depicted in Fig.~1.
By deforming periodically the holes in an optomechanical crystal,
an array of coupled defect cavities  can form, which strongly co-localize phonons and photons.
In each unit cell, there is a standard optomechanical interaction
$\hat{H}_n=\hbar\omega_c\hat{a}_n^\dag\hat{a}_n+\hbar\omega_M\hat{b}_n^\dag\hat{b}_n-\hbar g_0\hat{a}_n^\dag\hat{a}_n(\hat{b}^\dag_n+\hat{b}_n)$,
with $\hat{a}_n$ and $\hat{b}_n$ are annihilation operators for photonic and phononic modes, respectively.
The frequency of optical (acoustic) cavities $\omega_c$ ($\omega_M$)
and the coupling strength $g_0$ are assumed to be identical for each unit cell.
The optical cavities are driven by a laser with a site-dependent phase $e^{-in\theta}$,
with $\theta\in[0,2\pi)$ a tunable constant,
which can be utilized to break the time-reversal symmetry.
We derive the linearized interaction by replacing the cavity field
$\hat{a}_n\rightarrow\hat{a}_n+\alpha e^{-in\theta}e^{-i\omega_Lt}$,
and keeping the terms up to first order in $\alpha$,
with $\alpha$ the average light-field amplitude of the laser and $\omega_L$ the driving frequency.
For the intercell interaction, we assume the tight-binding model involving only the nearest-neighbor hoppings.
We have verified the validity of this basic model via a  full-wave simulation of the optomechanical system
\cite{SupplementalMaterial}.
Thus the whole Hamiltonian of the 1D optomechanical system in real space is given by ($\hbar=1$)
\begin{eqnarray}\label{ME1}
\hat{H}_{\text{OM}}&=&\Delta/2\sum_{n}\hat{a}_n^\dag\hat{a}_n+\omega_M/2\sum_n\hat{b}_n^\dag\hat{b}_n-G\sum_n e^{-in\theta}\hat{a}_n^\dag\hat{b}_n\notag\\
&-&J\sum_{n}\hat{a}_{n+1}^\dag\hat{a}_{n}-K\sum_{n}\hat{b}_{n+1}^\dag\hat{b}_{n}+\text{H.c.},
\end{eqnarray}
with detuning $\Delta=\omega_c-\omega_L$, enhanced coupling strength $G=g_0\alpha$, and
hopping rate $J$\,$(K)$ for photons (phonons).

\begin{figure}[t]
\includegraphics[scale=0.18]{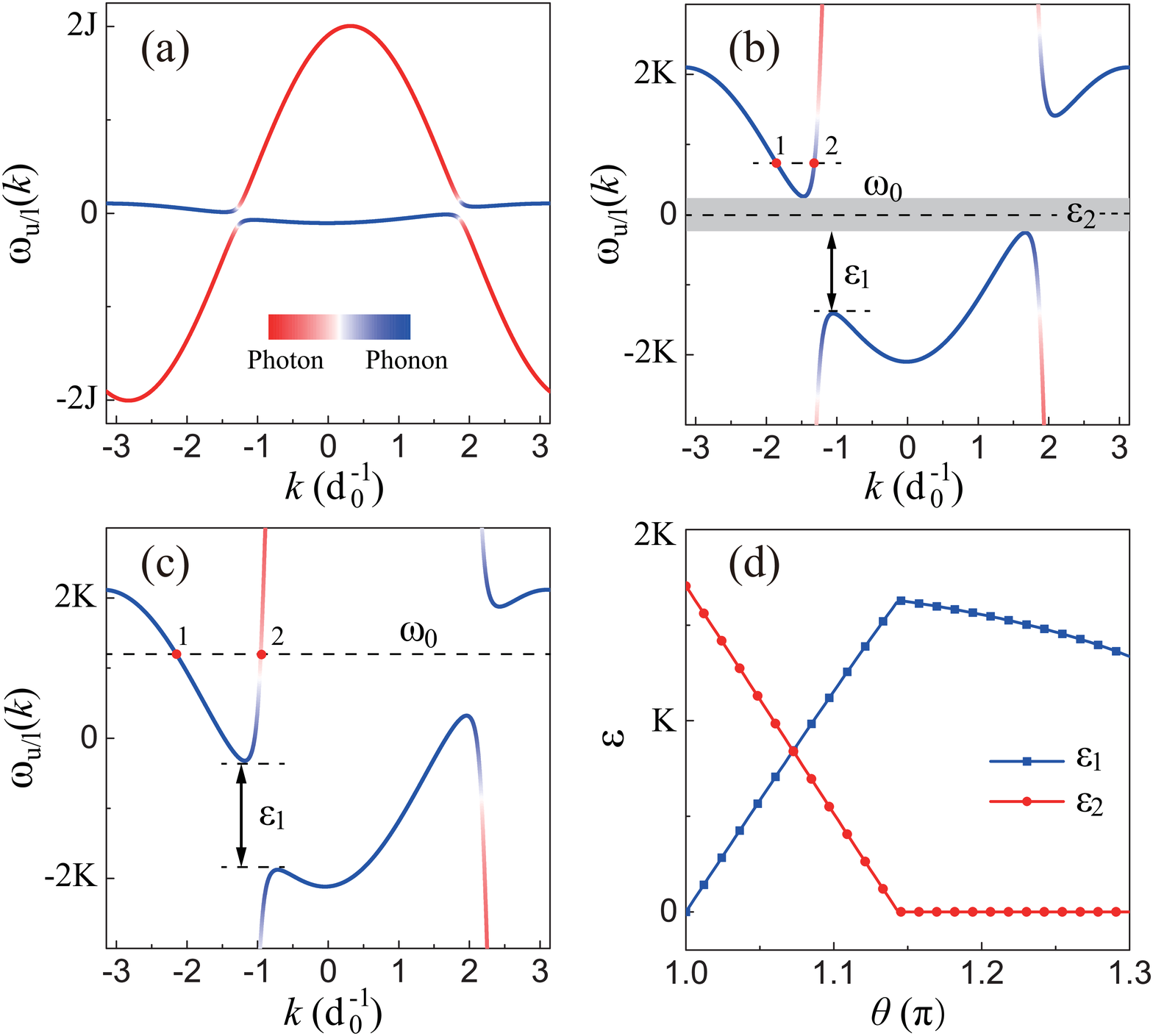}
\caption{\label{fig2}(Color online) Band structure $\omega(k)$ of a 1D optomechanical crystal under different system parameters.
(a), (b) $\theta=1.1\pi$. (c) $\theta=1.2\pi$.
(d) The size of bandgap $\varepsilon_2$ and asymmetric area $\varepsilon_1$ as a function of $\theta$.
Here, $J=20K$, $G=2K$. Note the relevant frequencies are shifted by $\Delta\approx\omega_M$.}
\end{figure}

By imposing periodic boundary conditions and introducing the Fourier transformation
$\hat{a}_k/\hat{b}_k=1/\sqrt{N}\sum_{n=1}^{N}e^{-iknd_{0}}\hat{a}_n/\hat{b}_n$ ($d_0$ is the lattice constant, and hereafter we choose $d_0=1$ for simplicity),
the optomechanical Hamiltonian in reciprocal space can be written as
$\hat{H}_{\text{OM}}=\sum_k\hat{V}_k^\dag\hat{H}(k)\hat{V}_k$,
with $\hat{V}_k^\dag=(\hat{a}_{k-\theta}^\dag,\hat{b}_k^\dag)$ and
\begin{eqnarray}\label{ME2}
\hat{H}(k)=
{\left( \begin{array}{ccc}
-2J(k) & -G\\
-G & -2K(k)
\end{array}\right )}.
\end{eqnarray}
Here $J(k)=J\cos(k-\theta)$, $K(k)=K\cos(k)$, and
$\Delta\approx\omega_M$ is taken as the energy reference.
Physically, a phonon with momentum $k$ is coupled to a photon with momentum $(k-\theta)$,
resulting in asymmetric hybridization polaritons, when $\theta\neq\{0,\pi\}$.
This Hamiltonian can be easily diagonalized as
$\hat{H}_{\text{OM}}=\sum_{k}(\omega_{u}(k)\hat{u}_{k}^\dag\hat{u}_{k}+\omega_{l}(k)\hat{l}_{k}^\dag\hat{l}_{k})$,
where the polariton operators $\hat{u}_{k}$ and $\hat{l}_{k}$ are related to $\hat{a}_{k-\theta}$
and $\hat{b}_{k}$ by means of a unitary transformation:
\begin{eqnarray}\label{ME3}
{\left( \begin{array}{ccc}
\hat{u}_{k}\\
\hat{l}_{k}
\end{array}\right )}=
{\left( \begin{array}{ccc}
-\sin\theta_k & \cos\theta_k\\
\cos\theta_k & \sin\theta_k
\end{array}\right )}
{\left( \begin{array}{ccc}
\hat{a}_{k-\theta}\\
\hat{b}_{k}
\end{array}\right )},
\end{eqnarray}
with $\sin\theta_k=-G/\sqrt{G^2+(\omega_{l}(k)+2K(k))^2}$,
$\cos\theta_k=-G/\sqrt{G^2+(\omega_{u}(k)+2K(k))^2}$ and the dispersion
\begin{eqnarray}\label{ME4}
\omega_{u/l}(k)=-J(k)-K(k)\pm\sqrt{(K(k)-J(k))^2+G^2}.
\end{eqnarray}
The dispersion of polaritonic energy bands can be tuned, and becomes either  symmetric or asymmetric
via changing  $\theta$, as shown in Figs.~2(a,b,c).
The bands are mostly optical and mechanical in nature except around the band-edges (coupling points).
Moreover, a sizable bandgap $\epsilon_2$ emerges, and can be tuned in a wide range of $\theta$ [see Fig. 2(b,d)]. In addition, the band asymmetry characterized by $\epsilon_1$ is also tunable  [see Fig. 2(b,c,d)].
The bandgap is maximized for $\theta=\pi$, and vanishes for $\theta\sim1.14\pi$, where
$\varepsilon_1$ reaches its maximum.

We now consider integrating a single or multiple solid-state spins into this 1D optomechanical crystal.
In this work, we take into account SiV centers, as shown in Fig.~1(c).
The levels $|g\rangle$ and $|e\rangle$ are  coupled to the acoustic modes
either (I) indirectly by a Raman process
or (II) directly by adding an off-axis magnetic field \cite{PhysRevLett.120.213603,PhysRevB.97.205444}.
Thus, in terms of spin operators $\{\hat{\sigma}_z,\hat{\sigma}_+,\hat{\sigma}_-\}$,
the free Hamiltonian is $\hat{H}_{\text{free}}=\omega_{0}/2\sum_{m}\hat{\sigma}_{z}^m$,
and the effective interaction Hamiltonian reads
$\hat{H}_{\text{int}}=g_{\text{eff}}\sum_{m}(\hat{\sigma}_{+}^m\hat{b}_{x_m}+\text{H.c.})$.
Here, $\omega_0$ is the transition frequency of the effective two-level system,
$g_{\text{eff}}$ is the effective spin-phonon coupling strength, and $x_m$ denotes the position at which the m\emph{th} spin is coupled to the phononic waveguide.
Lastly, working in  momentum space,
$\hat{H}_{\text{int}}$ becomes
\begin{eqnarray}\label{ME5}
\hat{H}_{\text{int}}=\frac{g_{\text{eff}}}{\sqrt{N}}\sum_{k,m}\hat{\sigma}_{+}^me^{ikx_m}(\cos\theta_k\hat{u}_k+\sin\theta_k\hat{l}_k)+\text{H.c.}.
\end{eqnarray}

\emph{Band regime.--}
We consider the spins resonant with only the one side of upper band at the wavevector $k_1$ and $k_2$ in Figs. 2(b,c).
The corresponding decay rates are $\gamma_1=g_\text{eff}^2\cos^2\theta_{k_1}/|v_g^1|$
and $\gamma_2=g_\text{eff}^2\cos^2\theta_{k_2}/|v_g^2|$, relating to
the left ($v_g^1<0$) and right ($v_g^2>0$) propagating acoustic waves,
with $\cos^2\theta_{k_{1,2}}$ the weights of the phononic components of the polaritons in the upper band,
and $v_g^{1,2}=\partial\omega/\partial k|_{\omega=\omega_0,k=k_{1,2}}$ the group velocity.
The condition $\gamma_1>\gamma_2$ is always maintained,
which reflects the chiral coupling between the spins and the acoustic modes in the optomechanical crystal.

For the case of $N$ SiV centers,
the optomechanical crystal-mediated interaction between the spins is realized by the emission
and reabsorption of real polaritons propagating in the crystal, thus inheriting the chiral properties of the emission.
However, in this system the exchange of virtual excitations between spins is possible even in the dissipative regime,
due to the interplay with the band-edges on the other side of the Brillouin zone.
Together with the band- and band-edge-induced interactions, the spin dissipative dynamics is quasi-chiral.
Unlike the unidirectional interaction exploiting helical topological edge states \cite{Barik666,JalaliMehrabad:20}, here the quasi-chiral interaction results from the breakdown of the time-reversal
symmetry in the topologically trivial acoustic band structure.

\begin{figure}
\includegraphics[scale=0.18]{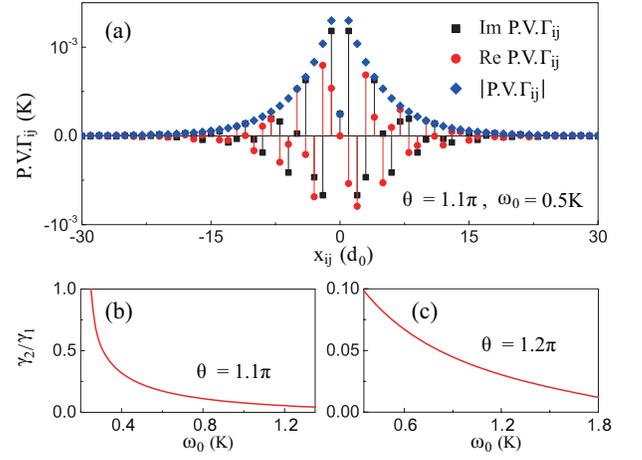}
\caption{\label{fig4}(Color online) (a) Cauchy's principal value of $\Gamma_{ij}$
versus position $x_{ij}$ with $\omega_0=0.5K$, and $\theta=1.1\pi$.
(b,c) The ratio $\gamma_2/\gamma_1$ of decay into the right and left propagating modes of the crystal
versus the spins frequency $\omega_0$.}
\end{figure}

To gain more insight into  this regime, we consider the Markovian approximation,
where the degrees of freedom of the optomechanical crystal can be adiabatically eliminated.
In this case,
the effective motion equation describing the spins' dynamics has the form
$d\hat{\rho}_s/dt=\sum_{i,j}\Gamma_{ij}(\hat{\sigma}_-^i\hat{\rho}_s\hat{\sigma}_+^j-\hat{\sigma}_+^j\hat{\sigma}_-^i\hat{\rho}_s)+\text{H.c.}$
\cite{breuer2002theory},
where $\hat{\rho}_s$ is the reduced density matrix for the spins, $\Gamma_{ij}$ is the optomechanical crystal-mediated interaction, $2\text{Re}(\Gamma_{ii})=\gamma_1+\gamma_2$
is the decay rate into the waveguide modes, and $\text{Im}(\Gamma_{ii})$ is the Lamb shift.
In this system, the collective decay rates have the following expression
\cite{SupplementalMaterial}
\begin{eqnarray}\label{ME6}
\Gamma_{ij}=\lim_{s\rightarrow0^+}\frac{g_\text{eff}^2}{2\pi}\!\!\!\int_{-\pi}^{\pi}\!\!\!\!dke^{ikx_{ij}}\!\!\!&\Big(&\!\!\!\frac{\cos^2\theta_k}{s-i(\omega_0-\omega_u(k))}\notag\\
&+&\frac{\sin^2\theta_k}{s-i(\omega_0-\omega_l(k))}\Big),
\end{eqnarray}
with $x_{ij}=x_j-x_i$ the distance between two distant spins.
We can divide $\Gamma_{ij}$ into three parts as
\begin{eqnarray}
\Gamma_{ij}&=&\text{P.V.}\Gamma_{ij}+\gamma_1e^{ik_1x_{ij}}\Theta(x_{ij}/v_g^1)+\gamma_2e^{ik_2x_{ij}}\Theta(x_{ij}/v_g^2),\nonumber
\end{eqnarray}
with \text{P.V.} being the Cauchy's principal value, and $\Theta$ the Heaviside function defined such that $\Theta(0)=1/2$.
The second and third parts describe the dynamics dominated by the resonant $k$-modes
that the polariton is emitted by spin $i$ and then is recaptured by spin $j$,
when $x_{ij}<0$ ($k_1$-mode) and $x_{ij}>0$ ($k_2$-mode), respectively.
The probability of these two processes are different, i.e., $\gamma_1\neq\gamma_2$, similar with
chiral quantum optics
\cite{PhysRevLett.113.237203,PhysRevA.91.042116,PhysRevA.93.062104}.
In order to reveal the physics behind the Cauchy's principal value,
we numerically plot $\text{P.V.}\Gamma_{ij}$ as a function of $x_{ij}$ in Fig.~3(a).
We find that it accounts for the long-range interaction induced by the band-edges,
with a strength $\max\{|g_\text{edges}^{ij}|\}<\gamma_1\sim g_\text{eff}^2/2K$,
and localized within $-15\lesssim x_{ij}\lesssim15$.
Beyond this range, the coupling between the spin and the optomechanical crystal is completely chiral.

To gain insight into this chirality,
we plot $\gamma_2/\gamma_1$ versus the spin's frequency throughout the asymmetric area
with different values of the phase gradient in Fig.~3(b,c).
For $\theta=1.1\pi$, we find $\gamma_2/\gamma_1\rightarrow1$ when $\omega_0$ is close to the band-edge.
Away from the band-edge, the ratio decreases very quickly.
The chirality can be tuned from bidirectional to completely chiral.
When $\theta=1.2\pi$, we show that $\gamma_2/\gamma_1\ll1$ over the frequency range
and in particular, we suggest a quasi-unidirectional spin-phonon coupling
when $\omega_0\gtrsim1.5K$, where $\gamma_2/\gamma_1\lesssim0.02$.
Actually, for the quasi-unidirectional channel in this case, $\sin^2\theta_{k_1}<0.01$ can be easily satisfied,
such that the polariton is almost phonon-like; which is more robust against the loss associated with optical decay.

\emph{Bandgap regime.--}
We now consider the situation where the middle bandgap is opened
and the spins frequency lies within this forbidden area for propagating photons and phonons.
In this case, there exists an exotic bound state formed by single spins and polariton excitions in the single-excitation subspace.
The bound state can be obtained by solving the secular equation
$\hat{H}|\psi\rangle=E_{BS}|\psi\rangle$
\cite{PhysRevA.78.063827,PhysRevA.81.042304},
with $\hat{H}=\hat{H}_{\text{OM}}+\hat{H}_{\text{free}}+\hat{H}_{\text{int}}$,
and the general form of the bound state
\begin{eqnarray}\label{ME7}
|\psi\rangle=(C_e\hat{\sigma}_{+}+\sum_{j=1}^{N}\sum_{\beta=a,b}C_{j,\beta}\beta_j^\dag)|g\rangle|\text{vac}\rangle,
\end{eqnarray}
where $|\text{vac}\rangle$ denotes the vacuum state of the 1D bath.
The coefficients $|C_e|^2$, $|C_{j,a}|^2$ and $|C_{j,b}|^2$ are  the probabilities for
finding the excitations in the spin excited state, photonic and phononic bound states at the $j$th cell,
when the spin is coupled to the acoustic cavity at the $j=0$ cell.

\begin{figure}
\includegraphics[scale=0.18]{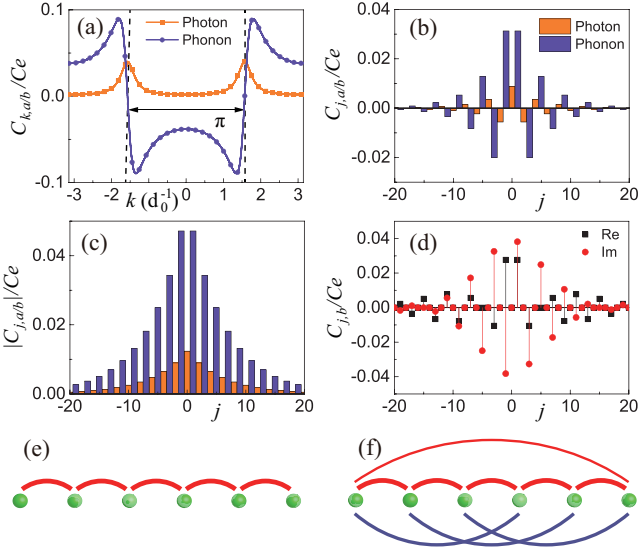}
\caption{\label{fig3}(Color online) (a,b)
Wave function distribution both in momentum $C_{k,a/b}$ and real space $C_{j,a/b}$ with $\theta=\pi$.
(c) Modulus of the wave distribution in real space $|C_{j,a/b}|$.
(d)  The real and imaginary parts of the wave function distribution for the acoustic phonons,
$\text{Re}(C_{j,b})$ and $\text{Im}(C_{j,b})$. $\theta=1.1\pi$ in (c,d).
(e,f) Long-range, tunable and odd-neighbor interactions in an array of spins,
with short (e) and long (f) localization length.
Here, $g_{\text{eff}}=0.08K$.}
\end{figure}

In Fig.~4, we plot the wave function distribution both in momentum and real space with $E_\text{BS}=0$.
First, we consider the symmetric case with the maximal value of the middle bandgap ($\theta=\pi$).
Figure~4(a) shows that the photon and phonon distributions in $k$-space are symmetric ($k\rightarrow-k$),
and more importantly, the conditions $C_{k,a}=C_{k+\pi,a}$ and $C_{k,b}=-C_{k+\pi,b}$ are satisfied,
which is crucial to the enhancement and suppression of the
wave function in certain neighbor sites in real space.
In real space, we observe several features of the bound state, as shown in Figs.~4(b,c,d).
First, the bound state is localized in the vicinity of the spin with an exponential envelope,
which is consistent with previous studies on band gap-induced bound states in optical (acoustic) lattices
\cite{PhysRevA.93.033833,BandPeng}.
The localization length is tunable when changing the system parameters such as the optomechanical coupling strength $G$.

The distinctive feature here is that the
photon component only appears in the even cavities while the
phonon component only appears in the odd cavities, with respect to the cavity to which the spin is coupled.
That is, the photon and phonon components of the bound state are staggered.
Also the extra factors $(-1)^{j/2}$ and $(-1)^{(j-1)/2}$ determine the sign of $C_{a,j}$ and $C_{b,j}$.
Moreover, the hybrid bound state obviously has a less photonic component,
which makes this system more resilient to the dissipation associated with photon decay.
For the asymmetric case ($\theta=1.1\pi$), the above features persist and
another distinctive feature appears: the coefficients  $C_{j,a}$ and $C_{j,b}$
are no longer real numbers but tunable complex numbers as a result of the asymmetric band structure,
i.e., these describe a bound state with a tunable phase.

Now we consider the situation involving two or multiple SiV centers.
When the bandgap is much larger than the spin-phonon coupling strength, (i.e.,
within the Markovian approximation), the effective spin interactions can be described as
\begin{eqnarray}\label{ME8}
\hat{H}_s=\sum_{i<j}(g_{ij}\hat{\sigma}_+^j\hat{\sigma}_-^i+g_{ij}^*\hat{\sigma}_+^i\hat{\sigma}_-^j),
\end{eqnarray}
which can be harnessed to simulate various quantum spin models
\cite{Quantum2015Douglas,Subwavelength2015AGT}.
Since spins are essentially coupled to phonon modes \cite{PhysRevLett.75.553},
the spin-spin coupling is only mediated by the phononic component of the bound state,
with $g_{ij}=g_\text{eff}\,C_{j-i,b}/C_e$.
From this expression, the spin-spin interactions can have a pattern similar to Fig.~4(d).
The inheritance of the features from the bound state consequently leads to
long-range tunable, odd-neighbor and complex interactions between spins, which is
still in the strong-coupling regime ($g_{ij}\gg\kappa_C\sum_j|C_{j,a}|^2$).
This type of interaction is shown schematically in Fig.~4(e)
with a short localization length involving only nearest-neighbor spins,
and in Fig.~4(f) with a longer localization length involving far away odd-neighbor spins.
To examine the validity of the Markovian approximation for deriving the spin interactions in Eq.~(\ref{ME8}), we perform numerical
simulations of the non-Markovian dynamics in the spin-optomechanical array system, and find that
the spin interactions can be well described by Eq.~(\ref{ME8})
(see Ref.~\cite{SupplementalMaterial} for details).
Since the method for producing this type of spin-spin interaction is general, these spin interactions could be realized in other systems,
such as
superconducting circuits \cite{PhysRevB.87.134504,Chiral2017Roushan,Superconducting2018Mirhosseini},
or photonic crystal platforms \cite{Realizing2012Fang,PhysRevB.86.195312,RevModPhys.87.347}.

\emph{Experimental feasibility.--} For actual implementations, we consider the
1D diamond optomechanical and photonic crystals with
embedded individual color centers, as experimentally
demonstrated in Refs.~\cite{Sipahigil847,Evans662}.
For the setups illustrated in Fig. 1,
the diamond optomechanical crystal has mechanical frequencies about
$2\pi\times46$ GHz (or few GHz)
and optical modes around $2\pi\times200$ THz \cite{Burek:16}.
The programmable hopping rate for photons and phonons are $J \sim 20K$ and $K/2\pi\sim 50$ MHz respectively
\cite{Generalized2017Fang,PhysRevX.5.031011,Mirhosseini2020Mirhosseini,PhysRevX.8.041027}.
In diamond nanostructures \cite{Burek:16}, the strong optomechanical coupling $G/2\pi\sim100$ MHz is realizable
\cite{PhysRevLett.112.153603}.
Also, lasers with position-dependent phase can be implemented on-chip by several methods
\cite{Thermal2018Seif,PhysRevLett.124.083601,Measurement2016Mittal,Generalized2017Fang}.
With a high-quality factor $Q\sim10^7$,
the optical and acoustic decay rates are $\kappa_C/2\pi\sim20$ MHz and $\kappa_M/2\pi\sim4.6$ kHz,
leading to low waveguide losses in the high-cooperativity regime ($G^2/\kappa_C\kappa_M\gg1$).
By targeted ion implantation, SiV centers can be accurately implanted into the diamond crystal.
At mK temperatures, the thermal excitations vanish and
the intrinsic dephasing rate of the SiV centers is $\gamma_s/2\pi\sim100$ Hz
\cite{PhysRevLett.119.223602}.
The strain-induced spin-phonon coupling strength can be calculated as $g_k/2\pi\sim30$ MHz
\cite{PhysRevLett.120.213603}.
When choosing $\delta/2\pi\sim450$ MHz and $\Omega/2\pi\sim60$ MHz
or an appropriate off-axis magnetic field \cite{PhysRevB.97.205444},
we have $g_\text{eff}/2\pi\sim 4$ MHz, $g_{12}/2\pi\sim150$ kHz and $\gamma_1/2\pi\sim200$ kHz.
In general, our system is still in the strong coupling regime.

\emph{Conclusion.--}
We have studied unconventional quantum sound-matter interactions in a  tunable solid-state device
with a combination of optomechanical crystals and SiV centers.
We predict the emergence of quasi-chiral sound-matter interactions for the case of spins resonantly coupled to the band.
We also show an exotic bound state with staggered photon and phonon components  in the bandgap regime,
which can be utilized to mediate long-range tunable, odd-neighbor and complex spin-spin interactions.
The work may be extended to higher dimensions or to other solid-state setups such as superconducting circuits.

\begin{acknowledgments}
We gratefully acknowledge the use of the open source
Python numerical packages QuTiP
\cite{JOHANSSON20121760,JOHANSSON20131234}.
P.B.L. is supported by the National Natural Science
Foundation of China under Grants No. 92065105 and No. 11774285 and Natural Science Basic Research Program of Shaanxi (Program No.
2020JC-02). F.N. is supported in part by:
Nippon Telegraph and Telephone Corporation (NTT) Research,
the Japan Science and Technology Agency (JST) [via
the Quantum Leap Flagship Program (Q-LEAP) program,
the Moonshot R\&D Grant Number JPMJMS2061, and
the Centers of Research Excellence in Science and Technology (CREST) Grant No. JPMJCR1676],
the Japan Society for the Promotion of Science (JSPS)
[via the Grants-in-Aid for Scientific Research (KAKENHI) Grant No. JP20H00134 and the
JSPS-RFBR Grant No. JPJSBP120194828],
the Army Research Office (ARO) (Grant No. W911NF-18-1-0358),
the Asian Office of Aerospace Research and Development (AOARD) (via Grant No. FA2386-20-1-4069), and
the Foundational Questions Institute Fund (FQXi) via Grant No. FQXi-IAF19-06.

\end{acknowledgments}


%

\onecolumngrid

\appendix

\clearpage

\section*{Supplemental Material:  }


\setcounter{equation}{0}
\setcounter{figure}{0}
\setcounter{table}{0}
\setcounter{page}{1}
\makeatletter
\renewcommand{\theequation}{S\arabic{equation}}
\renewcommand{\thefigure}{S\arabic{figure}}
\renewcommand{\bibnumfmt}[1]{[S#1]}
\renewcommand{\citenumfont}[1]{S#1}
\begin{quote}
In this Supplemental Material, we first present more details on the optomechanical systems,
including band properties versus related parameters, an accurate full-wave simulation for the optomechanical system,
and the realization of lasers with position-dependent phases.
The disorders in the optical and mechanical frequencies are considered as well.
Second, we discuss the properties of SiV centers and their strain coupling to the acoustic modes in optomechanical crystals.
Third, we introduce the master equation for the study of the Markovian dynamics of the system.
Fourth, we take a discussion of quasi-chiral sound-matter interactions and one of the applications
in quantum information processing, i.e., entangled state preparation.
Finally, we present more details on the photon-phonon bound states and odd-neighbor spin-spin interactions.
In particular, we consider the bound states and exact dynamics of spin interactions in a finite optomechanical array.
\end{quote}

\section{optomechanical crystals}

\renewcommand{\theequation}{S\arabic{equation}}
\renewcommand{\thefigure}{S\arabic{figure}}

\subsection{Properties of the band structure}

The optomechanical Hamiltonian in the main text ($\hbar=1$) is given by
\begin{eqnarray}\label{ME1}
\hat{H}_{\text{OM}}=\Delta/2\sum_{n}\hat{a}_n^\dag\hat{a}_n+\omega_M/2\sum_n\hat{b}_n^\dag\hat{b}_n-G\sum_n e^{-in\theta}\hat{a}_n^\dag\hat{b}_n-J\sum_{n}\hat{a}_{n+1}^\dag\hat{a}_{n}-K\sum_{n}\hat{b}_{n+1}^\dag\hat{b}_{n}+\text{H.c.}.
\end{eqnarray}
In the Fourier basis, it has the form of
\begin{eqnarray}\label{ME2}
\hat{H}(k)=
{\left( \begin{array}{ccc}
-2J\cos(k-\theta) & -G\\
-G & -2K\cos(k)
\end{array}\right )},
\end{eqnarray}
which shows a phonon with momentum $k$ is coupled to a photon with momentum $k-\theta$.
The eigenmodes are polaritons composed of photons and phonons.
We diagonalize the Hamiltonian by a unitary transformation
\begin{eqnarray}\label{ME3}
P_k=
{\left( \begin{array}{ccc}
-\sin\theta_k & \cos\theta_k\\
\cos\theta_k & \sin\theta_k
\end{array}\right )},
\end{eqnarray}
with $\sin^2\theta_k=G^2/[G^2+(\omega_{l}(k)+2K\cos(k))^2]$ the weight of photons (phonons) in upper (lower) band, and
$\cos^2\theta_k=G^2/[G^2+(\omega_{u}(k)+2K\cos(k))^2]$ the weight of phonons (photons) in upper (lower) band.
After diagonalization, the dispersions in the first Brillouin zone are
\begin{eqnarray}\label{ME4}
\omega_{u/l}(k)=-J\cos(k-\theta)-K\cos(k)\pm\sqrt{(K\cos(k)-J\cos(k-\theta))^2+G^2}.
\end{eqnarray}
Note that $\omega_u(k\pm\pi)=-\omega_l(k)$ and $\cos^2\theta_{k\pm\pi}=\sin^2\theta_k$.
These can be understood intuitively from the fact that the bare photonic and phononic bands are cosine functions
and $\cos(k\pm\pi)=-\cos(k)$. The parameter $\theta$ breaks the time-reversal symmetry,
that is, $\omega_{u/l}(k)\neq\omega_{u/l}(-k)$.

\begin{figure}[b]
\includegraphics[scale=0.25]{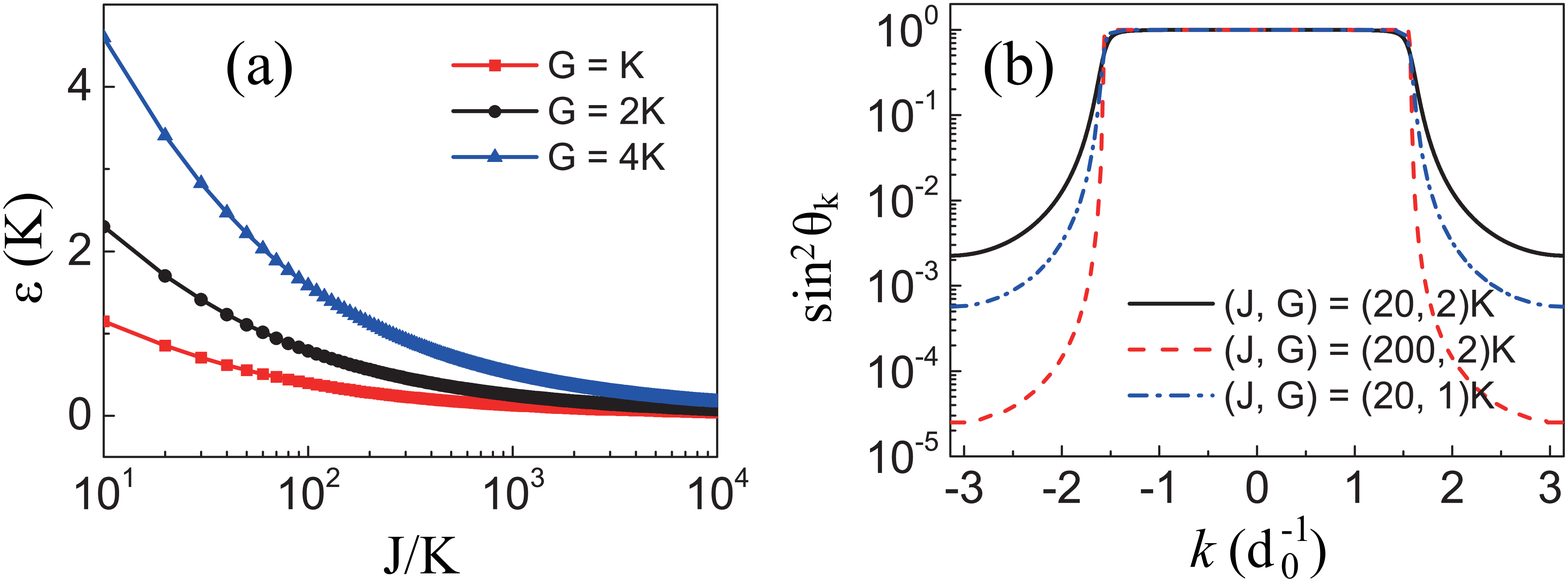}
\caption{\label{S1}(Color online)
(a) The size of the gap $\varepsilon$ versus the ratio $J/K$ with $G=1K$, $2K$ and $4K$.
(b) The weights of photons composing polaritons in the upper band for
$(J, G)=(20, 2)K$ in black solid line, $(J, G)=(200, 2)K$ in red dash line and $(J, G)=(20, 1)K$ in blue dash dot line.
Here, $\theta=\pi$.}
\end{figure}
The structure parameters $K, J$ and laser driving  parameters $G, \theta$
completely determine the size, shape and hybridization properties of the dispersions.
In our scheme, we mainly consider the values of the phase gradient such that the bare mechanical band
is bisected by the bare photon band. Since the effects of $\theta$ are discussed in the main text,
we here focus on the parameters $K, J, G$.
Actually, these parameters can impact  the bandgap and the weights of photons and phonons composing the polaritons.
Without loss of generality, we consider the parameter regime $\theta=\pi$.
We plot the size of the bandgap $\varepsilon$ ($\varepsilon=\varepsilon_1+\varepsilon_2$ and $\varepsilon_1=0$)
opened by the optomechanical interaction versus the ratio $J/K$ with $G=1K, 2K$ and $4K$ in Fig.~S1(a).
The  bandgap decreases when increasing the value of $J/K$ or decreasing the value of $G$.
As a result, the tunability of $\theta$ is reduced and
the system is more sensitive to the disorder which can close the bandgap.
We also plot the weights of photon components of polaritons in the upper band
for different values of $J$ and $G$ in Fig.~S1(b).
However, the result shows that the high purity of phonon-like excitations
requires larger photons hopping rate and smaller optomechanical coupling strength.
As a balance, we choose $J=20K$ and $G=2K$ in the main text such that the bandgap is large enough
and $\sin^2\theta_k<0.01$ can be easily satisfied.

\begin{figure}
\includegraphics[scale=0.25]{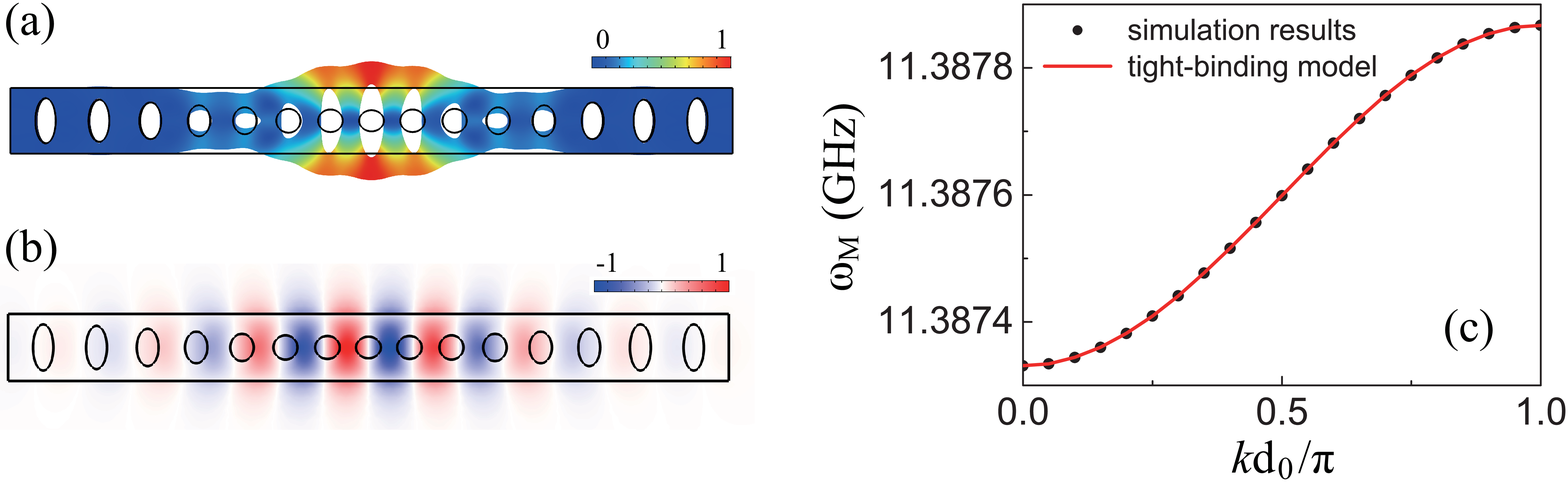}
\caption{\label{S2}(Color online) Finite-element (FEM) simulations of a single diamond optomechanical nanocavity,
which co-localizes (a) a phononic mode (displacement field $|Q|/|Q_{max}|$) and (b)
an optical mode ($y$ component of electric field $E_y/|E_{y\;max}|$).
The frequencies are $\omega_M/2\pi\sim11.4$ GHz and $\omega_c/2\pi\sim250$ THz.
(c) The phononic band close to the localized mode.
The dots indicate the simulation results and the solid line is from the tight-binding model with the expression of
$\omega_M(k)/2\pi=11.387599-0.000268\times\cos(kd_0)$ GHz.}
\end{figure}

\subsection{Implementation of the optomechanical Hamiltonian}

In this section, we simply discuss the implementation of the optomechanical system,
where we consider an array of co-localized phononic and optical cavities formed
in a diamond nanobeam and a tight-binding model is assumed for simplicity.
In addition, a position-dependent phase is introduced for each optomechanical interaction
through writing gradient phase in lasers.
Using the finite-element (FEM) simulation package COMSOL,
we present a  full-wave simulation for the optomechanical system in Fig.~S2,
with parameters of the nanocavity taken from related experimental work
\cite{doi:10.1063/1.4747726,MacCabe840}.
In Fig.~S2(a,b), we show the distribution of the displacement field $|Q|$ for localized mechanical modes
and the $y$ component of the electric field $E_y$ for localized optical modes in a unit cell (defect nanocavity).
The frequencies are around $2\pi\times11.4$ GHz for mechanical breathing modes and $2\pi\times250$ THz for optical modes.
In Fig.~S2(c), we compare the phononic dispersion near the localized mode
with the result predicted from the tight-binding model and find a good agreement,
which indicates the validity of the tight-binding approximation.

The position-dependent phase in laser drive can be implemented on chip.
Here, we follow closely the supplementary materials in Ref.~\cite{Thermal2018Seif},
which introduces a $1\times N$ multi-mode interferometer to realize it.
We use the $1\times N$ multi-mode interferometer to divide the power equally to $N$ optical fibers,
which are evanescently coupled to $N$ optomechanical nanocavities.
The first method is to vary the length of the fibers such that the
light propagating in different waveguides experiences different distances, thus acquiring a phase gradient.
The second one is to use heated zero-loss resonators as all-pass filters to pick up
phases which are related to resonators' resonance frequency tuned by temperature.

\begin{figure}
\includegraphics[scale=0.23]{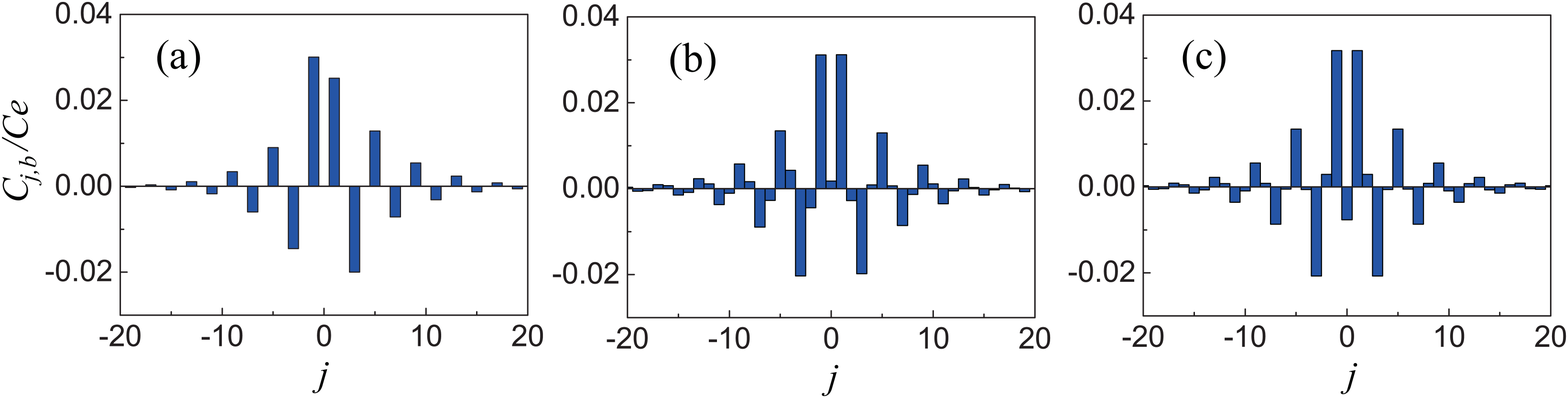}
\caption{\label{S3}(Color online)Wave function distribution of phonon bound state $C_{j,b}$
in real space with $\theta=\pi$ and $E_\text{BS}=0$.
(a) Off-diagonal disorder in both optical and mechanical frequencies with strength $W_C=0.5J$ and $W_M=0.5K$.
(b) On-site disorder in optical frequencies with strength $W_C=0.5J$. Here, $J=20K$, $G=2K$ and $g_\text{eff}=0.08K$.
(c) $E_\text{BS}=0.1\varepsilon$.}
\end{figure}

\subsection{Disorder}

We now consider the effects of the disorder on both the optical and mechanical frequencies,
which arises from the fabrication imperfection and is  the main experimental obstacle in optomechanical lattices.
The on-site disorder can cause an additional localization, i.e., Anderson localization.
As predicted in perturbation theory, the 1D chain with random disorders distributed in the interval $[-W/2,W/2]$
can have a localization length about $100J_c^2/W^2$
\cite{opto2013michael}, where $J_c$ is the hopping rate and $W$ is the disorder strength.
For the small disorder strength such that $W\lesssim J_c$, the localization length is hundred of sites,
which can be much larger than the size of the array.

Moreover, we expect that if the disorder is large enough
the focusing areas (gap and asymmetric area) will be smeared away.
We roughly estimate the critical disorder strength by considering the two limits that
 all optical (acoustic) cavities have a frequency offset of $\pm W_C/2$ ($\pm W_M/2$).  This will makes
the photonic (phononic) band and consequently the focusing area be shifted up and down.
Up to the specific value of $W_C$ ($W_M$), the focusing areas of these two limiting cases have no overlap.
This indicates that  the focusing area starts to close in the presence of the disorder.
We find this occurs when $W_C\sim J$ ($W_M\sim K$) in the case of $\varepsilon\sim K$, in line with the discussion in Ref.~\cite{Lemonde_2019}, where a topological acoustic bandgap is opened by optomechanical interactions.

As a example, we numerically give the bound state $(E_\text{BS}=0)$ in the presence of off-diagonal and on-site disorder,
which are shown in Fig.~S3(a,b). We consider off-diagonal and on-site disorder by adding random terms
$\sum_{n,\beta=a,b}(\xi_{n,\beta}\hat{\beta}_n^\dag\hat{\beta}_{n+1}+\text{H.c.})$ and $\sum_{n}\xi_{n}\hat{a}_n^\dag\hat{a}_n$ to Hamiltonian respectively.
The disorder strength is chosen in the range $\xi\in[-W/2, W/2]$,
with $W_C=J/2$ for optical frequencies and $W_M=K/2$ for mechanical frequencies.
In particular, we show that the feature of alternating photon and phonon components of
the bound state is robust against off-diagonal disorder.
The numerical results are consistent with the analysis above.

In general, about $10^{-6}$ precision for optical frequencies
and $10^{-3}$ precision for mechanical frequencies are enough to neglect the effects of the disorder,
under the parameters considered in our system.
Though it's challenge to fabricate such high-precision optical cavity array in experiment,
the previous work have demonstrated that through some approaches such as post-fabrication fine-tuning techniques
\cite{Sokolov:17,Sumetsky:12},
the required levels of accuracy can be reached.

\section{SiV centers}

SiV center's level structure and its strain coupling to phononic modes
in diamond crystal have been discussed in the previous work
\cite{PhysRevLett.120.213603}.
Here, for completeness, we make a simple discussion.
SiV centers are point defects in diamond with a silicon atom
lying in between two adjacent vacancies, whose electronic ground state is an unpaired hole of spin $S=1/2$.
In the present of spin-orbit coupling, the ground state is split into
two branches with a gap of $\Delta_{\text{SiV}}/2\pi\sim46$ GHz.
By further applying a magnetic field, each branches splits into two doublets,
labeled as $\{|g\rangle=|e_-\downarrow\rangle, |e\rangle=|e_+\uparrow\rangle\}$
and $\{|f\rangle=|e_+\downarrow\rangle, |d\rangle=|e_-\uparrow\rangle\}$, as shown in Fig.~1.
Here, $|e_\pm\rangle$ are eigenstates of the orbital angular momentum operator
$\hat{L}_z|e_\pm\rangle=\pm|e_\pm\rangle$.
We take the indirect coupling scheme via a Raman process for example.
We drive SiV centers by microwave fields with the frequency $\omega_L$ and the pump strength $\Omega$.
Thus, the Hamiltonian of single SiV centers is obtained
\begin{eqnarray}\label{ME5}
\hat{H}_{\text{SiV}}=\omega_B|e\rangle\langle e|+\Delta_\text{SiV}|f\rangle\langle f|+(\omega_B+\Delta_\text{SiV})|d\rangle\langle d|+[\Omega e^{i\omega_Lt}(|g\rangle\langle d|+|e\rangle\langle f|)+\text{H.c.}],
\end{eqnarray}
where $\omega_B$ is the Zeeman energy.

On the other hand, local lattice distortions associated with internal compression modes of the optomechanical
crystal affect the defect’s electronic structure, which induces a strain coupling between these phonons and the orbital
degrees of freedom of the center. The SiV-phonon coupling Hamiltonian is
\begin{eqnarray}\label{ME6}
\hat{H}_{\text{strain}}=g_k(|g\rangle\langle f|+|e\rangle\langle d|)\hat{b}+\text{H.c.},
\end{eqnarray}
with $\hat{b}$ the annihilation operator for the acoustic mode and $g_k$ the coupling strength, which can be expressed as
\begin{eqnarray}\label{ME7}
g_k=\frac{d}{v}\sqrt{\frac{\hbar\omega_k}{2\rho V}}\xi(\vec{r}_\text{SiV}).
\end{eqnarray}
Here, $d/2\pi\sim1$ PHz is the strain sensitivity, $v\sim10^4$ m/s is the group velocity of acoustic wave in diamonds,
$\omega_k\sim\Delta_\text{SiV}$, $\rho\sim3500$ kg/$\text{m}^3$,
$V$ is the volume of optomechanical nanocavity and $\xi(\vec{r}_\text{SiV})\sim1$
is the strain distribution at SiV centers position.
If we choose the size of cavities with length of several microns and cross section about $100$ nm $\times100$ nm,
the coupling strength can be calculated as $g_k/2\pi\sim30$ MHz.
This calculation is agreed with one in Ref.~\cite{PhysRevLett.120.213603} using finite-element simulations.

When the Zeeman frequency $\omega_B$ is large enough,
only the Raman channel $|g\rangle\rightarrow|f\rangle\rightarrow|e\rangle$ contributes.
By dropping the high frequency oscillation items and the constant items, we can
obtain the free Hamiltonian and effective SiV-phonon coupling Hamiltonian with many centers
\begin{eqnarray}\label{ME8}
\hat{H}_{\text{free}}&=&\omega_{0}/2\sum_{m}(|e\rangle^m\langle e|-|g\rangle^m\langle g|)\\
\hat{H}_{\text{int}}&=&g_{\text{eff}}\sum_{m}(|e\rangle^m\langle g|\hat{b}_{x_m}+\text{H.c.})\label{ME9},
\end{eqnarray}
where $\omega_0=\omega_B+\omega_L$ is the transition frequency of single spins,
$g_{\text{eff}}=g_k\Omega/(\Delta_\text{SiV}-\omega_0)$ is the effective spin-phonon coupling strength
and $x_m$ is the position of phonon cavity the $m$th spin coupled to.
Lastly, we switch the effective interaction to the reciprocal space,
in terms of polariton operators $\hat{u}_k$ and $\hat{l}_k$, it reads
\begin{eqnarray}\label{ME10}
\hat{H}_{\text{int}}=\frac{g_{\text{eff}}}{\sqrt{N}}\sum_{k,m} \hat{\sigma}_{+}^me^{ikx_m}(\cos\theta_k\hat{u}_k+\sin\theta_k\hat{l}_k)+\text{H.c.},
\end{eqnarray}
with $\hat{\sigma}^m_+=|e\rangle^m\langle g|$ the Pauli operator.

\section{Master equation}

Before the access to specific regimes for the study of sound-matter interactions,
we first introduce an outstanding approach in quantum optics, i.e., master equation approach.
This method allows us exploring the Markovian dynamics of system,
where the degree of freedom of bath can be traced out
\cite{breuer2002theory}
\begin{eqnarray}\label{ME11}
\frac{d\hat{\rho}_s}{dt}=\sum_{i,j}\Gamma_{ij}(\hat{\sigma}_-^i\hat{\rho}_s\hat{\sigma}_+^j-\hat{\sigma}_+^j\hat{\sigma}_-^i\hat{\rho}_s)+\text{H.c.},
\end{eqnarray}
with the reservoir-mediated coupling between spins
\begin{eqnarray}\label{ME12}
\Gamma_{ij}&=&\lim_{s\rightarrow0^+}\sum_{k}\sum_{\alpha=u,l}\frac{\langle\text{vac}|\langle g|\sigma_-^j\hat{H}_\text{int}\alpha_k^\dag\alpha_k\hat{H}_\text{int}\sigma_+^i|g\rangle|\text{vac}\rangle}{s-i(\omega_0-\omega_\alpha(k))}\notag\\
&=&\lim_{s\rightarrow0^+}\frac{g_\text{eff}^2}{2\pi}\!\!\!\int_{-\pi}^{\pi}\!\!\!\!dke^{ikx_{ij}}\big(\frac{\cos^2\theta_k}{s-i(\omega_0-\omega_u(k))}
+\frac{\sin^2\theta_k}{s-i(\omega_0-\omega_l(k))}\big).
\end{eqnarray}
By defining $\Gamma_{ij}=\gamma_{ij}+iJ_{ij}$, with
\begin{eqnarray}\label{ME13}
\gamma_{ij}&=&\frac{1}{2}(\Gamma_{ij}+\Gamma_{ji}^*)\notag\\
J_{ij}&=&\frac{1}{2i}(\Gamma_{ij}-\Gamma_{ji}^*),
\end{eqnarray}
we arrive to the following expression with a separated coherent and incoherent parts
\begin{eqnarray}\label{ME14}
\frac{d\hat{\rho}_s}{dt}=-i\sum_{i,j}J_{ij}[\hat\sigma_+^j\hat\sigma_-^i,\hat{\rho}_s]
+\sum_{i,j}\gamma_{ij}(2\hat\sigma_-^i\hat{\rho}_s\hat\sigma_+^j-\hat\sigma_+^j\hat\sigma_-^i\hat{\rho}_s-\hat{\rho}_s\hat\sigma_+^j\hat\sigma_-^i).
\end{eqnarray}

\section{band regime}
\subsection{Quasi-chiral spin dynamics}

\begin{figure}
\includegraphics[scale=0.5]{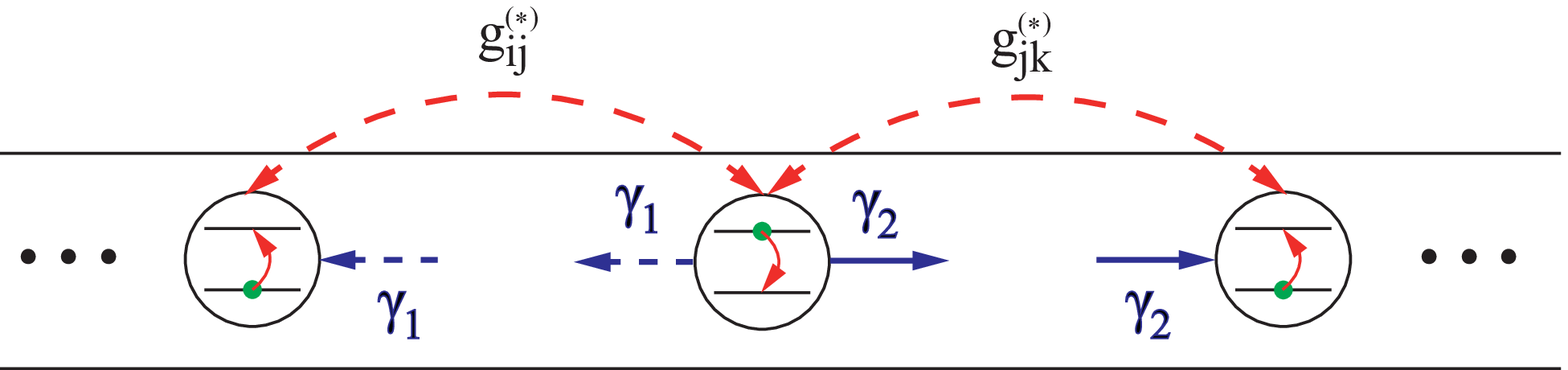}
\caption{\label{S4}(Color online)
Schematic of resonant sound-matter interactions in the optomechanical crystal.
The resonant spin-bath interactions are chiral, which is reflect by $\gamma_1\neq\gamma_2$,
with $\gamma_1$ and $\gamma_2$ the decay rate into the left- and right-moving reservoir modes.
The band-edge-induced long-range interactions are also given.}
\end{figure}

\begin{figure}
\includegraphics[scale=0.25]{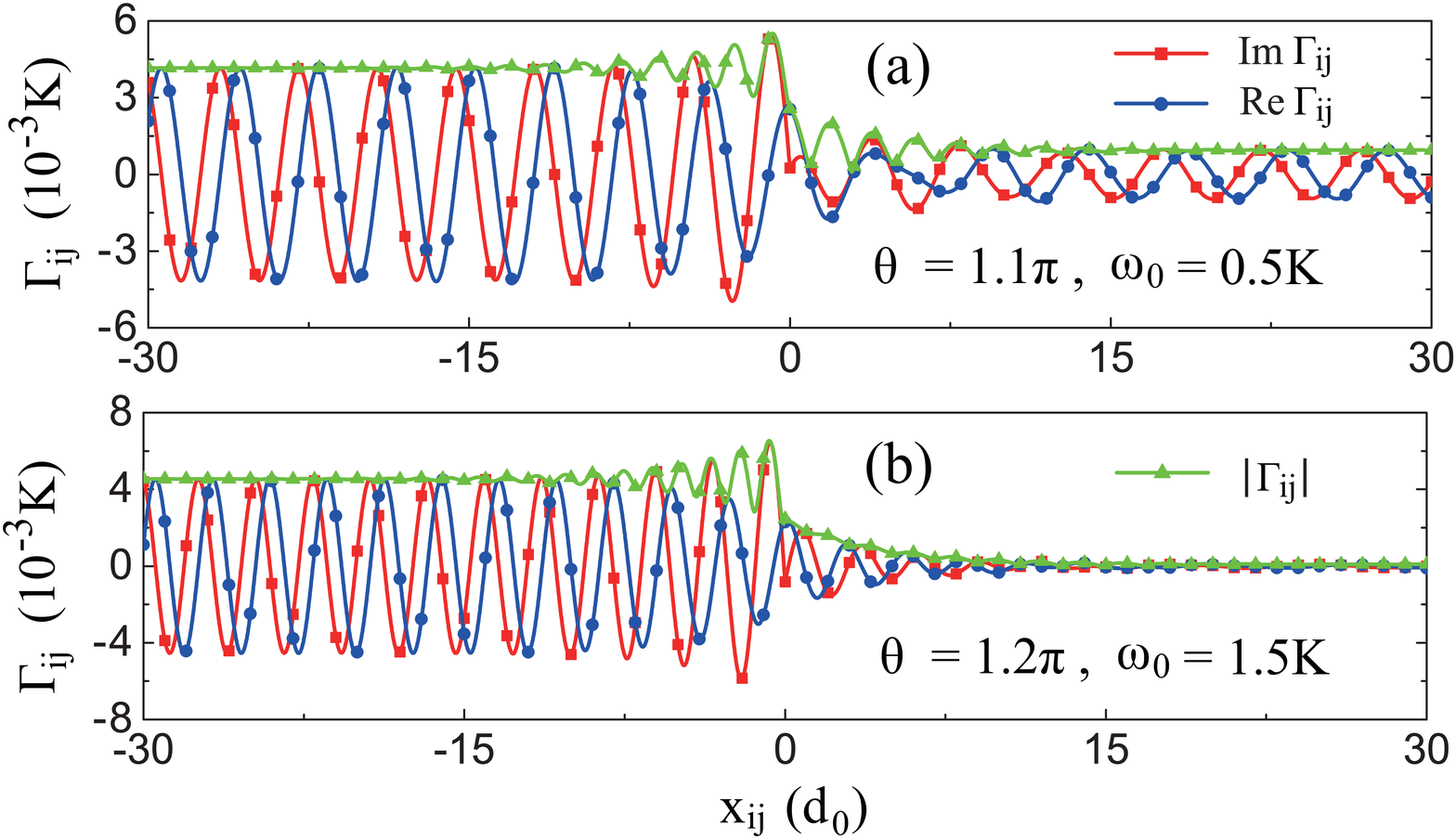}
\caption{\label{S5}(Color online) Collective coupling $\Gamma_{ij}$ as a function of
position $x_{ij}$ with (a) $\omega_0=0.5K$, $\theta=1.1\pi$ and (b) $\omega_0=1.5K$, $\theta=1.2\pi$.
In which, the symbols represent the values at lattice position.
The parameters are $g_\text{eff}=0.08K$, $J=20K$ and $G=2K$.}
\end{figure}

When the spins' frequency lies within the asymmetric area $\varepsilon_1$ in the main text,
the dynamics of spins is dominated by band-edges as well as two resonant k-modes with opposite group velocity,
leading to quasi-chiral sound-matter interactions.
The chiral spin-spin interactions arise from the chiral spin-bath interaction
that the possibility of spin decay into the left- and right-moving reservoir modes is unequal, which is shown in Fig.~S4.
To gain analytical intuition of this regime, we take the limit of weak spin-phonon coupling.
The dynamics is determined by Eq.~(\ref{ME11}).
In contrast to the main text, where we divide $\Gamma_{ij}$ into three parts,
we now calculate $\Gamma_{ij}$ directly.
The Eq.~(\ref{ME12}) can be unfolded as
\begin{eqnarray}\label{ME15}
\Gamma_{ij}(z)=\frac{ig_\text{eff}^2}{2\pi}\int_{-\pi}^{\pi}dke^{ikx_{ij}}\frac{z+2J\cos(k-\theta)}{z^2+2(J\cos(k-\theta)+t\cos(k))z+4JK\cos(k-\theta)\cos(k)-G^2},
\end{eqnarray}
By means of the change of variable $y\equiv e^{ik}$ for $x_{ij}\geq0$ and $y\equiv e^{-ik}$ for $x_{ij}<0$,
we can solve these integrals. Here, we take $x_{ij}\geq0$ as a example
\begin{eqnarray}\label{ME16}
\Gamma_{ij}(z)=\frac{g_\text{eff}^2}{2\pi} \oint_{|y|=1}dy\frac{y^{x_{ij}}[yz+J(y^2e^{-i\theta}+e^{i\theta})]}{ay^4+by^3+cy^2+dy+e}
=ig_\text{eff}^2\sum_{|y_l|<1}\frac{y_l^{x_{ij}}[y_lz+J(y_l^2e^{-i\theta}+e^{i\theta})]}{4ay_l^3+3by_l^2+2cy_l+d},
\end{eqnarray}
where we apply the Cauchy’s residue theorem and define
$a=e^*=JKe^{-i\theta}$, $b=d^*=z(Je^{-i\theta}+K)$, $c=z^2-G^2+2JK\cos(\theta)$.
The denominator is a quartic equation and the corresponding roots can be solved analytically or numerically.
Finally, taking $z=\omega_0+i0^+$ to the above equation and solving the integral in the interval $x_{ij}<0$,
we can obtain the required value of $\Gamma_{ij}$. We numerically plot $\Gamma_{ij}$
as a function of $x_{ij}$ with $\omega_0=0.5K$ and $\theta=1.1\pi$ in Fig.~S5(a),
and with $\omega_0=1.5K$ and $\theta=1.2\pi$ in Fig.~S5(b).
The function is continuous, where the symbols correspond to the values at lattice position.
We find that, in a limited area $-15\lesssim x_{ij}\lesssim15$, the coupling strength is oscillating
with a damped amplitude in the distance. Beyond this area, we observe the collective decay rates
are oscillating with certain but different amplitudes and periods,
$\Gamma_{ij}\approx\gamma_1e^{ik_1x_{ij}}$ and $\Gamma_{ij}\approx\gamma_2e^{ik_2x_{ij}}$ for respectively
$x_{ij}\lesssim-15$ and $x_{ij}\gtrsim15$.
Here, $\gamma_1=g_\text{eff}^2\cos^2\theta_{k_1}/|v_g^1|$
and $\gamma_2=g_\text{eff}^2\cos^2\theta_{k_2}/|v_g^2|$, relating to
the left ($v_g^1<0$) and right ($v_g^2>0$) propagating acoustic waves,
with $\cos^2\theta_{k_{1,2}}$ the weights of phononic components of the polaritons in the upper band
and $v_g^{1,2}=\partial\omega/\partial k|_{\omega=\omega_0,k=k_{1,2}}$ the group velocity.
As discussed in the main text,
the quasi-chiral interactions contain band-induced interaction part $\sum_{l=1,2}\gamma_le^{ik_lx_{ij}}\Theta(x_{ij}/v_g^l)$,
and band-edge-induced interaction part $\text{P.V.}\Gamma_{ij}\sim Ce^{-|x_{ij}|/\xi}e^{ikx_{ij}}$,
which decays exponentially (see below).
In particular, for the parameters used in Fig.~S5(b), the value of $\gamma_2$ is small enough
such that for $x_{ij}>0$, $|\Gamma_{ij}|\approx|\text{P.V.}\Gamma_{ij}|\sim Ce^{-|x_{ij}|/\xi}$,
i.e.,  the scaling of the self-energy is exponential decay.

Mathematically,  Eq.~(\ref{ME16}) can be rewritten in a more compact form as
$\Gamma_{ij}(z)=\sum_{|y_l|<1}C(y_l)e^{-|x_{ij}|/\xi_l}e^{ik_lx_{ij}}$,
which is a superposition of exponential functions.
When the spin's frequency lies within the band $\omega_0=\omega(k)$, there exist poles in the unit circle
and the integral should be done above the real axis ($z=\omega_0+i0^+$).
It is clear that the contributions from such poles are oscillatory solutions expressed as
$\sum_l\gamma_le^{ik_l(\omega_0)x_{ij}}$ ($|y_l|=e^{-1/\xi_l}=1$), in keeping with that in the 1D standard bath.
However, for band-edge-induced part (analogous to $\omega_0\neq\omega(k)$),
the poles are all inside or outside the integral circle,
making the self-energy decay by summing several exponentials ($\xi_l>0$) \cite{PhysRevA.102.013709}.

\subsection{Entangled state preparation}

The chiral phonons channel can enable quantum state transfer and entangled states preparation.
In this section, we prepare unique entangled steady states via a quasi-unidirectional phonon channel,
following closely in Ref.~\cite{PhysRevA.91.042116}.
Though the quasi-chiral interactions can do this too, it requires a more careful treatment of optical dissipation
(since the optical fraction of the polaritons at $k_2$-mode is large).
We drive every spins by classical fields at a common frequency $\nu$
with the amplitude $\Omega_j$ and the detuning $\delta_j=\omega_{0j}-\nu$,
with $\omega_{0j}$ the transition frequency of the $j$th spin.
Thus, in the rotating frame with respect to the driving frequency $\nu$
and under the rotating wave approximation, the Hamiltonian of spins is transferred to
\begin{eqnarray}\label{ME17}
\hat{H}_{\text{spins}}=\sum_{j=1}^{N_s}(-\delta_j/2\hat{\sigma}_z^j+\Omega_j\hat{\sigma}_{-}^j+\Omega_j^*\hat{\sigma}_{+}^j).
\end{eqnarray}
When neglecting the level shifts and separating the spins by a proper distance
such that $(k_1-k_2)x_{ij}=2\pi n$ ($n$ integer), we can obtain a generalized master equation
\begin{eqnarray}\label{ME18}
\frac{d\hat{\rho}_s}{dt}&=&-i[\hat{H}_\text{edges}+\hat{H}_\text{spins},\hat{\rho}_s]+(\gamma_1+\gamma_2) \sum_{j} D[\hat{\sigma}_-^j]\hat{\rho}_s+\gamma_{s}\sum_jD[\hat{\sigma}_{z}^j]\hat{\rho}_s\notag\\
&+&\gamma_1(1-\eta_1)\sum_{i>j}([\hat{\sigma}_-^i\hat{\rho}_s,\hat{\sigma}_+^j]+[\hat{\sigma}_-^j,\hat{\rho}_s\hat{\sigma}_+^i])
+\gamma_2(1-\eta_2)\sum_{i<j}([\hat{\sigma}_-^i\hat{\rho}_s,\hat{\sigma}_+^j]+[\hat{\sigma}_-^j,\hat{\rho}_s\hat{\sigma}_+^i]),
\end{eqnarray}
with $\gamma_s$ the spins dephasing rate and $D[\hat{O}]\hat{\rho}_s=\hat{O}\hat{\rho}_s\hat{O}^\dag
-\frac{1}{2}\hat{O}^\dag\hat{O}\hat{\rho}_s-\frac{1}{2}\hat{\rho}_s\hat{O}^\dag\hat{O}$ for a given operator $\hat{O}$.
Here, $\hat{H}_\text{edges}=g_s\sum_{j}(\hat{\sigma}_j\hat{\sigma}_{j+1}^\dag+\text{H.c.})$
is the coherent interaction only retained up to the nearest-neighbour.
The factor of $1-\eta_{1,2}$ are to model the propagation losses
in optomechanical waveguide for respectively right- and left-moving excitation,
with $\eta_{1,2}\sim N_p\kappa_C\sin^2\theta_{k_{1,2}}/v_g^{1,2}$
the amplitudes that get loss between the spins, $N_p$ the number of traveled cavities
and $\kappa_C$ the decay rate of the optical cavities.
For the quasi-unidirectional channel in this system, $\sin^2\theta_{k_1}\sim0.003$ can be realized,
leading to a low optical loss $\eta_1\sim0.01$.
Note that if the value of $\gamma_2/\gamma_1$ is small enough such that we can exclude the last term in Eq.~(\ref{ME18}),
the condition $(k_1-k_2)x_{ij}=2\pi n$ is not necessary.

Ideally, the dynamics evolution can be towards a pure entangled steady-state
when the detuning pattern is designed in pairs, for instance,
one with the detuning $\delta_i=\delta_1$ and another with the detuning $\delta_j=-\delta_1$.
Here, we assume the frequency difference is tiny enough such that spins are identical
with regard to reservoir interaction.
In the simplest case of two spins coupled to the bath, the target state has the form of
\cite{Stannigel_2012,PhysRevA.91.042116}
\begin{eqnarray}\label{ME19}
|\psi_{\text{dimer}}\rangle=\frac{1}{\sqrt{1+|\alpha|^2}}(|gg\rangle+\alpha|S\rangle),
\end{eqnarray}
with $\alpha=2\sqrt{2}\Omega_0/[i(\gamma_1-\gamma_2)+2\delta_1]$ and the singlet $|S\rangle=(|eg\rangle-|ge\rangle)/\sqrt{2}$.
The time scale of system to reach the steady state is
\begin{eqnarray}\label{ME20}
\tau=\frac{\pi[(\gamma_1-\gamma_2)^2/4+\delta_1^2+2\Omega_0^2]}{(\gamma_1+\gamma_2)[(\gamma_1-\gamma_2)^2/4+\delta_1^2]}.
\end{eqnarray}
We numerically plot the fidelity using Eq.~(\ref{ME18}) in Fig.~S6(a), where the spin dephasing effect, waveguide loss and band-edge-induced coherent interaction are taken into account. We show the high fidelity is still available.
We note that high fidelity does not mean high concurrence
since high concurrence requires larger pumping strength $\Omega_0$ which,
however, gives rise to longer time to reach the steady state (see Eq.~(\ref{ME20})).
Actually, for the parameters used in Fig.~S6(a), the concurrence of ideal target state is about $0.68$
while the concurrence of the steady state under realistic conditions is about 0.59.

\begin{figure}
\includegraphics[scale=0.25]{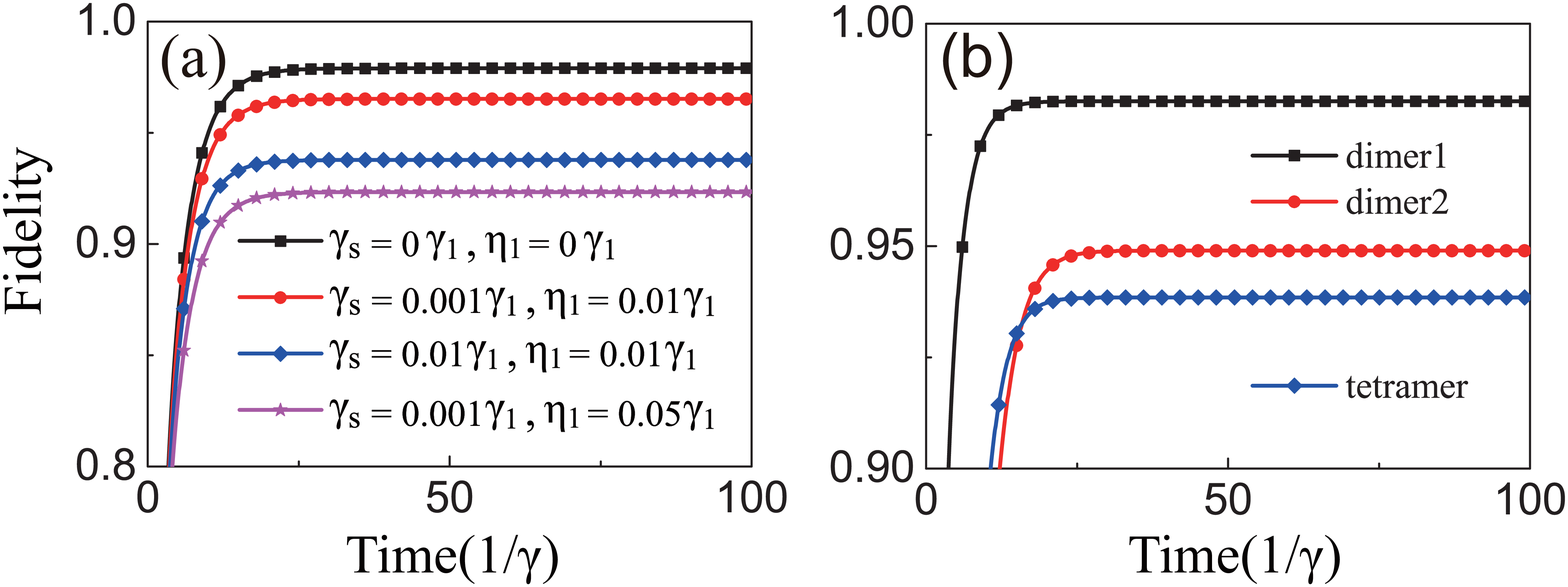}
\caption{\label{S6}(Color online)
(a) Time evolution of the fidelity with different noise strength ($\eta_1,\gamma_s$)
when $(\delta_1,\delta_2)=(0,0)\gamma_1$.
(b) Time evolution of the fidelity for four spins at ($\eta_1,\gamma_s$)=(0.01,0.001)$\gamma_1$.
Two dimers at $(\delta_1,\delta_2,\delta_3,\delta_4)=(0.4,-0.4,0.4,-0.4)\gamma_1$ (the black and red lines).
A tetramer at $(\delta_1,\delta_2,\delta_3,\delta_4)=(0.6,0.4,-0.6,-0.4)\gamma_1$ (the blue line).
Here, $\Omega_j$=0.5$\gamma_1$, $\gamma_2$=0.02$\gamma_1$, $\eta_2=1$, $g_s=0.01\gamma_1$.}
\end{figure}

Further, we study the situation where the number of coupled spins is increased to four.
There are two cases, one is to form two dimers with the detuning profile
$(\delta_1,-\delta_1,\delta_2,-\delta_2)$ and the dark state
$|\psi_\text{dimer}^{(12)}\rangle|\psi_\text{dimer}^{(34)}\rangle$,
the other is to form a four-particle entangled state with the detuning profile
$(\delta_1,\delta_2,-\delta_1,-\delta_2)$ (or $(\delta_1,\delta_2,-\delta_2,-\delta_1)$) and the dark state
\cite{PhysRevA.91.042116}
\begin{eqnarray}\label{ME21}
|\psi_\text{tetramer}\rangle&\propto&|gggg\rangle+a_{12}|S\rangle_{12}|gg\rangle_{34}+a_{34}|S\rangle_{34}|gg\rangle_{12}\notag\\
&+&a_{13}(|S\rangle_{13}|gg\rangle_{24}+|S\rangle_{14}|gg\rangle_{23}+|S\rangle_{23}|gg\rangle_{14}+|S\rangle_{24}|gg\rangle_{13})\notag\\
&+&a_{1234}|S\rangle_{12}|S\rangle_{34}+a_{1324}(|S\rangle_{13}|S\rangle_{24}+|S\rangle_{14}|S\rangle_{23}).
\end{eqnarray}
In terms of $Z\equiv-i(\gamma_1-\gamma_2)/2$, the five coefficients read
\begin{eqnarray}\label{ME22}
a_{12}&=&\frac{-\Omega_0[2Z^2+2\delta_1\delta_2-(Z+\delta_1)(\delta_1+\delta_2)]}{\sqrt{2}(Z-\delta_1)^2(Z-\delta_2)}\\
a_{34}&=&\frac{-\Omega_0(Z-\delta_1+\delta_2)}{\sqrt{2}(Z-\delta_1)(Z-\delta_2)}\label{ME23}\\
a_{13}&=&\frac{\Omega_0(\delta_1+\delta_2)}{2\sqrt{2}(Z-\delta_1)(Z-\delta_2)}\label{ME24}\\
a_{1324}&=&\frac{-2\sqrt{2}\Omega_0a_{13}}{2Z-\delta_1-\delta_2}\\\label{ME25}
a_{1234}&=&\frac{\Omega_0^2(\delta_1+\delta_2-4Z)}{(Z-\delta_1)(Z-\delta_2)(\delta_1+\delta_2-2Z)}\label{ME26}.
\end{eqnarray}
In Fig.~S6(b), we plot the fidelity of these two type of entangled states preparation
and show the high fidelity ($>0.9$) is realizable in this system under realistic conditions.

\section{bandgap regime}

\subsection{Photon-phonon bound state and its robustness}

When spins' frequency lies within the acoustic bandgap,
the bound state can form and exponentially localize in the vicinity of the cavity to which spin is coupled.
The bound state is obtained by solving
$\hat{H}|\psi\rangle=E_{BS}|\psi\rangle$,
with $\hat{H}=\hat{H}_{\text{OM}}+\hat{H}_{\text{free}}+\hat{H}_{\text{int}}$,
and the general form of the bound state in momentum space
\begin{eqnarray}\label{ME27}
|\psi\rangle=(C_e\hat{\sigma}_{+}+\sum_{k}\sum_{\beta=a,b}C_{k,\beta}\beta_k^\dag)|g\rangle|\text{vac}\rangle.
\end{eqnarray}
Here, $|C_e|^2$ is the probability of finding the excitation in spin excited state,
which can be obtained by imposing the normalization condition.
$C_{k,a}$ and $C_{k,b}$ are photon and phonon distributions of bound state in k-space.
The energy of bound state $E_\text{BS}$ can be solved by pole equation $E_{BS}=\omega_0+\Sigma_e(E_{BS})$,
with $\Sigma_e$ the self-energy\cite{doi:10.1002/sca.4950140612}.
After some algebra, we arrive to the following expression
\begin{eqnarray}\label{ME28}
C_{k,a}&=&g_\text{eff}C_e\big(-\frac{\sin\theta_k\cos\theta_k}{E_{\text{BS}}-\omega_u(k)}+\frac{\sin\theta_k\cos\theta_k}{E_{\text{BS}}-\omega_l(k)}\big)\\
C_{k,b}&=&g_\text{eff}C_e\big(\frac{\cos^2\theta_k}{E_{\text{BS}}-\omega_u(k)}+\frac{\sin^2\theta_k}{E_{\text{BS}}-\omega_l(k)}\big)\label{ME29}.
\end{eqnarray}
We focus on the specific situation where the middle bound state forms.
If choosing $E_\text{BS}=0$, the conditions $C_{k,a}=C_{k+\pi,a}$ and $C_{k,b}=-C_{k+\pi,b}$
are satisfied regardless the value of $\theta$ we choose, as discussed before.
The wave function distributions of bound state in real space can be obtained by fourier transform
\begin{eqnarray}\label{ME30}
C_{j,a/b}=\frac{1}{2\pi}\!\!\!\int_{-\pi}^{\pi}\!\!\!\!dke^{ikj}C_{k,a/b}.
\end{eqnarray}
Since there is no singularity, we can integrate it directly.
Intuitively, when $j=2n$ ($n$ integer), $e^{i(k+\pi)j}C_{k+\pi,a}=e^{ikj}C_{k,a}$
and $e^{i(k+\pi)j}C_{k+\pi,b}=-e^{ikj}C_{k,b}$. The situation is opposite when $j=2n+1$.
Thus we obtain the main feature of this bound state that the photon and phonon components are alternating,
i.e., $C_{2n,b}=0$ and $C_{2n+1,a}=0$. Actually, it takes advantage of the feature of cosine function.
Moreover, if the bands are asymmetry when $\theta\neq\pi$,
the distributions in real space can acquire a tunable phase
(as a result of the asymmetric band structure), which can also be understood from
the formula ($C_{k,a/b}\neq\pm C_{-k,a/b}$).

We also consider the robustness of this exotic bound state by plotting the phonon component of bound state.
First, for varying spins' frequency around the optimal value, the bound state is still robust when $\omega_0\lesssim0.1\varepsilon$, as shown in Fig.~S3(c).
Second, the off-diagonal and on-site disorder are considered, which are shown in Fig.~S3(a) and S3(b).
We show that the middle bound state is robust against two type of disorders and
the feature of alternating photon and phonon components of
the bound state is robust against off-diagonal disorder.

\subsection{Tunable odd-neighbor spin-spin interactions}

When considering the situation that two or multiple SiV centers are coupled to
the optomechanical crystal, the bound state can mediate spin-spin interactions,
which can be harnessed to simulate spin models.
The dynamics in Markovian limit is described in Eq.~(\ref{ME11}).
Since $\omega_0\neq\omega_{u/l}(k)$, $\Gamma_{ij}=-(\Gamma_{ji})^*$ and we can simplify the Eq.~(\ref{ME11}) to
(using Eq.~(\ref{ME13}) and Eq.~(\ref{ME14}))
\begin{eqnarray}\label{ME31}
\frac{d\hat{\rho}_s}{dt}=\sum_{i,j}-\Gamma_{ij}(\omega_0)[\hat{\sigma}_+^j\hat{\sigma}_-^i,\hat{\rho}_s].
\end{eqnarray}
It shows coherent spin-spin interactions with a coupling strength of $-i\Gamma_{ij}(\omega_0)$.
Comparing Eq.~(\ref{ME12}) with Eq.~(\ref{ME29}),
we show the spin-spin interactions are indeed mediated by the phonon component of the bound states.
Since the bound state is alternating for photon and phonon components, the mediated spin-spin interactions is odd-neighbor.

There are several applications in quantum simulation and quantum information processing
such as the simulation of spin models, state transfer and
long-distance entanglement of many qubits through an auxiliary one
(the auxiliary spin in odd site and others in even sites).

\begin{figure}[t]
\includegraphics[scale=0.23]{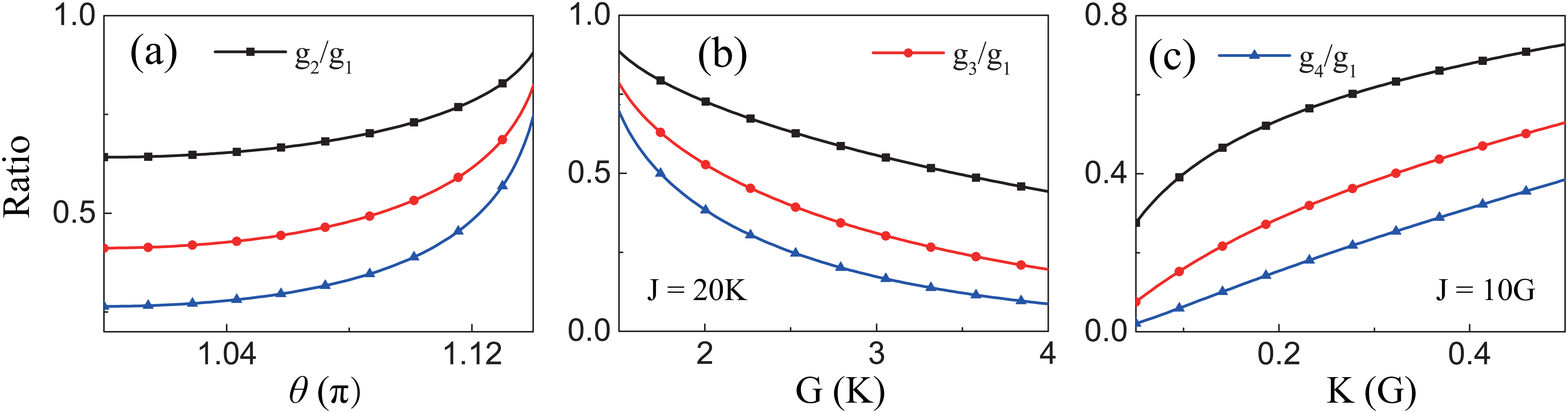}
\caption{\label{S7}(Color online) The ratios of the coupling strength of the first four interacting spins.
(a) $J=20K$ and $G=2K$ are fixed while $\theta$ varies from $\pi$ to $1.14\pi$.
(b) $J=20K$ and $\theta=1.1\pi$ are fixed while $G$ varies from $1.5K$ to $4K$.
(c) $J=10G$ and $\theta=1.1\pi$ are fixed while $K$ varies from $0.05G$ to $0.5G$.}
\end{figure}
Besides, we also consider the tunability of the localization length of this long-range interactions
with respect to the system parameters $\theta, G, K$.
The ratios of the coupling strength of the first four interacting spins are shown in Fig.~S7,
which reflects the localization length.
We show the localization length can be tuned and made larger by closing the bandgap,
where we fix $G, K$ and vary $\theta$ or fix $\theta, K$ and vary $G$, as shown in Fig.~S7(a) and S7(b).
Also, a short localization length involving only nearest-neighbor spins can achieve
when fixing the values of $G, J$ and reducing the value of phonon hopping rate $K$, as shown in Fig.~S7(c).

\subsection{Bound states and spin interactions in a finite optomechanical array}

In the Markovian limit, we predict the spin-spin coupling strength as $g_{ij}=-i\Gamma_{ij}(\omega_0)$,
or equivalently as $g_{ij}=g_\text{eff}\,C_{j-i,b}/C_e$.
The first expression can only be calculated with periodic boundary conditions,
while the second expression is also applicable to the case of open boundary conditions.
Here, we consider a finite optomechanical array with   size $N=7$
and $N=41$, for $\theta=1.1\pi$, respectively.
We label the cells from the left to right as $-(N-1)/2,-(N-1)/2+1,\cdots,0,\cdots,(N-1)/2$
such that the laser phase in the middle nanocavity is $e^{-i*0*\theta}$ for symmetry.
We first focus on the case of a single spin placed in the middle nanocavity.
We find the exotic spin-polariton bound state with alternating photon and phonon components persists,
even in an optomechanical array with small size, as shown in Fig.~S8(a,d).
The energy of the bound state is the same as the bare spin
and is robust against to spin-phonon coupling strength \cite{PhysRevLett.126.063601},
which can be   verified by solving $\hat{H}|\psi\rangle=\omega_0|\psi\rangle$ for $N=7$,
with $\hat{H}=\hat{H}_\text{OM}+\hat{H}_\text{free}+\hat{H}_\text{int}$
(see concrete expressions in Eq.~(\ref{ME1}), Eq.~(\ref{ME8}) and Eq.~(\ref{ME9})).
This bound state can mediate odd-neighbor and complex spin-spin interactions.
Figures S8 (a) and (d) display the bound state with alternating photon and phonon components.

\begin{figure}
\includegraphics[scale=0.23]{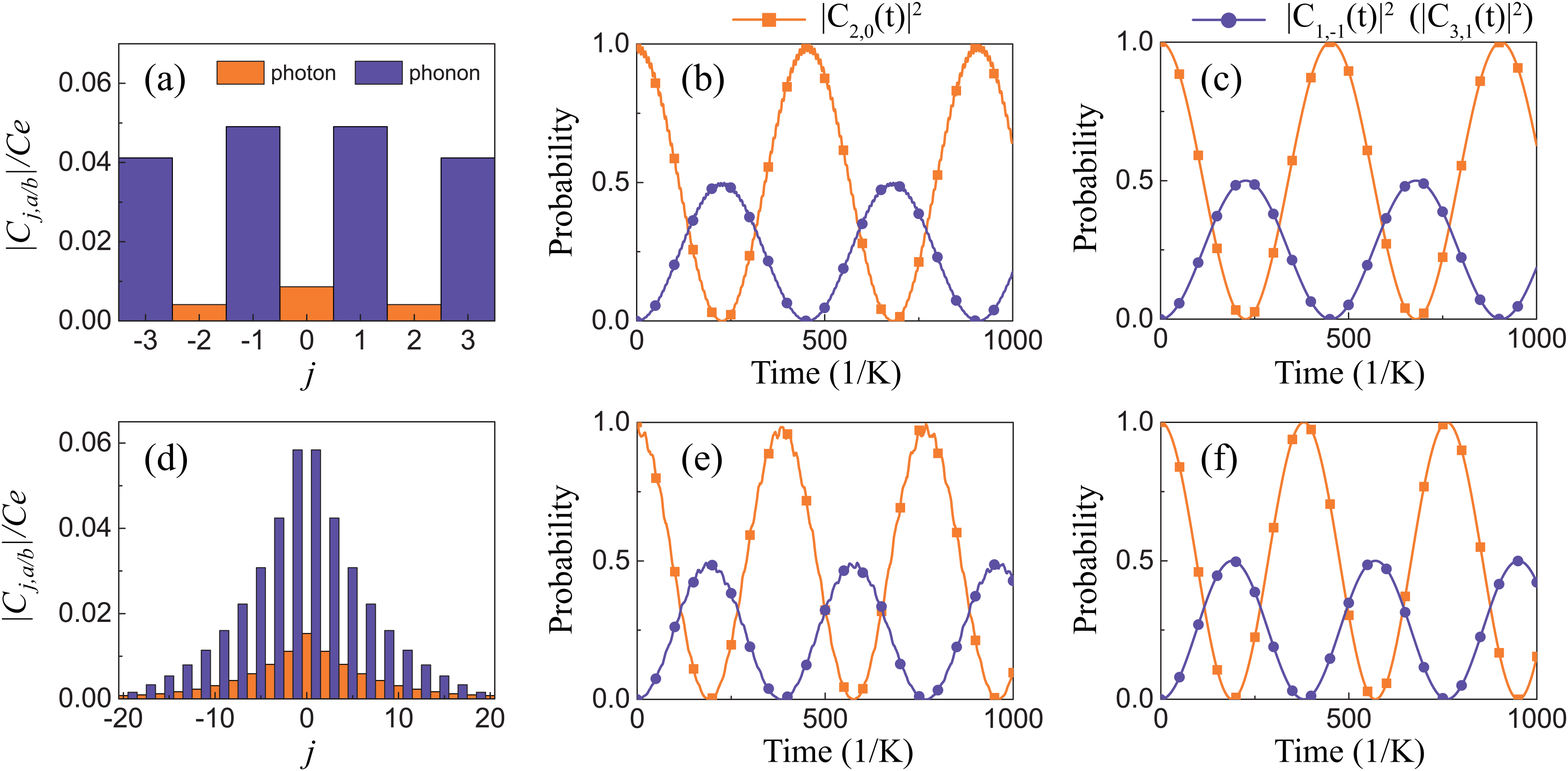}
\caption{\label{S8}(Color online) Bound state and spin interactions in a finite optomechanical system
with (a,b,c) $N=7$ and (d,e,f) $N=41$.
(a,d) Photon-phonon bound state with energy $E_\text{BS}=\omega_0=\omega_M$,
which can be used to predict spin-spin coupling strength as $g_{ij}=g_\text{eff}\,C_{j-i,b}/C_e$.
(a) and (d): The bound state with alternating photon and phonon components.
(b,c,e,f): Dynamics of three spins resonant with the middle three mechanical modes,
with $|C_{i,j}(t)|^2$ the probability of spin $i$ (at the $j$th lattice site) being in the upper state.
The spin's dynamics is governed by (b,e) total Hamiltonian
$\hat{H}=\hat{H}_\text{OM}+\hat{H}_\text{free}+\hat{H}_\text{int}$ and
(c,f) spin interactions $\hat{H}_{s}$ (Eq.~(\ref{ME33})),
with $g_\text{eff}=0.1K$ and initial state $|\psi(0)\rangle=\hat\sigma^2_+|ggg\rangle|\text{vac}\rangle$.
Here, we set $\theta=1.1\pi$, $J=20K$ and $G=2K$.}
\end{figure}
In the following, we exam the spin interactions in Eq.~(8) beyond the Markovian approximation.
Without loss of generality,
we first consider three spins coupled to the mechanical modes of  three middle  nanocavities ($j=0,\pm1$)
and single excitations in the system with the ansatz
\begin{eqnarray}\label{ME32}
|\psi(t)\rangle=(C_{1,-1}(t)\hat\sigma_+^1+C_{2,0}(t)\hat\sigma_+^2+C_{3,1}(t)\hat\sigma_+^3)|ggg\rangle|\text{vac}\rangle
+\sum_{j=-(N-1)/2}^{(N-1)/2}\sum_{\beta=a,b}C_{j,\beta}(t)\hat{\beta}^\dag_j|ggg\rangle|\text{vac}\rangle,
\end{eqnarray}
where $|C_{i,j}(t)|^2$ is the probability of spin $i$ being in the upper state
with subscript $j$ denoting the position of spin at the optomechanical array,
and $|ggg\rangle$ represents three spins being in the lower level.
Within Markovian approximation, the dynamics is governed by odd-neighbor and complex spin-spin interactions modeled as
\begin{eqnarray}\label{ME33}
\hat{H}_{s}=g_{12}(\hat{\sigma}_+^2\hat{\sigma}_-^1+\hat{\sigma}_+^3\hat{\sigma}_-^2)
+g_{12}^*(\hat{\sigma}_+^1\hat{\sigma}_-^2+\hat{\sigma}_+^2\hat{\sigma}_-^3),
\end{eqnarray}
with $g_{12}=g_\text{eff}\,C_{1,b}/C_e=g_{32}^*$.
As discussed in the main text, this approximation is valid in the weak-coupling limit $g_\text{eff}\ll K$.
To exam the validity, we plot the time evolution of probability $|C_{i,j}(t)|^2$
in Fig.~S8(b,c) for $N=7$ and in Fig.~S8(e,f) for $N=41$,
with $g_\text{eff}/K=0.1$ and the initial state $|\psi(0)\rangle=\hat\sigma^2_+|ggg\rangle|\text{vac}\rangle$.
The exact spin dynamics is directly calculated by using $\hat{H}$, as shown in Fig.~S8(b,e).
As a comparison, the Markovian dynamics dominated by Eq.~(\ref{ME33}) is given in Fig.~S8(c,f).
We show the exact evolution is in line with the prediction from the Markovian approximation.
Thus, for the parameters used in the main text, the Markovian approximation is valid.

\begin{figure}
\includegraphics[scale=0.23]{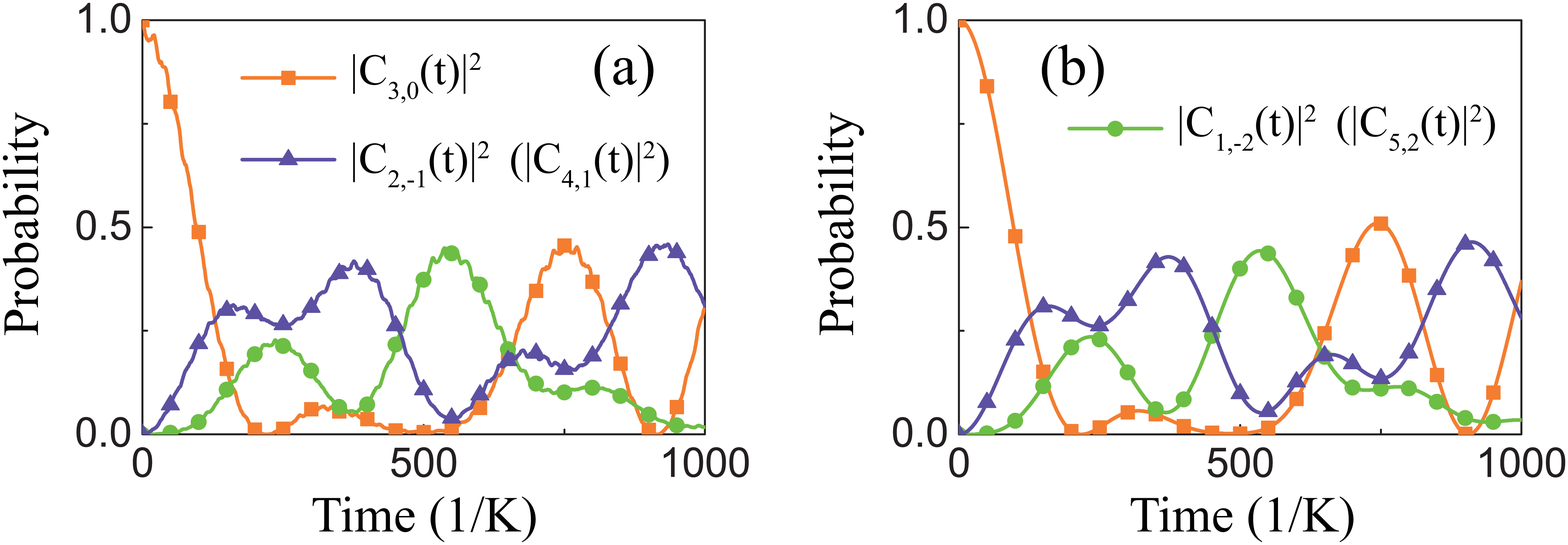}
\caption{\label{S9}(Color online) Dynamics of five spins resonant with the middle five mechanical modes
of a finite optomechanical system with $N=41$.
$|C_{i,j}(t)|^2$ is the probability of spin $i$ (at the $j$th lattice site) being in the upper state.
(a) Non-Markovian dynamics governed by total Hamiltonian
$\hat{H}=\hat{H}_\text{OM}+\hat{H}_\text{free}+\hat{H}_\text{int}$.
(b) Markovian dynamics governed by $\hat{H}^2_{s}$ (Eq.~(\ref{ME35})).
Here, $g_\text{eff}=0.1K$, $|\psi^2(0)\rangle=\hat\sigma^3_+|ggggg\rangle|\text{vac}\rangle$,
$\theta=1.1\pi$, $J=20K$ and $G=2K$.}
\end{figure}

To see the properties of the spin interactions more clearly, we now investigate the case of
five spins placed in the middle five cavities of a $N=41$ optomechanical array.
The single-excitation ansatz is
\begin{eqnarray}\label{ME34}
|\psi^2(t)\rangle&=&(C_{1,-2}(t)\hat\sigma_+^1+C_{2,1}(t)\hat\sigma_+^2+C_{3,0}(t)\hat\sigma_+^3
+C_{4,1}(t)\hat\sigma_+^4+C_{5,2}(t)\hat\sigma_+^5)|ggggg\rangle|\text{vac}\rangle\notag\\
&+&\sum_{j=-(N-1)/2}^{(N-1)/2}\sum_{\beta=a,b}C_{j,\beta}(t)\hat{\beta}^\dag_j|ggggg\rangle|\text{vac}\rangle,
\end{eqnarray}
and the Markovian dynamics is governed by the effective Hamiltonian
\begin{eqnarray}\label{ME35}
\hat{H}_{s}^2=g_{12}(\hat{\sigma}_+^2\hat{\sigma}_-^1+\hat{\sigma}_+^3\hat{\sigma}_-^2+\hat{\sigma}_+^4\hat{\sigma}_-^3+\hat{\sigma}_+^5\hat{\sigma}_-^4)
+g_{14}(\hat{\sigma}_+^4\hat{\sigma}_-^1+\hat{\sigma}_+^5\hat{\sigma}_-^2)+\text{H.c.},
\end{eqnarray}
with $g_{12}=g_\text{eff}\,C_{1,b}/C_e$ and $g_{14}=g_\text{eff}\,C_{3,b}/C_e$.
Clearly, this Hamiltonian does not involve interactions between even-neighbor spins.
We plot spin dynamics both within and beyond the Markovian approximation in  Fig.~S9,
with $g_\text{eff}/K=0.1$ and the  initial state
$|\psi(0)\rangle=\hat\sigma^3_+|ggggg\rangle|\text{vac}\rangle$.
Also, the agreement between Fig.~S9(a) and Fig.~S9(b) indicates that the Markovian approximation is valid.


\begin{thebibliography}{105}%
\makeatletter
\providecommand \@ifxundefined [1]{%
 \@ifx{#1\undefined}
}%
\providecommand \@ifnum [1]{%
 \ifnum #1\expandafter \@firstoftwo
 \else \expandafter \@secondoftwo
 \fi
}%
\providecommand \@ifx [1]{%
 \ifx #1\expandafter \@firstoftwo
 \else \expandafter \@secondoftwo
 \fi
}%
\providecommand \natexlab [1]{#1}%
\providecommand \enquote  [1]{``#1''}%
\providecommand \bibnamefont  [1]{#1}%
\providecommand \bibfnamefont [1]{#1}%
\providecommand \citenamefont [1]{#1}%
\providecommand \href@noop [0]{\@secondoftwo}%
\providecommand \href [0]{\begingroup \@sanitize@url \@href}%
\providecommand \@href[1]{\@@startlink{#1}\@@href}%
\providecommand \@@href[1]{\endgroup#1\@@endlink}%
\providecommand \@sanitize@url [0]{\catcode `\\12\catcode `\$12\catcode
  `\&12\catcode `\#12\catcode `\^12\catcode `\_12\catcode `\%12\relax}%
\providecommand \@@startlink[1]{}%
\providecommand \@@endlink[0]{}%
\providecommand \url  [0]{\begingroup\@sanitize@url \@url }%
\providecommand \@url [1]{\endgroup\@href {#1}{\urlprefix }}%
\providecommand \urlprefix  [0]{URL }%
\providecommand \Eprint [0]{\href }%
\providecommand \doibase [0]{https://doi.org/}%
\providecommand \selectlanguage [0]{\@gobble}%
\providecommand \bibinfo  [0]{\@secondoftwo}%
\providecommand \bibfield  [0]{\@secondoftwo}%
\providecommand \translation [1]{[#1]}%
\providecommand \BibitemOpen [0]{}%
\providecommand \bibitemStop [0]{}%
\providecommand \bibitemNoStop [0]{.\EOS\space}%
\providecommand \EOS [0]{\spacefactor3000\relax}%
\providecommand \BibitemShut  [1]{\csname bibitem#1\endcsname}%
\let\auto@bib@innerbib\@empty
\bibitem [{\citenamefont {Chang}\ \emph {et~al.}(2018)\citenamefont {Chang},
  \citenamefont {Douglas}, \citenamefont {Gonz\'alez-Tudela}, \citenamefont
  {Hung},\ and\ \citenamefont {Kimble}}]{RevModPhys.90.031002}%
  \BibitemOpen
  \bibfield  {author} {\bibinfo {author} {\bibfnamefont {D.~E.}\ \bibnamefont
  {Chang}}, \bibinfo {author} {\bibfnamefont {J.~S.}\ \bibnamefont {Douglas}},
  \bibinfo {author} {\bibfnamefont {A.}~\bibnamefont {Gonz\'alez-Tudela}},
  \bibinfo {author} {\bibfnamefont {C.-L.}\ \bibnamefont {Hung}},\ and\
  \bibinfo {author} {\bibfnamefont {H.~J.}\ \bibnamefont {Kimble}},\ }\bibfield
   {title} {\bibinfo {title} {Colloquium: Quantum matter built from nanoscopic
  lattices of atoms and photons},\ }\href
  {https://doi.org/10.1103/RevModPhys.90.031002} {\bibfield  {journal}
  {\bibinfo  {journal} {Rev. Mod. Phys.}\ }\textbf {\bibinfo {volume} {90}},\
  \bibinfo {pages} {031002} (\bibinfo {year} {2018})}\BibitemShut {NoStop}%
\bibitem [{\citenamefont {Mitsch}\ \emph {et~al.}(2014)\citenamefont {Mitsch},
  \citenamefont {Sayrin}, \citenamefont {Albrecht}, \citenamefont
  {Schneeweiss},\ and\ \citenamefont {Rauschenbeutel}}]{Quantum2014Mitsch}%
  \BibitemOpen
  \bibfield  {author} {\bibinfo {author} {\bibfnamefont {R.}~\bibnamefont
  {Mitsch}}, \bibinfo {author} {\bibfnamefont {C.}~\bibnamefont {Sayrin}},
  \bibinfo {author} {\bibfnamefont {B.}~\bibnamefont {Albrecht}}, \bibinfo
  {author} {\bibfnamefont {P.}~\bibnamefont {Schneeweiss}},\ and\ \bibinfo
  {author} {\bibfnamefont {A.}~\bibnamefont {Rauschenbeutel}},\ }\bibfield
  {title} {\bibinfo {title} {Quantum state-controlled directional spontaneous
  emission of photons into a nanophotonic waveguide},\ }\href
  {https://doi.org/10.1038/ncomms6713} {\bibfield  {journal} {\bibinfo
  {journal} {Nat. Commun.}\ }\textbf {\bibinfo {volume} {5}},\ \bibinfo {pages}
  {5713} (\bibinfo {year} {2014})}\BibitemShut {NoStop}%
\bibitem [{\citenamefont {Ramos}\ \emph {et~al.}(2014)\citenamefont {Ramos},
  \citenamefont {Pichler}, \citenamefont {Daley},\ and\ \citenamefont
  {Zoller}}]{PhysRevLett.113.237203}%
  \BibitemOpen
  \bibfield  {author} {\bibinfo {author} {\bibfnamefont {T.}~\bibnamefont
  {Ramos}}, \bibinfo {author} {\bibfnamefont {H.}~\bibnamefont {Pichler}},
  \bibinfo {author} {\bibfnamefont {A.~J.}\ \bibnamefont {Daley}},\ and\
  \bibinfo {author} {\bibfnamefont {P.}~\bibnamefont {Zoller}},\ }\bibfield
  {title} {\bibinfo {title} {Quantum spin dimers from chiral dissipation in
  cold-atom chains},\ }\href {https://doi.org/10.1103/PhysRevLett.113.237203}
  {\bibfield  {journal} {\bibinfo  {journal} {Phys. Rev. Lett.}\ }\textbf
  {\bibinfo {volume} {113}},\ \bibinfo {pages} {237203} (\bibinfo {year}
  {2014})}\BibitemShut {NoStop}%
\bibitem [{\citenamefont {Bliokh}\ \emph
  {et~al.}(2015{\natexlab{a}})\citenamefont {Bliokh}, \citenamefont
  {Smirnova},\ and\ \citenamefont {Nori}}]{Bliokh1448}%
  \BibitemOpen
  \bibfield  {author} {\bibinfo {author} {\bibfnamefont {K.~Y.}\ \bibnamefont
  {Bliokh}}, \bibinfo {author} {\bibfnamefont {D.}~\bibnamefont {Smirnova}},\
  and\ \bibinfo {author} {\bibfnamefont {F.}~\bibnamefont {Nori}},\ }\bibfield
  {title} {\bibinfo {title} {Quantum spin {H}all effect of light},\ }\href
  {https://doi.org/10.1126/science.aaa9519} {\bibfield  {journal} {\bibinfo
  {journal} {Science}\ }\textbf {\bibinfo {volume} {348}},\ \bibinfo {pages}
  {1448} (\bibinfo {year} {2015}{\natexlab{a}})}\BibitemShut {NoStop}%
\bibitem [{\citenamefont {Bliokh}\ \emph
  {et~al.}(2015{\natexlab{b}})\citenamefont {Bliokh}, \citenamefont
  {Rodr\'iguez-Fortu\~no}, \citenamefont {Nori},\ and\ \citenamefont
  {Zayats}}]{Spin2015Bliokh}%
  \BibitemOpen
  \bibfield  {author} {\bibinfo {author} {\bibfnamefont {K.~Y.}\ \bibnamefont
  {Bliokh}}, \bibinfo {author} {\bibfnamefont {F.~J.}\ \bibnamefont
  {Rodr\'iguez-Fortu\~no}}, \bibinfo {author} {\bibfnamefont {F.}~\bibnamefont
  {Nori}},\ and\ \bibinfo {author} {\bibfnamefont {A.~V.}\ \bibnamefont
  {Zayats}},\ }\bibfield  {title} {\bibinfo {title} {Spin-orbit interactions of
  light},\ }\href {https://doi.org/10.1038/nphoton.2015.201} {\bibfield
  {journal} {\bibinfo  {journal} {Nat. photonics}\ }\textbf {\bibinfo {volume}
  {9}},\ \bibinfo {pages} {796} (\bibinfo {year}
  {2015}{\natexlab{b}})}\BibitemShut {NoStop}%
\bibitem [{\citenamefont {Lodahl}\ \emph {et~al.}(2017)\citenamefont {Lodahl},
  \citenamefont {Mahmoodian}, \citenamefont {Stobbe}, \citenamefont
  {Rauschenbeutel}, \citenamefont {Schneeweiss}, \citenamefont {Volz},
  \citenamefont {Pichler},\ and\ \citenamefont {Zoller}}]{Chiral2017Lodahl}%
  \BibitemOpen
  \bibfield  {author} {\bibinfo {author} {\bibfnamefont {P.}~\bibnamefont
  {Lodahl}}, \bibinfo {author} {\bibfnamefont {S.}~\bibnamefont {Mahmoodian}},
  \bibinfo {author} {\bibfnamefont {S.}~\bibnamefont {Stobbe}}, \bibinfo
  {author} {\bibfnamefont {A.}~\bibnamefont {Rauschenbeutel}}, \bibinfo
  {author} {\bibfnamefont {P.}~\bibnamefont {Schneeweiss}}, \bibinfo {author}
  {\bibfnamefont {J.}~\bibnamefont {Volz}}, \bibinfo {author} {\bibfnamefont
  {H.}~\bibnamefont {Pichler}},\ and\ \bibinfo {author} {\bibfnamefont
  {P.}~\bibnamefont {Zoller}},\ }\bibfield  {title} {\bibinfo {title} {Chiral
  quantum optics},\ }\href {https://doi.org/10.1038/nature21037} {\bibfield
  {journal} {\bibinfo  {journal} {Nature}\ }\textbf {\bibinfo {volume} {541}},\
  \bibinfo {pages} {473} (\bibinfo {year} {2017})}\BibitemShut {NoStop}%
\bibitem [{\citenamefont {Triolo}\ \emph {et~al.}(2017)\citenamefont {Triolo},
  \citenamefont {Cacciola}, \citenamefont {Patan\`e}, \citenamefont {Saija},
  \citenamefont {Savasta},\ and\ \citenamefont {Nori}}]{triolo2017spin}%
  \BibitemOpen
  \bibfield  {author} {\bibinfo {author} {\bibfnamefont {C.}~\bibnamefont
  {Triolo}}, \bibinfo {author} {\bibfnamefont {A.}~\bibnamefont {Cacciola}},
  \bibinfo {author} {\bibfnamefont {S.}~\bibnamefont {Patan\`e}}, \bibinfo
  {author} {\bibfnamefont {R.}~\bibnamefont {Saija}}, \bibinfo {author}
  {\bibfnamefont {S.}~\bibnamefont {Savasta}},\ and\ \bibinfo {author}
  {\bibfnamefont {F.}~\bibnamefont {Nori}},\ }\bibfield  {title} {\bibinfo
  {title} {Spin-momentum locking in the near field of metal nanoparticles},\
  }\href {https://doi.org/10.1021/acsphotonics.7b00436} {\bibfield  {journal}
  {\bibinfo  {journal} {ACS Photonics}\ }\textbf {\bibinfo {volume} {4}},\
  \bibinfo {pages} {2242} (\bibinfo {year} {2017})}\BibitemShut {NoStop}%
\bibitem [{\citenamefont {S\'anchez-Burillo}\ \emph
  {et~al.}(2020{\natexlab{a}})\citenamefont {S\'anchez-Burillo}, \citenamefont
  {Wan}, \citenamefont {Zueco},\ and\ \citenamefont
  {Gonz\'alez-Tudela}}]{PhysRevResearch.2.023003}%
  \BibitemOpen
  \bibfield  {author} {\bibinfo {author} {\bibfnamefont {E.}~\bibnamefont
  {S\'anchez-Burillo}}, \bibinfo {author} {\bibfnamefont {C.}~\bibnamefont
  {Wan}}, \bibinfo {author} {\bibfnamefont {D.}~\bibnamefont {Zueco}},\ and\
  \bibinfo {author} {\bibfnamefont {A.}~\bibnamefont {Gonz\'alez-Tudela}},\
  }\bibfield  {title} {\bibinfo {title} {Chiral quantum optics in photonic
  sawtooth lattices},\ }\href
  {https://doi.org/10.1103/PhysRevResearch.2.023003} {\bibfield  {journal}
  {\bibinfo  {journal} {Phys. Rev. Research}\ }\textbf {\bibinfo {volume}
  {2}},\ \bibinfo {pages} {023003} (\bibinfo {year}
  {2020}{\natexlab{a}})}\BibitemShut {NoStop}%
\bibitem [{\citenamefont {Gonz\'alez-Tudela}\ \emph {et~al.}(2015)\citenamefont
  {Gonz\'alez-Tudela}, \citenamefont {Hung}, \citenamefont {Chang},
  \citenamefont {Cirac},\ and\ \citenamefont {Kimble}}]{Subwavelength2015AGT}%
  \BibitemOpen
  \bibfield  {author} {\bibinfo {author} {\bibfnamefont {A.}~\bibnamefont
  {Gonz\'alez-Tudela}}, \bibinfo {author} {\bibfnamefont {C.-L.}\ \bibnamefont
  {Hung}}, \bibinfo {author} {\bibfnamefont {D.~E.}\ \bibnamefont {Chang}},
  \bibinfo {author} {\bibfnamefont {J.~I.}\ \bibnamefont {Cirac}},\ and\
  \bibinfo {author} {\bibfnamefont {H.~J.}\ \bibnamefont {Kimble}},\ }\bibfield
   {title} {\bibinfo {title} {Subwavelength vacuum lattices and atom-atom
  interactions in two-dimensional photonic crystals},\ }\href
  {https://doi.org/10.1038/nphoton.2015.54} {\bibfield  {journal} {\bibinfo
  {journal} {Nat. Photonics}\ }\textbf {\bibinfo {volume} {9}},\ \bibinfo
  {pages} {320} (\bibinfo {year} {2015})}\BibitemShut {NoStop}%
\bibitem [{\citenamefont {Douglas}\ \emph {et~al.}(2015)\citenamefont
  {Douglas}, \citenamefont {Habibian}, \citenamefont {Hung}, \citenamefont
  {Gorshkov}, \citenamefont {Kimble},\ and\ \citenamefont
  {Chang}}]{Quantum2015Douglas}%
  \BibitemOpen
  \bibfield  {author} {\bibinfo {author} {\bibfnamefont {J.~S.}\ \bibnamefont
  {Douglas}}, \bibinfo {author} {\bibfnamefont {H.}~\bibnamefont {Habibian}},
  \bibinfo {author} {\bibfnamefont {C.-L.}\ \bibnamefont {Hung}}, \bibinfo
  {author} {\bibfnamefont {A.~V.}\ \bibnamefont {Gorshkov}}, \bibinfo {author}
  {\bibfnamefont {H.~J.}\ \bibnamefont {Kimble}},\ and\ \bibinfo {author}
  {\bibfnamefont {D.~E.}\ \bibnamefont {Chang}},\ }\bibfield  {title} {\bibinfo
  {title} {Quantum many-body models with cold atoms coupled to photonic
  crystals},\ }\href {https://doi.org/10.1038/nphoton.2015.57} {\bibfield
  {journal} {\bibinfo  {journal} {Nat. Photonics}\ }\textbf {\bibinfo {volume}
  {9}},\ \bibinfo {pages} {326} (\bibinfo {year} {2015})}\BibitemShut {NoStop}%
\bibitem [{\citenamefont {Hood}\ \emph {et~al.}(2016)\citenamefont {Hood},
  \citenamefont {Goban}, \citenamefont {Asenjo-Garcia}, \citenamefont {Lu},
  \citenamefont {Yu}, \citenamefont {Chang},\ and\ \citenamefont
  {Kimble}}]{Hood10507}%
  \BibitemOpen
  \bibfield  {author} {\bibinfo {author} {\bibfnamefont {J.~D.}\ \bibnamefont
  {Hood}}, \bibinfo {author} {\bibfnamefont {A.}~\bibnamefont {Goban}},
  \bibinfo {author} {\bibfnamefont {A.}~\bibnamefont {Asenjo-Garcia}}, \bibinfo
  {author} {\bibfnamefont {M.}~\bibnamefont {Lu}}, \bibinfo {author}
  {\bibfnamefont {S.-P.}\ \bibnamefont {Yu}}, \bibinfo {author} {\bibfnamefont
  {D.~E.}\ \bibnamefont {Chang}},\ and\ \bibinfo {author} {\bibfnamefont
  {H.~J.}\ \bibnamefont {Kimble}},\ }\bibfield  {title} {\bibinfo {title}
  {Atom{\textendash}atom interactions around the band edge of a photonic
  crystal waveguide},\ }\href {https://doi.org/10.1073/pnas.1603788113}
  {\bibfield  {journal} {\bibinfo  {journal} {Proc. Natl. Acad. Sci. U.S.A.}\
  }\textbf {\bibinfo {volume} {113}},\ \bibinfo {pages} {10507} (\bibinfo
  {year} {2016})}\BibitemShut {NoStop}%
\bibitem [{\citenamefont {Gonz\'alez-Tudela}\ and\ \citenamefont
  {Galve}(2019)}]{doi:10.1021/acsphotonics.8b01455}%
  \BibitemOpen
  \bibfield  {author} {\bibinfo {author} {\bibfnamefont {A.}~\bibnamefont
  {Gonz\'alez-Tudela}}\ and\ \bibinfo {author} {\bibfnamefont {F.}~\bibnamefont
  {Galve}},\ }\bibfield  {title} {\bibinfo {title} {Anisotropic quantum emitter
  interactions in two-dimensional photonic-crystal baths},\ }\href
  {https://doi.org/10.1021/acsphotonics.8b01455} {\bibfield  {journal}
  {\bibinfo  {journal} {ACS Photonics}\ }\textbf {\bibinfo {volume} {6}},\
  \bibinfo {pages} {221} (\bibinfo {year} {2019})}\BibitemShut {NoStop}%
\bibitem [{\citenamefont {Bello}\ \emph {et~al.}(2019)\citenamefont {Bello},
  \citenamefont {Platero}, \citenamefont {Cirac},\ and\ \citenamefont
  {Gonz{\'a}lez-Tudela}}]{Belloeaaw0297}%
  \BibitemOpen
  \bibfield  {author} {\bibinfo {author} {\bibfnamefont {M.}~\bibnamefont
  {Bello}}, \bibinfo {author} {\bibfnamefont {G.}~\bibnamefont {Platero}},
  \bibinfo {author} {\bibfnamefont {J.~I.}\ \bibnamefont {Cirac}},\ and\
  \bibinfo {author} {\bibfnamefont {A.}~\bibnamefont {Gonz{\'a}lez-Tudela}},\
  }\bibfield  {title} {\bibinfo {title} {Unconventional quantum optics in
  topological waveguide \text{QED}},\ }\href
  {https://doi.org/10.1126/sciadv.aaw0297} {\bibfield  {journal} {\bibinfo
  {journal} {Sci. Adv.}\ }\textbf {\bibinfo {volume} {5}},\ \bibinfo {pages}
  {eaaw0297} (\bibinfo {year} {2019})}\BibitemShut {NoStop}%
\bibitem [{\citenamefont {Perczel}\ \emph {et~al.}(2017)\citenamefont
  {Perczel}, \citenamefont {Borregaard}, \citenamefont {Chang}, \citenamefont
  {Pichler}, \citenamefont {Yelin}, \citenamefont {Zoller},\ and\ \citenamefont
  {Lukin}}]{PhysRevLett.119.023603}%
  \BibitemOpen
  \bibfield  {author} {\bibinfo {author} {\bibfnamefont {J.}~\bibnamefont
  {Perczel}}, \bibinfo {author} {\bibfnamefont {J.}~\bibnamefont {Borregaard}},
  \bibinfo {author} {\bibfnamefont {D.~E.}\ \bibnamefont {Chang}}, \bibinfo
  {author} {\bibfnamefont {H.}~\bibnamefont {Pichler}}, \bibinfo {author}
  {\bibfnamefont {S.~F.}\ \bibnamefont {Yelin}}, \bibinfo {author}
  {\bibfnamefont {P.}~\bibnamefont {Zoller}},\ and\ \bibinfo {author}
  {\bibfnamefont {M.~D.}\ \bibnamefont {Lukin}},\ }\bibfield  {title} {\bibinfo
  {title} {Topological quantum optics in two-dimensional atomic arrays},\
  }\href {https://doi.org/10.1103/PhysRevLett.119.023603} {\bibfield  {journal}
  {\bibinfo  {journal} {Phys. Rev. Lett.}\ }\textbf {\bibinfo {volume} {119}},\
  \bibinfo {pages} {023603} (\bibinfo {year} {2017})}\BibitemShut {NoStop}%
\bibitem [{\citenamefont {Bliokh}\ \emph {et~al.}(2019)\citenamefont {Bliokh},
  \citenamefont {Leykam}, \citenamefont {Lein},\ and\ \citenamefont
  {Nori}}]{Topological2019Bliokh}%
  \BibitemOpen
  \bibfield  {author} {\bibinfo {author} {\bibfnamefont {K.~Y.}\ \bibnamefont
  {Bliokh}}, \bibinfo {author} {\bibfnamefont {D.}~\bibnamefont {Leykam}},
  \bibinfo {author} {\bibfnamefont {M.}~\bibnamefont {Lein}},\ and\ \bibinfo
  {author} {\bibfnamefont {F.}~\bibnamefont {Nori}},\ }\bibfield  {title}
  {\bibinfo {title} {Topological non-hermitian origin of surface {M}axwell
  waves},\ }\href {https://doi.org/10.1038/s41467-019-08397-6} {\bibfield
  {journal} {\bibinfo  {journal} {Nat. Commun.}\ }\textbf {\bibinfo {volume}
  {10}},\ \bibinfo {pages} {580} (\bibinfo {year} {2019})}\BibitemShut
  {NoStop}%
\bibitem [{\citenamefont {Ozawa}\ \emph {et~al.}(2019)\citenamefont {Ozawa},
  \citenamefont {Price}, \citenamefont {Amo}, \citenamefont {Goldman},
  \citenamefont {Hafezi}, \citenamefont {Lu}, \citenamefont {Rechtsman},
  \citenamefont {Schuster}, \citenamefont {Simon}, \citenamefont {Zilberberg},\
  and\ \citenamefont {Carusotto}}]{RevModPhys.91.015006}%
  \BibitemOpen
  \bibfield  {author} {\bibinfo {author} {\bibfnamefont {T.}~\bibnamefont
  {Ozawa}}, \bibinfo {author} {\bibfnamefont {H.~M.}\ \bibnamefont {Price}},
  \bibinfo {author} {\bibfnamefont {A.}~\bibnamefont {Amo}}, \bibinfo {author}
  {\bibfnamefont {N.}~\bibnamefont {Goldman}}, \bibinfo {author} {\bibfnamefont
  {M.}~\bibnamefont {Hafezi}}, \bibinfo {author} {\bibfnamefont
  {L.}~\bibnamefont {Lu}}, \bibinfo {author} {\bibfnamefont {M.~C.}\
  \bibnamefont {Rechtsman}}, \bibinfo {author} {\bibfnamefont {D.}~\bibnamefont
  {Schuster}}, \bibinfo {author} {\bibfnamefont {J.}~\bibnamefont {Simon}},
  \bibinfo {author} {\bibfnamefont {O.}~\bibnamefont {Zilberberg}},\ and\
  \bibinfo {author} {\bibfnamefont {I.}~\bibnamefont {Carusotto}},\ }\bibfield
  {title} {\bibinfo {title} {Topological photonics},\ }\href
  {https://doi.org/10.1103/RevModPhys.91.015006} {\bibfield  {journal}
  {\bibinfo  {journal} {Rev. Mod. Phys.}\ }\textbf {\bibinfo {volume} {91}},\
  \bibinfo {pages} {015006} (\bibinfo {year} {2019})}\BibitemShut {NoStop}%
\bibitem [{\citenamefont {Perczel}\ \emph {et~al.}(2020)\citenamefont
  {Perczel}, \citenamefont {Borregaard}, \citenamefont {Chang}, \citenamefont
  {Yelin},\ and\ \citenamefont {Lukin}}]{PhysRevLett.124.083603}%
  \BibitemOpen
  \bibfield  {author} {\bibinfo {author} {\bibfnamefont {J.}~\bibnamefont
  {Perczel}}, \bibinfo {author} {\bibfnamefont {J.}~\bibnamefont {Borregaard}},
  \bibinfo {author} {\bibfnamefont {D.~E.}\ \bibnamefont {Chang}}, \bibinfo
  {author} {\bibfnamefont {S.~F.}\ \bibnamefont {Yelin}},\ and\ \bibinfo
  {author} {\bibfnamefont {M.~D.}\ \bibnamefont {Lukin}},\ }\bibfield  {title}
  {\bibinfo {title} {Topological quantum optics using atomlike emitter arrays
  coupled to photonic crystals},\ }\href
  {https://doi.org/10.1103/PhysRevLett.124.083603} {\bibfield  {journal}
  {\bibinfo  {journal} {Phys. Rev. Lett.}\ }\textbf {\bibinfo {volume} {124}},\
  \bibinfo {pages} {083603} (\bibinfo {year} {2020})}\BibitemShut {NoStop}%
\bibitem [{\citenamefont {You}\ and\ \citenamefont
  {Nori}(2011)}]{Atomic2011You}%
  \BibitemOpen
  \bibfield  {author} {\bibinfo {author} {\bibfnamefont {J.~Q.}\ \bibnamefont
  {You}}\ and\ \bibinfo {author} {\bibfnamefont {F.}~\bibnamefont {Nori}},\
  }\bibfield  {title} {\bibinfo {title} {Atomic physics and quantum optics
  using superconducting circuits},\ }\href
  {https://doi.org/10.1038/nature10122} {\bibfield  {journal} {\bibinfo
  {journal} {Nature}\ }\textbf {\bibinfo {volume} {474}},\ \bibinfo {pages}
  {589} (\bibinfo {year} {2011})}\BibitemShut {NoStop}%
\bibitem [{\citenamefont {van Loo}\ \emph {et~al.}(2013)\citenamefont {van
  Loo}, \citenamefont {Fedorov}, \citenamefont {Lalumi{\`e}re}, \citenamefont
  {Sanders}, \citenamefont {Blais},\ and\ \citenamefont
  {Wallraff}}]{vanLoo1494}%
  \BibitemOpen
  \bibfield  {author} {\bibinfo {author} {\bibfnamefont {A.~F.}\ \bibnamefont
  {van Loo}}, \bibinfo {author} {\bibfnamefont {A.}~\bibnamefont {Fedorov}},
  \bibinfo {author} {\bibfnamefont {K.}~\bibnamefont {Lalumi{\`e}re}}, \bibinfo
  {author} {\bibfnamefont {B.~C.}\ \bibnamefont {Sanders}}, \bibinfo {author}
  {\bibfnamefont {A.}~\bibnamefont {Blais}},\ and\ \bibinfo {author}
  {\bibfnamefont {A.}~\bibnamefont {Wallraff}},\ }\bibfield  {title} {\bibinfo
  {title} {Photon-mediated interactions between distant artificial atoms},\
  }\href {https://doi.org/10.1126/science.1244324} {\bibfield  {journal}
  {\bibinfo  {journal} {Science}\ }\textbf {\bibinfo {volume} {342}},\ \bibinfo
  {pages} {1494} (\bibinfo {year} {2013})}\BibitemShut {NoStop}%
\bibitem [{\citenamefont {Gu}\ \emph {et~al.}(2017)\citenamefont {Gu},
  \citenamefont {Kockum}, \citenamefont {Miranowicz}, \citenamefont {xi~Liu},\
  and\ \citenamefont {Nori}}]{GU20171}%
  \BibitemOpen
  \bibfield  {author} {\bibinfo {author} {\bibfnamefont {X.}~\bibnamefont
  {Gu}}, \bibinfo {author} {\bibfnamefont {A.~F.}\ \bibnamefont {Kockum}},
  \bibinfo {author} {\bibfnamefont {A.}~\bibnamefont {Miranowicz}}, \bibinfo
  {author} {\bibfnamefont {Y.}~\bibnamefont {xi~Liu}},\ and\ \bibinfo {author}
  {\bibfnamefont {F.}~\bibnamefont {Nori}},\ }\bibfield  {title} {\bibinfo
  {title} {Microwave photonics with superconducting quantum circuits},\ }\href
  {https://doi.org/https://doi.org/10.1016/j.physrep.2017.10.002} {\bibfield
  {journal} {\bibinfo  {journal} {Phys. Rep.}\ }\textbf {\bibinfo {volume}
  {718}},\ \bibinfo {pages} {1} (\bibinfo {year} {2017})}\BibitemShut {NoStop}%
\bibitem [{\citenamefont {Nie}\ \emph {et~al.}(2020)\citenamefont {Nie},
  \citenamefont {Peng}, \citenamefont {Nori},\ and\ \citenamefont
  {Liu}}]{PhysRevLett.124.023603}%
  \BibitemOpen
  \bibfield  {author} {\bibinfo {author} {\bibfnamefont {W.}~\bibnamefont
  {Nie}}, \bibinfo {author} {\bibfnamefont {Z.~H.}\ \bibnamefont {Peng}},
  \bibinfo {author} {\bibfnamefont {F.}~\bibnamefont {Nori}},\ and\ \bibinfo
  {author} {\bibfnamefont {Y.-x.}\ \bibnamefont {Liu}},\ }\bibfield  {title}
  {\bibinfo {title} {Topologically protected quantum coherence in a
  superatom},\ }\href {https://doi.org/10.1103/PhysRevLett.124.023603}
  {\bibfield  {journal} {\bibinfo  {journal} {Phys. Rev. Lett.}\ }\textbf
  {\bibinfo {volume} {124}},\ \bibinfo {pages} {023603} (\bibinfo {year}
  {2020})}\BibitemShut {NoStop}%
\bibitem [{\citenamefont {Habraken}\ \emph {et~al.}(2012)\citenamefont
  {Habraken}, \citenamefont {Stannigel}, \citenamefont {Lukin}, \citenamefont
  {Zoller},\ and\ \citenamefont {Rabl}}]{Habraken_2012}%
  \BibitemOpen
  \bibfield  {author} {\bibinfo {author} {\bibfnamefont {S.~J.~M.}\
  \bibnamefont {Habraken}}, \bibinfo {author} {\bibfnamefont {K.}~\bibnamefont
  {Stannigel}}, \bibinfo {author} {\bibfnamefont {M.~D.}\ \bibnamefont
  {Lukin}}, \bibinfo {author} {\bibfnamefont {P.}~\bibnamefont {Zoller}},\ and\
  \bibinfo {author} {\bibfnamefont {P.}~\bibnamefont {Rabl}},\ }\bibfield
  {title} {\bibinfo {title} {Continuous mode cooling and phonon routers for
  phononic quantum networks},\ }\href
  {https://doi.org/10.1088/1367-2630/14/11/115004} {\bibfield  {journal}
  {\bibinfo  {journal} {New J. Phys.}\ }\textbf {\bibinfo {volume} {14}},\
  \bibinfo {pages} {115004} (\bibinfo {year} {2012})}\BibitemShut {NoStop}%
\bibitem [{\citenamefont {Aspelmeyer}\ \emph {et~al.}(2014)\citenamefont
  {Aspelmeyer}, \citenamefont {Kippenberg},\ and\ \citenamefont
  {Marquardt}}]{RevModPhys.86.1391}%
  \BibitemOpen
  \bibfield  {author} {\bibinfo {author} {\bibfnamefont {M.}~\bibnamefont
  {Aspelmeyer}}, \bibinfo {author} {\bibfnamefont {T.~J.}\ \bibnamefont
  {Kippenberg}},\ and\ \bibinfo {author} {\bibfnamefont {F.}~\bibnamefont
  {Marquardt}},\ }\bibfield  {title} {\bibinfo {title} {Cavity optomechanics},\
  }\href {https://doi.org/10.1103/RevModPhys.86.1391} {\bibfield  {journal}
  {\bibinfo  {journal} {Rev. Mod. Phys.}\ }\textbf {\bibinfo {volume} {86}},\
  \bibinfo {pages} {1391} (\bibinfo {year} {2014})}\BibitemShut {NoStop}%
\bibitem [{\citenamefont {Peano}\ \emph {et~al.}(2015)\citenamefont {Peano},
  \citenamefont {Brendel}, \citenamefont {Schmidt},\ and\ \citenamefont
  {Marquardt}}]{PhysRevX.5.031011}%
  \BibitemOpen
  \bibfield  {author} {\bibinfo {author} {\bibfnamefont {V.}~\bibnamefont
  {Peano}}, \bibinfo {author} {\bibfnamefont {C.}~\bibnamefont {Brendel}},
  \bibinfo {author} {\bibfnamefont {M.}~\bibnamefont {Schmidt}},\ and\ \bibinfo
  {author} {\bibfnamefont {F.}~\bibnamefont {Marquardt}},\ }\bibfield  {title}
  {\bibinfo {title} {Topological phases of sound and light},\ }\href
  {https://doi.org/10.1103/PhysRevX.5.031011} {\bibfield  {journal} {\bibinfo
  {journal} {Phys. Rev. X}\ }\textbf {\bibinfo {volume} {5}},\ \bibinfo {pages}
  {031011} (\bibinfo {year} {2015})}\BibitemShut {NoStop}%
\bibitem [{\citenamefont {Kim}\ \emph {et~al.}(2017)\citenamefont {Kim},
  \citenamefont {Xu}, \citenamefont {Taylor},\ and\ \citenamefont
  {Bahl}}]{Dynamically2017Kim}%
  \BibitemOpen
  \bibfield  {author} {\bibinfo {author} {\bibfnamefont {S.}~\bibnamefont
  {Kim}}, \bibinfo {author} {\bibfnamefont {X.}~\bibnamefont {Xu}}, \bibinfo
  {author} {\bibfnamefont {J.~M.}\ \bibnamefont {Taylor}},\ and\ \bibinfo
  {author} {\bibfnamefont {G.}~\bibnamefont {Bahl}},\ }\bibfield  {title}
  {\bibinfo {title} {Dynamically induced robust phonon transport and chiral
  cooling in an optomechanical system},\ }\href
  {https://doi.org/10.1038/s41467-017-00247-7} {\bibfield  {journal} {\bibinfo
  {journal} {Nat. Commun.}\ }\textbf {\bibinfo {volume} {8}},\ \bibinfo {pages}
  {205} (\bibinfo {year} {2017})}\BibitemShut {NoStop}%
\bibitem [{\citenamefont {Verhagen}\ and\ \citenamefont
  {Al\`u}(2017)}]{Optomechanical2017Verhagen}%
  \BibitemOpen
  \bibfield  {author} {\bibinfo {author} {\bibfnamefont {E.}~\bibnamefont
  {Verhagen}}\ and\ \bibinfo {author} {\bibfnamefont {A.}~\bibnamefont
  {Al\`u}},\ }\bibfield  {title} {\bibinfo {title} {Optomechanical
  nonreciprocity},\ }\href {https://doi.org/10.1038/nphys4283} {\bibfield
  {journal} {\bibinfo  {journal} {Nat. Phys.}\ }\textbf {\bibinfo {volume}
  {13}},\ \bibinfo {pages} {922} (\bibinfo {year} {2017})}\BibitemShut
  {NoStop}%
\bibitem [{\citenamefont {Seif}\ \emph {et~al.}(2018)\citenamefont {Seif},
  \citenamefont {DeGottardi}, \citenamefont {Esfarjani},\ and\ \citenamefont
  {Hafezi}}]{Thermal2018Seif}%
  \BibitemOpen
  \bibfield  {author} {\bibinfo {author} {\bibfnamefont {A.}~\bibnamefont
  {Seif}}, \bibinfo {author} {\bibfnamefont {W.}~\bibnamefont {DeGottardi}},
  \bibinfo {author} {\bibfnamefont {K.}~\bibnamefont {Esfarjani}},\ and\
  \bibinfo {author} {\bibfnamefont {M.}~\bibnamefont {Hafezi}},\ }\bibfield
  {title} {\bibinfo {title} {Thermal management and non-reciprocal control of
  phonon flow via optomechanics},\ }\href
  {https://doi.org/10.1038/s41467-018-03624-y} {\bibfield  {journal} {\bibinfo
  {journal} {Nat. Commun.}\ }\textbf {\bibinfo {volume} {9}},\ \bibinfo {pages}
  {1207} (\bibinfo {year} {2018})}\BibitemShut {NoStop}%
\bibitem [{\citenamefont {Rakich}\ and\ \citenamefont
  {Marquardt}(2018)}]{Rakich_2018}%
  \BibitemOpen
  \bibfield  {author} {\bibinfo {author} {\bibfnamefont {P.}~\bibnamefont
  {Rakich}}\ and\ \bibinfo {author} {\bibfnamefont {F.}~\bibnamefont
  {Marquardt}},\ }\bibfield  {title} {\bibinfo {title} {Quantum theory of
  continuum optomechanics},\ }\href {https://doi.org/10.1088/1367-2630/aaac4f}
  {\bibfield  {journal} {\bibinfo  {journal} {New J. Phys.}\ }\textbf {\bibinfo
  {volume} {20}},\ \bibinfo {pages} {045005} (\bibinfo {year}
  {2018})}\BibitemShut {NoStop}%
\bibitem [{\citenamefont {Xu}\ \emph {et~al.}(2019)\citenamefont {Xu},
  \citenamefont {Jiang}, \citenamefont {Clerk},\ and\ \citenamefont
  {Harris}}]{Nonreciprocal2019Xu}%
  \BibitemOpen
  \bibfield  {author} {\bibinfo {author} {\bibfnamefont {H.}~\bibnamefont
  {Xu}}, \bibinfo {author} {\bibfnamefont {L.}~\bibnamefont {Jiang}}, \bibinfo
  {author} {\bibfnamefont {A.~A.}\ \bibnamefont {Clerk}},\ and\ \bibinfo
  {author} {\bibfnamefont {J.~G.~E.}\ \bibnamefont {Harris}},\ }\bibfield
  {title} {\bibinfo {title} {Nonreciprocal control and cooling of phonon modes
  in an optomechanical system},\ }\href
  {https://doi.org/10.1038/s41586-019-1061-2} {\bibfield  {journal} {\bibinfo
  {journal} {Nature}\ }\textbf {\bibinfo {volume} {568}},\ \bibinfo {pages}
  {65} (\bibinfo {year} {2019})}\BibitemShut {NoStop}%
\bibitem [{\citenamefont {Sanavio}\ \emph {et~al.}(2020)\citenamefont
  {Sanavio}, \citenamefont {Peano},\ and\ \citenamefont
  {Xuereb}}]{PhysRevB.101.085108}%
  \BibitemOpen
  \bibfield  {author} {\bibinfo {author} {\bibfnamefont {C.}~\bibnamefont
  {Sanavio}}, \bibinfo {author} {\bibfnamefont {V.}~\bibnamefont {Peano}},\
  and\ \bibinfo {author} {\bibfnamefont {A.}~\bibnamefont {Xuereb}},\
  }\bibfield  {title} {\bibinfo {title} {Nonreciprocal topological phononics in
  optomechanical arrays},\ }\href {https://doi.org/10.1103/PhysRevB.101.085108}
  {\bibfield  {journal} {\bibinfo  {journal} {Phys. Rev. B}\ }\textbf {\bibinfo
  {volume} {101}},\ \bibinfo {pages} {085108} (\bibinfo {year}
  {2020})}\BibitemShut {NoStop}%
\bibitem [{\citenamefont {Denis}\ \emph {et~al.}(2020)\citenamefont {Denis},
  \citenamefont {Biella}, \citenamefont {Favero},\ and\ \citenamefont
  {Ciuti}}]{PhysRevLett.124.083601}%
  \BibitemOpen
  \bibfield  {author} {\bibinfo {author} {\bibfnamefont {Z.}~\bibnamefont
  {Denis}}, \bibinfo {author} {\bibfnamefont {A.}~\bibnamefont {Biella}},
  \bibinfo {author} {\bibfnamefont {I.}~\bibnamefont {Favero}},\ and\ \bibinfo
  {author} {\bibfnamefont {C.}~\bibnamefont {Ciuti}},\ }\bibfield  {title}
  {\bibinfo {title} {Permanent directional heat currents in lattices of
  optomechanical resonators},\ }\href
  {https://doi.org/10.1103/PhysRevLett.124.083601} {\bibfield  {journal}
  {\bibinfo  {journal} {Phys. Rev. Lett.}\ }\textbf {\bibinfo {volume} {124}},\
  \bibinfo {pages} {083601} (\bibinfo {year} {2020})}\BibitemShut {NoStop}%
\bibitem [{\citenamefont {Gustafsson}\ \emph {et~al.}(2014)\citenamefont
  {Gustafsson}, \citenamefont {Aref}, \citenamefont {Kockum}, \citenamefont
  {Ekstr{\"o}m}, \citenamefont {Johansson},\ and\ \citenamefont
  {Delsing}}]{Gustafsson207}%
  \BibitemOpen
  \bibfield  {author} {\bibinfo {author} {\bibfnamefont {M.~V.}\ \bibnamefont
  {Gustafsson}}, \bibinfo {author} {\bibfnamefont {T.}~\bibnamefont {Aref}},
  \bibinfo {author} {\bibfnamefont {A.~F.}\ \bibnamefont {Kockum}}, \bibinfo
  {author} {\bibfnamefont {M.~K.}\ \bibnamefont {Ekstr{\"o}m}}, \bibinfo
  {author} {\bibfnamefont {G.}~\bibnamefont {Johansson}},\ and\ \bibinfo
  {author} {\bibfnamefont {P.}~\bibnamefont {Delsing}},\ }\bibfield  {title}
  {\bibinfo {title} {Propagating phonons coupled to an artificial atom},\
  }\href {https://doi.org/10.1126/science.1257219} {\bibfield  {journal}
  {\bibinfo  {journal} {Science}\ }\textbf {\bibinfo {volume} {346}},\ \bibinfo
  {pages} {207} (\bibinfo {year} {2014})}\BibitemShut {NoStop}%
\bibitem [{\citenamefont {Schuetz}\ \emph {et~al.}(2015)\citenamefont
  {Schuetz}, \citenamefont {Kessler}, \citenamefont {Giedke}, \citenamefont
  {Vandersypen}, \citenamefont {Lukin},\ and\ \citenamefont
  {Cirac}}]{PhysRevX.5.031031}%
  \BibitemOpen
  \bibfield  {author} {\bibinfo {author} {\bibfnamefont {M.~J.~A.}\
  \bibnamefont {Schuetz}}, \bibinfo {author} {\bibfnamefont {E.~M.}\
  \bibnamefont {Kessler}}, \bibinfo {author} {\bibfnamefont {G.}~\bibnamefont
  {Giedke}}, \bibinfo {author} {\bibfnamefont {L.~M.~K.}\ \bibnamefont
  {Vandersypen}}, \bibinfo {author} {\bibfnamefont {M.~D.}\ \bibnamefont
  {Lukin}},\ and\ \bibinfo {author} {\bibfnamefont {J.~I.}\ \bibnamefont
  {Cirac}},\ }\bibfield  {title} {\bibinfo {title} {Universal quantum
  transducers based on surface acoustic waves},\ }\href
  {https://doi.org/10.1103/PhysRevX.5.031031} {\bibfield  {journal} {\bibinfo
  {journal} {Phys. Rev. X}\ }\textbf {\bibinfo {volume} {5}},\ \bibinfo {pages}
  {031031} (\bibinfo {year} {2015})}\BibitemShut {NoStop}%
\bibitem [{\citenamefont {Kuzyk}\ and\ \citenamefont
  {Wang}(2018)}]{PhysRevX.8.041027}%
  \BibitemOpen
  \bibfield  {author} {\bibinfo {author} {\bibfnamefont {M.~C.}\ \bibnamefont
  {Kuzyk}}\ and\ \bibinfo {author} {\bibfnamefont {H.}~\bibnamefont {Wang}},\
  }\bibfield  {title} {\bibinfo {title} {Scaling phononic quantum networks of
  solid-state spins with closed mechanical subsystems},\ }\href
  {https://doi.org/10.1103/PhysRevX.8.041027} {\bibfield  {journal} {\bibinfo
  {journal} {Phys. Rev. X}\ }\textbf {\bibinfo {volume} {8}},\ \bibinfo {pages}
  {041027} (\bibinfo {year} {2018})}\BibitemShut {NoStop}%
\bibitem [{\citenamefont {Bienfait}\ \emph {et~al.}(2019)\citenamefont
  {Bienfait}, \citenamefont {Satzinger}, \citenamefont {Zhong}, \citenamefont
  {Chang}, \citenamefont {Chou}, \citenamefont {Conner}, \citenamefont {Dumur},
  \citenamefont {Grebel}, \citenamefont {Peairs}, \citenamefont {Povey},\ and\
  \citenamefont {Cleland}}]{Bienfait368}%
  \BibitemOpen
  \bibfield  {author} {\bibinfo {author} {\bibfnamefont {A.}~\bibnamefont
  {Bienfait}}, \bibinfo {author} {\bibfnamefont {K.~J.}\ \bibnamefont
  {Satzinger}}, \bibinfo {author} {\bibfnamefont {Y.~P.}\ \bibnamefont
  {Zhong}}, \bibinfo {author} {\bibfnamefont {H.-S.}\ \bibnamefont {Chang}},
  \bibinfo {author} {\bibfnamefont {M.-H.}\ \bibnamefont {Chou}}, \bibinfo
  {author} {\bibfnamefont {C.~R.}\ \bibnamefont {Conner}}, \bibinfo {author}
  {\bibfnamefont {{\'E}.}~\bibnamefont {Dumur}}, \bibinfo {author}
  {\bibfnamefont {J.}~\bibnamefont {Grebel}}, \bibinfo {author} {\bibfnamefont
  {G.~A.}\ \bibnamefont {Peairs}}, \bibinfo {author} {\bibfnamefont {R.~G.}\
  \bibnamefont {Povey}},\ and\ \bibinfo {author} {\bibfnamefont {A.~N.}\
  \bibnamefont {Cleland}},\ }\bibfield  {title} {\bibinfo {title}
  {Phonon-mediated quantum state transfer and remote qubit entanglement},\
  }\href {https://doi.org/10.1126/science.aaw8415} {\bibfield  {journal}
  {\bibinfo  {journal} {Science}\ }\textbf {\bibinfo {volume} {364}},\ \bibinfo
  {pages} {368} (\bibinfo {year} {2019})}\BibitemShut {NoStop}%
\bibitem [{\citenamefont {Bin}\ \emph {et~al.}(2020)\citenamefont {Bin},
  \citenamefont {L\"u}, \citenamefont {Laussy}, \citenamefont {Nori},\ and\
  \citenamefont {Wu}}]{PhysRevLett.124.053601}%
  \BibitemOpen
  \bibfield  {author} {\bibinfo {author} {\bibfnamefont {Q.}~\bibnamefont
  {Bin}}, \bibinfo {author} {\bibfnamefont {X.-Y.}\ \bibnamefont {L\"u}},
  \bibinfo {author} {\bibfnamefont {F.~P.}\ \bibnamefont {Laussy}}, \bibinfo
  {author} {\bibfnamefont {F.}~\bibnamefont {Nori}},\ and\ \bibinfo {author}
  {\bibfnamefont {Y.}~\bibnamefont {Wu}},\ }\bibfield  {title} {\bibinfo
  {title} {\emph{N}-phonon bundle emission via the {S}tokes process},\ }\href
  {https://doi.org/10.1103/PhysRevLett.124.053601} {\bibfield  {journal}
  {\bibinfo  {journal} {Phys. Rev. Lett.}\ }\textbf {\bibinfo {volume} {124}},\
  \bibinfo {pages} {053601} (\bibinfo {year} {2020})}\BibitemShut {NoStop}%
\bibitem [{\citenamefont {Xiang}\ \emph {et~al.}(2013)\citenamefont {Xiang},
  \citenamefont {Ashhab}, \citenamefont {You},\ and\ \citenamefont
  {Nori}}]{RevModPhys.85.623}%
  \BibitemOpen
  \bibfield  {author} {\bibinfo {author} {\bibfnamefont {Z.-L.}\ \bibnamefont
  {Xiang}}, \bibinfo {author} {\bibfnamefont {S.}~\bibnamefont {Ashhab}},
  \bibinfo {author} {\bibfnamefont {J.~Q.}\ \bibnamefont {You}},\ and\ \bibinfo
  {author} {\bibfnamefont {F.}~\bibnamefont {Nori}},\ }\bibfield  {title}
  {\bibinfo {title} {Hybrid quantum circuits: Superconducting circuits
  interacting with other quantum systems},\ }\href
  {https://doi.org/10.1103/RevModPhys.85.623} {\bibfield  {journal} {\bibinfo
  {journal} {Rev. Mod. Phys.}\ }\textbf {\bibinfo {volume} {85}},\ \bibinfo
  {pages} {623} (\bibinfo {year} {2013})}\BibitemShut {NoStop}%
\bibitem [{\citenamefont {Chu}\ \emph {et~al.}(2017)\citenamefont {Chu},
  \citenamefont {Kharel}, \citenamefont {Renninger}, \citenamefont {Burkhart},
  \citenamefont {Frunzio}, \citenamefont {Rakich},\ and\ \citenamefont
  {Schoelkopf}}]{Chu199}%
  \BibitemOpen
  \bibfield  {author} {\bibinfo {author} {\bibfnamefont {Y.}~\bibnamefont
  {Chu}}, \bibinfo {author} {\bibfnamefont {P.}~\bibnamefont {Kharel}},
  \bibinfo {author} {\bibfnamefont {W.~H.}\ \bibnamefont {Renninger}}, \bibinfo
  {author} {\bibfnamefont {L.~D.}\ \bibnamefont {Burkhart}}, \bibinfo {author}
  {\bibfnamefont {L.}~\bibnamefont {Frunzio}}, \bibinfo {author} {\bibfnamefont
  {P.~T.}\ \bibnamefont {Rakich}},\ and\ \bibinfo {author} {\bibfnamefont
  {R.~J.}\ \bibnamefont {Schoelkopf}},\ }\bibfield  {title} {\bibinfo {title}
  {Quantum acoustics with superconducting qubits},\ }\href
  {https://doi.org/10.1126/science.aao1511} {\bibfield  {journal} {\bibinfo
  {journal} {Science}\ }\textbf {\bibinfo {volume} {358}},\ \bibinfo {pages}
  {199} (\bibinfo {year} {2017})}\BibitemShut {NoStop}%
\bibitem [{\citenamefont {Manenti}\ \emph {et~al.}(2017)\citenamefont
  {Manenti}, \citenamefont {Kockum}, \citenamefont {Patterson}, \citenamefont
  {Behrle}, \citenamefont {Rahamim}, \citenamefont {Tancredi}, \citenamefont
  {Nori},\ and\ \citenamefont {Leek}}]{Circuit2017Manenti}%
  \BibitemOpen
  \bibfield  {author} {\bibinfo {author} {\bibfnamefont {R.}~\bibnamefont
  {Manenti}}, \bibinfo {author} {\bibfnamefont {A.~F.}\ \bibnamefont {Kockum}},
  \bibinfo {author} {\bibfnamefont {A.}~\bibnamefont {Patterson}}, \bibinfo
  {author} {\bibfnamefont {T.}~\bibnamefont {Behrle}}, \bibinfo {author}
  {\bibfnamefont {J.}~\bibnamefont {Rahamim}}, \bibinfo {author} {\bibfnamefont
  {G.}~\bibnamefont {Tancredi}}, \bibinfo {author} {\bibfnamefont
  {F.}~\bibnamefont {Nori}},\ and\ \bibinfo {author} {\bibfnamefont {P.~J.}\
  \bibnamefont {Leek}},\ }\bibfield  {title} {\bibinfo {title} {Circuit quantum
  acoustodynamics with surface acoustic waves},\ }\href
  {https://doi.org/10.1038/s41467-017-01063-9} {\bibfield  {journal} {\bibinfo
  {journal} {Nat. Commun.}\ }\textbf {\bibinfo {volume} {8}},\ \bibinfo {pages}
  {975} (\bibinfo {year} {2017})}\BibitemShut {NoStop}%
\bibitem [{\citenamefont {Satzinger}\ \emph {et~al.}(2018)\citenamefont
  {Satzinger}, \citenamefont {Zhong}, \citenamefont {Chang}, \citenamefont
  {Peairs}, \citenamefont {Bienfait}, \citenamefont {Chou}, \citenamefont
  {Cleland}, \citenamefont {Conner}, \citenamefont {Dumur}, \citenamefont
  {Grebel}, \citenamefont {Gutierrez}, \citenamefont {November}, \citenamefont
  {Povey}, \citenamefont {Whiteley}, \citenamefont {Awschalom}, \citenamefont
  {Schuster},\ and\ \citenamefont {Cleland}}]{Quantum2018Satzinger}%
  \BibitemOpen
  \bibfield  {author} {\bibinfo {author} {\bibfnamefont {K.~J.}\ \bibnamefont
  {Satzinger}}, \bibinfo {author} {\bibfnamefont {Y.~P.}\ \bibnamefont
  {Zhong}}, \bibinfo {author} {\bibfnamefont {H.-S.}\ \bibnamefont {Chang}},
  \bibinfo {author} {\bibfnamefont {G.~A.}\ \bibnamefont {Peairs}}, \bibinfo
  {author} {\bibfnamefont {A.}~\bibnamefont {Bienfait}}, \bibinfo {author}
  {\bibfnamefont {M.-H.}\ \bibnamefont {Chou}}, \bibinfo {author}
  {\bibfnamefont {A.~Y.}\ \bibnamefont {Cleland}}, \bibinfo {author}
  {\bibfnamefont {C.~R.}\ \bibnamefont {Conner}}, \bibinfo {author}
  {\bibfnamefont {{\'E}.}~\bibnamefont {Dumur}}, \bibinfo {author}
  {\bibfnamefont {J.}~\bibnamefont {Grebel}}, \bibinfo {author} {\bibfnamefont
  {I.}~\bibnamefont {Gutierrez}}, \bibinfo {author} {\bibfnamefont {B.~H.}\
  \bibnamefont {November}}, \bibinfo {author} {\bibfnamefont {R.~G.}\
  \bibnamefont {Povey}}, \bibinfo {author} {\bibfnamefont {S.~J.}\ \bibnamefont
  {Whiteley}}, \bibinfo {author} {\bibfnamefont {D.~D.}\ \bibnamefont
  {Awschalom}}, \bibinfo {author} {\bibfnamefont {D.~I.}\ \bibnamefont
  {Schuster}},\ and\ \bibinfo {author} {\bibfnamefont {A.~N.}\ \bibnamefont
  {Cleland}},\ }\bibfield  {title} {\bibinfo {title} {Quantum control of
  surface acoustic-wave phonons},\ }\href
  {https://doi.org/10.1038/s41586-018-0719-5} {\bibfield  {journal} {\bibinfo
  {journal} {Nature}\ }\textbf {\bibinfo {volume} {563}},\ \bibinfo {pages}
  {661} (\bibinfo {year} {2018})}\BibitemShut {NoStop}%
\bibitem [{\citenamefont {Metcalfe}\ \emph {et~al.}(2010)\citenamefont
  {Metcalfe}, \citenamefont {Carr}, \citenamefont {Muller}, \citenamefont
  {Solomon},\ and\ \citenamefont {Lawall}}]{PhysRevLett.105.037401}%
  \BibitemOpen
  \bibfield  {author} {\bibinfo {author} {\bibfnamefont {M.}~\bibnamefont
  {Metcalfe}}, \bibinfo {author} {\bibfnamefont {S.~M.}\ \bibnamefont {Carr}},
  \bibinfo {author} {\bibfnamefont {A.}~\bibnamefont {Muller}}, \bibinfo
  {author} {\bibfnamefont {G.~S.}\ \bibnamefont {Solomon}},\ and\ \bibinfo
  {author} {\bibfnamefont {J.}~\bibnamefont {Lawall}},\ }\bibfield  {title}
  {\bibinfo {title} {Resolved sideband emission of
  $\mathrm{InAs}/\mathrm{GaAs}$ quantum dots strained by surface acoustic
  waves},\ }\href {https://doi.org/10.1103/PhysRevLett.105.037401} {\bibfield
  {journal} {\bibinfo  {journal} {Phys. Rev. Lett.}\ }\textbf {\bibinfo
  {volume} {105}},\ \bibinfo {pages} {037401} (\bibinfo {year}
  {2010})}\BibitemShut {NoStop}%
\bibitem [{\citenamefont {Kolkowitz}\ \emph {et~al.}(2012)\citenamefont
  {Kolkowitz}, \citenamefont {Bleszynski~Jayich}, \citenamefont
  {Unterreithmeier}, \citenamefont {Bennett}, \citenamefont {Rabl},
  \citenamefont {Harris},\ and\ \citenamefont {Lukin}}]{Kolkowitz1603}%
  \BibitemOpen
  \bibfield  {author} {\bibinfo {author} {\bibfnamefont {S.}~\bibnamefont
  {Kolkowitz}}, \bibinfo {author} {\bibfnamefont {A.~C.}\ \bibnamefont
  {Bleszynski~Jayich}}, \bibinfo {author} {\bibfnamefont {Q.~P.}\ \bibnamefont
  {Unterreithmeier}}, \bibinfo {author} {\bibfnamefont {S.~D.}\ \bibnamefont
  {Bennett}}, \bibinfo {author} {\bibfnamefont {P.}~\bibnamefont {Rabl}},
  \bibinfo {author} {\bibfnamefont {J.~G.~E.}\ \bibnamefont {Harris}},\ and\
  \bibinfo {author} {\bibfnamefont {M.~D.}\ \bibnamefont {Lukin}},\ }\bibfield
  {title} {\bibinfo {title} {Coherent sensing of a mechanical resonator with a
  single-spin qubit},\ }\href {https://doi.org/10.1126/science.1216821}
  {\bibfield  {journal} {\bibinfo  {journal} {Science}\ }\textbf {\bibinfo
  {volume} {335}},\ \bibinfo {pages} {1603} (\bibinfo {year}
  {2012})}\BibitemShut {NoStop}%
\bibitem [{\citenamefont {Hepp}\ \emph {et~al.}(2014)\citenamefont {Hepp},
  \citenamefont {M\"uller}, \citenamefont {Waselowski}, \citenamefont {Becker},
  \citenamefont {Pingault}, \citenamefont {Sternschulte}, \citenamefont
  {Steinm\"uller-Nethl}, \citenamefont {Gali}, \citenamefont {Maze},
  \citenamefont {Atat\"ure},\ and\ \citenamefont
  {Becher}}]{PhysRevLett.112.036405}%
  \BibitemOpen
  \bibfield  {author} {\bibinfo {author} {\bibfnamefont {C.}~\bibnamefont
  {Hepp}}, \bibinfo {author} {\bibfnamefont {T.}~\bibnamefont {M\"uller}},
  \bibinfo {author} {\bibfnamefont {V.}~\bibnamefont {Waselowski}}, \bibinfo
  {author} {\bibfnamefont {J.~N.}\ \bibnamefont {Becker}}, \bibinfo {author}
  {\bibfnamefont {B.}~\bibnamefont {Pingault}}, \bibinfo {author}
  {\bibfnamefont {H.}~\bibnamefont {Sternschulte}}, \bibinfo {author}
  {\bibfnamefont {D.}~\bibnamefont {Steinm\"uller-Nethl}}, \bibinfo {author}
  {\bibfnamefont {A.}~\bibnamefont {Gali}}, \bibinfo {author} {\bibfnamefont
  {J.~R.}\ \bibnamefont {Maze}}, \bibinfo {author} {\bibfnamefont
  {M.}~\bibnamefont {Atat\"ure}},\ and\ \bibinfo {author} {\bibfnamefont
  {C.}~\bibnamefont {Becher}},\ }\bibfield  {title} {\bibinfo {title}
  {Electronic structure of the silicon vacancy color center in diamond},\
  }\href {https://doi.org/10.1103/PhysRevLett.112.036405} {\bibfield  {journal}
  {\bibinfo  {journal} {Phys. Rev. Lett.}\ }\textbf {\bibinfo {volume} {112}},\
  \bibinfo {pages} {036405} (\bibinfo {year} {2014})}\BibitemShut {NoStop}%
\bibitem [{\citenamefont {Li}\ \emph {et~al.}(2015)\citenamefont {Li},
  \citenamefont {Liu}, \citenamefont {Gao}, \citenamefont {Xiang},
  \citenamefont {Rabl}, \citenamefont {Xiao},\ and\ \citenamefont
  {Li}}]{PhysRevApplied.4.044003}%
  \BibitemOpen
  \bibfield  {author} {\bibinfo {author} {\bibfnamefont {P.-B.}\ \bibnamefont
  {Li}}, \bibinfo {author} {\bibfnamefont {Y.-C.}\ \bibnamefont {Liu}},
  \bibinfo {author} {\bibfnamefont {S.-Y.}\ \bibnamefont {Gao}}, \bibinfo
  {author} {\bibfnamefont {Z.-L.}\ \bibnamefont {Xiang}}, \bibinfo {author}
  {\bibfnamefont {P.}~\bibnamefont {Rabl}}, \bibinfo {author} {\bibfnamefont
  {Y.-F.}\ \bibnamefont {Xiao}},\ and\ \bibinfo {author} {\bibfnamefont
  {F.-L.}\ \bibnamefont {Li}},\ }\bibfield  {title} {\bibinfo {title} {Hybrid
  quantum device based on \text{NV} centers in diamond nanomechanical
  resonators plus superconducting waveguide cavities},\ }\href
  {https://doi.org/10.1103/PhysRevApplied.4.044003} {\bibfield  {journal}
  {\bibinfo  {journal} {Phys. Rev. Applied}\ }\textbf {\bibinfo {volume} {4}},\
  \bibinfo {pages} {044003} (\bibinfo {year} {2015})}\BibitemShut {NoStop}%
\bibitem [{\citenamefont {Golter}\ \emph
  {et~al.}(2016{\natexlab{a}})\citenamefont {Golter}, \citenamefont {Oo},
  \citenamefont {Amezcua}, \citenamefont {Stewart},\ and\ \citenamefont
  {Wang}}]{PhysRevLett.116.143602}%
  \BibitemOpen
  \bibfield  {author} {\bibinfo {author} {\bibfnamefont {D.~A.}\ \bibnamefont
  {Golter}}, \bibinfo {author} {\bibfnamefont {T.}~\bibnamefont {Oo}}, \bibinfo
  {author} {\bibfnamefont {M.}~\bibnamefont {Amezcua}}, \bibinfo {author}
  {\bibfnamefont {K.~A.}\ \bibnamefont {Stewart}},\ and\ \bibinfo {author}
  {\bibfnamefont {H.}~\bibnamefont {Wang}},\ }\bibfield  {title} {\bibinfo
  {title} {Optomechanical quantum control of a nitrogen-vacancy center in
  diamond},\ }\href {https://doi.org/10.1103/PhysRevLett.116.143602} {\bibfield
   {journal} {\bibinfo  {journal} {Phys. Rev. Lett.}\ }\textbf {\bibinfo
  {volume} {116}},\ \bibinfo {pages} {143602} (\bibinfo {year}
  {2016}{\natexlab{a}})}\BibitemShut {NoStop}%
\bibitem [{\citenamefont {Li}\ \emph {et~al.}(2016)\citenamefont {Li},
  \citenamefont {Xiang}, \citenamefont {Rabl},\ and\ \citenamefont
  {Nori}}]{PhysRevLett.117.015502}%
  \BibitemOpen
  \bibfield  {author} {\bibinfo {author} {\bibfnamefont {P.-B.}\ \bibnamefont
  {Li}}, \bibinfo {author} {\bibfnamefont {Z.-L.}\ \bibnamefont {Xiang}},
  \bibinfo {author} {\bibfnamefont {P.}~\bibnamefont {Rabl}},\ and\ \bibinfo
  {author} {\bibfnamefont {F.}~\bibnamefont {Nori}},\ }\bibfield  {title}
  {\bibinfo {title} {Hybrid quantum device with nitrogen-vacancy centers in
  diamond coupled to carbon nanotubes},\ }\href
  {https://doi.org/10.1103/PhysRevLett.117.015502} {\bibfield  {journal}
  {\bibinfo  {journal} {Phys. Rev. Lett.}\ }\textbf {\bibinfo {volume} {117}},\
  \bibinfo {pages} {015502} (\bibinfo {year} {2016})}\BibitemShut {NoStop}%
\bibitem [{\citenamefont {Golter}\ \emph
  {et~al.}(2016{\natexlab{b}})\citenamefont {Golter}, \citenamefont {Oo},
  \citenamefont {Amezcua}, \citenamefont {Lekavicius}, \citenamefont
  {Stewart},\ and\ \citenamefont {Wang}}]{PhysRevX.6.041060}%
  \BibitemOpen
  \bibfield  {author} {\bibinfo {author} {\bibfnamefont {D.~A.}\ \bibnamefont
  {Golter}}, \bibinfo {author} {\bibfnamefont {T.}~\bibnamefont {Oo}}, \bibinfo
  {author} {\bibfnamefont {M.}~\bibnamefont {Amezcua}}, \bibinfo {author}
  {\bibfnamefont {I.}~\bibnamefont {Lekavicius}}, \bibinfo {author}
  {\bibfnamefont {K.~A.}\ \bibnamefont {Stewart}},\ and\ \bibinfo {author}
  {\bibfnamefont {H.}~\bibnamefont {Wang}},\ }\bibfield  {title} {\bibinfo
  {title} {Coupling a surface acoustic wave to an electron spin in diamond via
  a dark state},\ }\href {https://doi.org/10.1103/PhysRevX.6.041060} {\bibfield
   {journal} {\bibinfo  {journal} {Phys. Rev. X}\ }\textbf {\bibinfo {volume}
  {6}},\ \bibinfo {pages} {041060} (\bibinfo {year}
  {2016}{\natexlab{b}})}\BibitemShut {NoStop}%
\bibitem [{\citenamefont {Lee}\ \emph {et~al.}(2017)\citenamefont {Lee},
  \citenamefont {Lee}, \citenamefont {Cady}, \citenamefont {Ovartchaiyapong},\
  and\ \citenamefont {Jayich}}]{Lee2017Topical}%
  \BibitemOpen
  \bibfield  {author} {\bibinfo {author} {\bibfnamefont {D.}~\bibnamefont
  {Lee}}, \bibinfo {author} {\bibfnamefont {K.~W.}\ \bibnamefont {Lee}},
  \bibinfo {author} {\bibfnamefont {J.~V.}\ \bibnamefont {Cady}}, \bibinfo
  {author} {\bibfnamefont {P.}~\bibnamefont {Ovartchaiyapong}},\ and\ \bibinfo
  {author} {\bibfnamefont {A.~C.~B.}\ \bibnamefont {Jayich}},\ }\bibfield
  {title} {\bibinfo {title} {Topical review: Spins and mechanics in diamond},\
  }\href {https://doi.org/10.1088/2040-8986/aa52cd} {\bibfield  {journal}
  {\bibinfo  {journal} {J. Opt.}\ }\textbf {\bibinfo {volume} {19}},\ \bibinfo
  {pages} {033001} (\bibinfo {year} {2017})}\BibitemShut {NoStop}%
\bibitem [{\citenamefont {Bhaskar}\ \emph {et~al.}(2017)\citenamefont
  {Bhaskar}, \citenamefont {Sukachev}, \citenamefont {Sipahigil}, \citenamefont
  {Evans}, \citenamefont {Burek}, \citenamefont {Nguyen}, \citenamefont
  {Rogers}, \citenamefont {Siyushev}, \citenamefont {Metsch}, \citenamefont
  {Park}, \citenamefont {Jelezko}, \citenamefont {Lon\ifmmode~\check{c}\else
  \v{c}\fi{}ar},\ and\ \citenamefont {Lukin}}]{PhysRevLett.118.223603}%
  \BibitemOpen
  \bibfield  {author} {\bibinfo {author} {\bibfnamefont {M.~K.}\ \bibnamefont
  {Bhaskar}}, \bibinfo {author} {\bibfnamefont {D.~D.}\ \bibnamefont
  {Sukachev}}, \bibinfo {author} {\bibfnamefont {A.}~\bibnamefont {Sipahigil}},
  \bibinfo {author} {\bibfnamefont {R.~E.}\ \bibnamefont {Evans}}, \bibinfo
  {author} {\bibfnamefont {M.~J.}\ \bibnamefont {Burek}}, \bibinfo {author}
  {\bibfnamefont {C.~T.}\ \bibnamefont {Nguyen}}, \bibinfo {author}
  {\bibfnamefont {L.~J.}\ \bibnamefont {Rogers}}, \bibinfo {author}
  {\bibfnamefont {P.}~\bibnamefont {Siyushev}}, \bibinfo {author}
  {\bibfnamefont {M.~H.}\ \bibnamefont {Metsch}}, \bibinfo {author}
  {\bibfnamefont {H.}~\bibnamefont {Park}}, \bibinfo {author} {\bibfnamefont
  {F.}~\bibnamefont {Jelezko}}, \bibinfo {author} {\bibfnamefont
  {M.}~\bibnamefont {Lon\ifmmode~\check{c}\else \v{c}\fi{}ar}},\ and\ \bibinfo
  {author} {\bibfnamefont {M.~D.}\ \bibnamefont {Lukin}},\ }\bibfield  {title}
  {\bibinfo {title} {Quantum nonlinear optics with a germanium-vacancy color
  center in a nanoscale diamond waveguide},\ }\href
  {https://doi.org/10.1103/PhysRevLett.118.223603} {\bibfield  {journal}
  {\bibinfo  {journal} {Phys. Rev. Lett.}\ }\textbf {\bibinfo {volume} {118}},\
  \bibinfo {pages} {223603} (\bibinfo {year} {2017})}\BibitemShut {NoStop}%
\bibitem [{\citenamefont {Meesala}\ \emph {et~al.}(2018)\citenamefont
  {Meesala}, \citenamefont {Sohn}, \citenamefont {Pingault}, \citenamefont
  {Shao}, \citenamefont {Atikian}, \citenamefont {Holzgrafe}, \citenamefont
  {G\"undo\ifmmode~\breve{g}\else \u{g}\fi{}an}, \citenamefont {Stavrakas},
  \citenamefont {Sipahigil}, \citenamefont {Chia}, \citenamefont {Evans},
  \citenamefont {Burek}, \citenamefont {Zhang}, \citenamefont {Wu},
  \citenamefont {Pacheco}, \citenamefont {Abraham}, \citenamefont {Bielejec},
  \citenamefont {Lukin}, \citenamefont {Atat\"ure},\ and\ \citenamefont
  {Lon\ifmmode~\check{c}\else \v{c}\fi{}ar}}]{PhysRevB.97.205444}%
  \BibitemOpen
  \bibfield  {author} {\bibinfo {author} {\bibfnamefont {S.}~\bibnamefont
  {Meesala}}, \bibinfo {author} {\bibfnamefont {Y.-I.}\ \bibnamefont {Sohn}},
  \bibinfo {author} {\bibfnamefont {B.}~\bibnamefont {Pingault}}, \bibinfo
  {author} {\bibfnamefont {L.}~\bibnamefont {Shao}}, \bibinfo {author}
  {\bibfnamefont {H.~A.}\ \bibnamefont {Atikian}}, \bibinfo {author}
  {\bibfnamefont {J.}~\bibnamefont {Holzgrafe}}, \bibinfo {author}
  {\bibfnamefont {M.}~\bibnamefont {G\"undo\ifmmode~\breve{g}\else
  \u{g}\fi{}an}}, \bibinfo {author} {\bibfnamefont {C.}~\bibnamefont
  {Stavrakas}}, \bibinfo {author} {\bibfnamefont {A.}~\bibnamefont
  {Sipahigil}}, \bibinfo {author} {\bibfnamefont {C.}~\bibnamefont {Chia}},
  \bibinfo {author} {\bibfnamefont {R.}~\bibnamefont {Evans}}, \bibinfo
  {author} {\bibfnamefont {M.~J.}\ \bibnamefont {Burek}}, \bibinfo {author}
  {\bibfnamefont {M.}~\bibnamefont {Zhang}}, \bibinfo {author} {\bibfnamefont
  {L.}~\bibnamefont {Wu}}, \bibinfo {author} {\bibfnamefont {J.~L.}\
  \bibnamefont {Pacheco}}, \bibinfo {author} {\bibfnamefont {J.}~\bibnamefont
  {Abraham}}, \bibinfo {author} {\bibfnamefont {E.}~\bibnamefont {Bielejec}},
  \bibinfo {author} {\bibfnamefont {M.~D.}\ \bibnamefont {Lukin}}, \bibinfo
  {author} {\bibfnamefont {M.}~\bibnamefont {Atat\"ure}},\ and\ \bibinfo
  {author} {\bibfnamefont {M.}~\bibnamefont {Lon\ifmmode~\check{c}\else
  \v{c}\fi{}ar}},\ }\bibfield  {title} {\bibinfo {title} {Strain engineering of
  the silicon-vacancy center in diamond},\ }\href
  {https://doi.org/10.1103/PhysRevB.97.205444} {\bibfield  {journal} {\bibinfo
  {journal} {Phys. Rev. B}\ }\textbf {\bibinfo {volume} {97}},\ \bibinfo
  {pages} {205444} (\bibinfo {year} {2018})}\BibitemShut {NoStop}%
\bibitem [{\citenamefont {Li}\ and\ \citenamefont
  {Nori}(2018)}]{PhysRevApplied.10.024011}%
  \BibitemOpen
  \bibfield  {author} {\bibinfo {author} {\bibfnamefont {P.-B.}\ \bibnamefont
  {Li}}\ and\ \bibinfo {author} {\bibfnamefont {F.}~\bibnamefont {Nori}},\
  }\bibfield  {title} {\bibinfo {title} {Hybrid quantum system with
  nitrogen-vacancy centers in diamond coupled to surface-phonon polaritons in
  piezomagnetic superlattices},\ }\href
  {https://doi.org/10.1103/PhysRevApplied.10.024011} {\bibfield  {journal}
  {\bibinfo  {journal} {Phys. Rev. Applied}\ }\textbf {\bibinfo {volume}
  {10}},\ \bibinfo {pages} {024011} (\bibinfo {year} {2018})}\BibitemShut
  {NoStop}%
\bibitem [{\citenamefont {Maity}\ \emph {et~al.}(2020)\citenamefont {Maity},
  \citenamefont {Shao}, \citenamefont {Bogdanovi\'c}, \citenamefont {Meesala},
  \citenamefont {Sohn}, \citenamefont {Sinclair}, \citenamefont {Pingault},
  \citenamefont {Chalupnik}, \citenamefont {Chia}, \citenamefont {Zheng},
  \citenamefont {Lai},\ and\ \citenamefont {Lon\v{c}ar}}]{Coherent2020Maity}%
  \BibitemOpen
  \bibfield  {author} {\bibinfo {author} {\bibfnamefont {S.}~\bibnamefont
  {Maity}}, \bibinfo {author} {\bibfnamefont {L.}~\bibnamefont {Shao}},
  \bibinfo {author} {\bibfnamefont {S.}~\bibnamefont {Bogdanovi\'c}}, \bibinfo
  {author} {\bibfnamefont {S.}~\bibnamefont {Meesala}}, \bibinfo {author}
  {\bibfnamefont {Y.-I.}\ \bibnamefont {Sohn}}, \bibinfo {author}
  {\bibfnamefont {N.}~\bibnamefont {Sinclair}}, \bibinfo {author}
  {\bibfnamefont {B.}~\bibnamefont {Pingault}}, \bibinfo {author}
  {\bibfnamefont {M.}~\bibnamefont {Chalupnik}}, \bibinfo {author}
  {\bibfnamefont {C.}~\bibnamefont {Chia}}, \bibinfo {author} {\bibfnamefont
  {L.}~\bibnamefont {Zheng}}, \bibinfo {author} {\bibfnamefont
  {K.}~\bibnamefont {Lai}},\ and\ \bibinfo {author} {\bibfnamefont
  {M.}~\bibnamefont {Lon\v{c}ar}},\ }\bibfield  {title} {\bibinfo {title}
  {Coherent acoustic control of a single silicon vacancy spin in diamond},\
  }\href {https://doi.org/10.1038/s41467-019-13822-x} {\bibfield  {journal}
  {\bibinfo  {journal} {Nat. Commun.}\ }\textbf {\bibinfo {volume} {11}},\
  \bibinfo {pages} {193} (\bibinfo {year} {2020})}\BibitemShut {NoStop}%
\bibitem [{\citenamefont {Balasubramanian}\ \emph {et~al.}(2009)\citenamefont
  {Balasubramanian}, \citenamefont {Neumann}, \citenamefont {Twitchen},
  \citenamefont {Markham}, \citenamefont {Kolesov}, \citenamefont {Mizuochi},
  \citenamefont {Isoya}, \citenamefont {Achard}, \citenamefont {Beck},
  \citenamefont {Tissler}, \citenamefont {Jacques}, \citenamefont {Hemmer},
  \citenamefont {Jelezko},\ and\ \citenamefont
  {Wrachtrup}}]{Balasubramanian2009Ultralong}%
  \BibitemOpen
  \bibfield  {author} {\bibinfo {author} {\bibfnamefont {G.}~\bibnamefont
  {Balasubramanian}}, \bibinfo {author} {\bibfnamefont {P.}~\bibnamefont
  {Neumann}}, \bibinfo {author} {\bibfnamefont {D.}~\bibnamefont {Twitchen}},
  \bibinfo {author} {\bibfnamefont {M.}~\bibnamefont {Markham}}, \bibinfo
  {author} {\bibfnamefont {R.}~\bibnamefont {Kolesov}}, \bibinfo {author}
  {\bibfnamefont {N.}~\bibnamefont {Mizuochi}}, \bibinfo {author}
  {\bibfnamefont {J.}~\bibnamefont {Isoya}}, \bibinfo {author} {\bibfnamefont
  {J.}~\bibnamefont {Achard}}, \bibinfo {author} {\bibfnamefont
  {J.}~\bibnamefont {Beck}}, \bibinfo {author} {\bibfnamefont {J.}~\bibnamefont
  {Tissler}}, \bibinfo {author} {\bibfnamefont {V.}~\bibnamefont {Jacques}},
  \bibinfo {author} {\bibfnamefont {P.~R.}\ \bibnamefont {Hemmer}}, \bibinfo
  {author} {\bibfnamefont {F.}~\bibnamefont {Jelezko}},\ and\ \bibinfo {author}
  {\bibfnamefont {J.}~\bibnamefont {Wrachtrup}},\ }\bibfield  {title} {\bibinfo
  {title} {Ultralong spin coherence time in isotopically engineered diamond},\
  }\href {https://doi.org/10.1038/nmat2420} {\bibfield  {journal} {\bibinfo
  {journal} {Nat. Mater.}\ }\textbf {\bibinfo {volume} {8}},\ \bibinfo {pages}
  {383} (\bibinfo {year} {2009})}\BibitemShut {NoStop}%
\bibitem [{\citenamefont {Rabl}\ \emph {et~al.}(2010)\citenamefont {Rabl},
  \citenamefont {Kolkowitz}, \citenamefont {Koppens}, \citenamefont {Harris},
  \citenamefont {Zoller},\ and\ \citenamefont {Lukin}}]{Rabl2010A}%
  \BibitemOpen
  \bibfield  {author} {\bibinfo {author} {\bibfnamefont {P.}~\bibnamefont
  {Rabl}}, \bibinfo {author} {\bibfnamefont {S.~J.}\ \bibnamefont {Kolkowitz}},
  \bibinfo {author} {\bibfnamefont {F.~H.~L.}\ \bibnamefont {Koppens}},
  \bibinfo {author} {\bibfnamefont {J.~G.~E.}\ \bibnamefont {Harris}}, \bibinfo
  {author} {\bibfnamefont {P.}~\bibnamefont {Zoller}},\ and\ \bibinfo {author}
  {\bibfnamefont {M.~D.}\ \bibnamefont {Lukin}},\ }\bibfield  {title} {\bibinfo
  {title} {A quantum spin transducer based on nanoelectromechanical resonator
  arrays},\ }\href {https://doi.org/10.1038/NPHYS1679} {\bibfield  {journal}
  {\bibinfo  {journal} {Nat. Phys.}\ }\textbf {\bibinfo {volume} {6}},\
  \bibinfo {pages} {602} (\bibinfo {year} {2010})}\BibitemShut {NoStop}%
\bibitem [{\citenamefont {Bennett}\ \emph {et~al.}(2013)\citenamefont
  {Bennett}, \citenamefont {Yao}, \citenamefont {Otterbach}, \citenamefont
  {Zoller}, \citenamefont {Rabl},\ and\ \citenamefont
  {Lukin}}]{PhysRevLett.110.156402}%
  \BibitemOpen
  \bibfield  {author} {\bibinfo {author} {\bibfnamefont {S.~D.}\ \bibnamefont
  {Bennett}}, \bibinfo {author} {\bibfnamefont {N.~Y.}\ \bibnamefont {Yao}},
  \bibinfo {author} {\bibfnamefont {J.}~\bibnamefont {Otterbach}}, \bibinfo
  {author} {\bibfnamefont {P.}~\bibnamefont {Zoller}}, \bibinfo {author}
  {\bibfnamefont {P.}~\bibnamefont {Rabl}},\ and\ \bibinfo {author}
  {\bibfnamefont {M.~D.}\ \bibnamefont {Lukin}},\ }\bibfield  {title} {\bibinfo
  {title} {Phonon-induced spin-spin interactions in diamond nanostructures:
  Application to spin squeezing},\ }\href
  {https://doi.org/10.1103/PhysRevLett.110.156402} {\bibfield  {journal}
  {\bibinfo  {journal} {Phys. Rev. Lett.}\ }\textbf {\bibinfo {volume} {110}},\
  \bibinfo {pages} {156402} (\bibinfo {year} {2013})}\BibitemShut {NoStop}%
\bibitem [{\citenamefont {Bar-Gill}\ \emph {et~al.}(2013)\citenamefont
  {Bar-Gill}, \citenamefont {Pham}, \citenamefont {Jarmola}, \citenamefont
  {Budker},\ and\ \citenamefont {Walsworth}}]{Bar2013Solid}%
  \BibitemOpen
  \bibfield  {author} {\bibinfo {author} {\bibfnamefont {N.}~\bibnamefont
  {Bar-Gill}}, \bibinfo {author} {\bibfnamefont {L.}~\bibnamefont {Pham}},
  \bibinfo {author} {\bibfnamefont {A.}~\bibnamefont {Jarmola}}, \bibinfo
  {author} {\bibfnamefont {D.}~\bibnamefont {Budker}},\ and\ \bibinfo {author}
  {\bibfnamefont {R.}~\bibnamefont {Walsworth}},\ }\bibfield  {title} {\bibinfo
  {title} {Solid-state electronic spin coherence time approaching one second},\
  }\href {https://doi.org/10.1038/ncomms2771} {\bibfield  {journal} {\bibinfo
  {journal} {Nat. Commun.}\ }\textbf {\bibinfo {volume} {4}},\ \bibinfo {pages}
  {1743} (\bibinfo {year} {2013})}\BibitemShut {NoStop}%
\bibitem [{\citenamefont {Sukachev}\ \emph {et~al.}(2017)\citenamefont
  {Sukachev}, \citenamefont {Sipahigil}, \citenamefont {Nguyen}, \citenamefont
  {Bhaskar}, \citenamefont {Evans}, \citenamefont {Jelezko},\ and\
  \citenamefont {Lukin}}]{PhysRevLett.119.223602}%
  \BibitemOpen
  \bibfield  {author} {\bibinfo {author} {\bibfnamefont {D.~D.}\ \bibnamefont
  {Sukachev}}, \bibinfo {author} {\bibfnamefont {A.}~\bibnamefont {Sipahigil}},
  \bibinfo {author} {\bibfnamefont {C.~T.}\ \bibnamefont {Nguyen}}, \bibinfo
  {author} {\bibfnamefont {M.~K.}\ \bibnamefont {Bhaskar}}, \bibinfo {author}
  {\bibfnamefont {R.~E.}\ \bibnamefont {Evans}}, \bibinfo {author}
  {\bibfnamefont {F.}~\bibnamefont {Jelezko}},\ and\ \bibinfo {author}
  {\bibfnamefont {M.~D.}\ \bibnamefont {Lukin}},\ }\bibfield  {title} {\bibinfo
  {title} {Silicon-vacancy spin qubit in diamond: A quantum memory exceeding 10
  ms with single-shot state readout},\ }\href
  {https://doi.org/10.1103/PhysRevLett.119.223602} {\bibfield  {journal}
  {\bibinfo  {journal} {Phys. Rev. Lett.}\ }\textbf {\bibinfo {volume} {119}},\
  \bibinfo {pages} {223602} (\bibinfo {year} {2017})}\BibitemShut {NoStop}%
\bibitem [{\citenamefont {Lemonde}\ \emph {et~al.}(2018)\citenamefont
  {Lemonde}, \citenamefont {Meesala}, \citenamefont {Sipahigil}, \citenamefont
  {Schuetz}, \citenamefont {Lukin}, \citenamefont {Loncar},\ and\ \citenamefont
  {Rabl}}]{PhysRevLett.120.213603}%
  \BibitemOpen
  \bibfield  {author} {\bibinfo {author} {\bibfnamefont {M.-A.}\ \bibnamefont
  {Lemonde}}, \bibinfo {author} {\bibfnamefont {S.}~\bibnamefont {Meesala}},
  \bibinfo {author} {\bibfnamefont {A.}~\bibnamefont {Sipahigil}}, \bibinfo
  {author} {\bibfnamefont {M.~J.~A.}\ \bibnamefont {Schuetz}}, \bibinfo
  {author} {\bibfnamefont {M.~D.}\ \bibnamefont {Lukin}}, \bibinfo {author}
  {\bibfnamefont {M.}~\bibnamefont {Loncar}},\ and\ \bibinfo {author}
  {\bibfnamefont {P.}~\bibnamefont {Rabl}},\ }\bibfield  {title} {\bibinfo
  {title} {Phonon networks with silicon-vacancy centers in diamond
  waveguides},\ }\href {https://doi.org/10.1103/PhysRevLett.120.213603}
  {\bibfield  {journal} {\bibinfo  {journal} {Phys. Rev. Lett.}\ }\textbf
  {\bibinfo {volume} {120}},\ \bibinfo {pages} {213603} (\bibinfo {year}
  {2018})}\BibitemShut {NoStop}%
\bibitem [{\citenamefont {S\'anchez Mu\~noz}\ \emph {et~al.}(2018)\citenamefont
  {S\'anchez Mu\~noz}, \citenamefont {Lara}, \citenamefont {Puebla},\ and\
  \citenamefont {Nori}}]{PhysRevLett.121.123604}%
  \BibitemOpen
  \bibfield  {author} {\bibinfo {author} {\bibfnamefont {C.}~\bibnamefont
  {S\'anchez Mu\~noz}}, \bibinfo {author} {\bibfnamefont {A.}~\bibnamefont
  {Lara}}, \bibinfo {author} {\bibfnamefont {J.}~\bibnamefont {Puebla}},\ and\
  \bibinfo {author} {\bibfnamefont {F.}~\bibnamefont {Nori}},\ }\bibfield
  {title} {\bibinfo {title} {Hybrid systems for the generation of nonclassical
  mechanical states via quadratic interactions},\ }\href
  {https://doi.org/10.1103/PhysRevLett.121.123604} {\bibfield  {journal}
  {\bibinfo  {journal} {Phys. Rev. Lett.}\ }\textbf {\bibinfo {volume} {121}},\
  \bibinfo {pages} {123604} (\bibinfo {year} {2018})}\BibitemShut {NoStop}%
\bibitem [{\citenamefont {Li}\ \emph {et~al.}(2020{\natexlab{a}})\citenamefont
  {Li}, \citenamefont {Zhou}, \citenamefont {Gao},\ and\ \citenamefont
  {Nori}}]{PhysRevLett.125.153602}%
  \BibitemOpen
  \bibfield  {author} {\bibinfo {author} {\bibfnamefont {P.-B.}\ \bibnamefont
  {Li}}, \bibinfo {author} {\bibfnamefont {Y.}~\bibnamefont {Zhou}}, \bibinfo
  {author} {\bibfnamefont {W.-B.}\ \bibnamefont {Gao}},\ and\ \bibinfo {author}
  {\bibfnamefont {F.}~\bibnamefont {Nori}},\ }\bibfield  {title} {\bibinfo
  {title} {Enhancing spin-phonon and spin-spin interactions using linear
  resources in a hybrid quantum system},\ }\href
  {https://doi.org/10.1103/PhysRevLett.125.153602} {\bibfield  {journal}
  {\bibinfo  {journal} {Phys. Rev. Lett.}\ }\textbf {\bibinfo {volume} {125}},\
  \bibinfo {pages} {153602} (\bibinfo {year} {2020}{\natexlab{a}})}\BibitemShut
  {NoStop}%
\bibitem [{\citenamefont {Li}\ \emph {et~al.}()\citenamefont {Li},
  \citenamefont {Li},\ and\ \citenamefont {Nori}}]{BandPeng}%
  \BibitemOpen
  \bibfield  {author} {\bibinfo {author} {\bibfnamefont {P.-B.}\ \bibnamefont
  {Li}}, \bibinfo {author} {\bibfnamefont {X.-X.}\ \bibnamefont {Li}},\ and\
  \bibinfo {author} {\bibfnamefont {F.}~\bibnamefont {Nori}},\
  }\href@noop {} {\bibinfo {title} {Band-gap-engineered spin-phonon, and spin-spin
  interactions with defect centers in diamond coupled to phononic crystals}},\ \Eprint
  {https://arxiv.org/abs/1901.04650} {arXiv:1901.04650} \BibitemShut {NoStop}%
\bibitem [{\citenamefont {Lemonde}\ \emph {et~al.}(2019)\citenamefont
  {Lemonde}, \citenamefont {Peano}, \citenamefont {Rabl},\ and\ \citenamefont
  {Angelakis}}]{Lemonde_2019}%
  \BibitemOpen
\bibfield  {journal} {  }\bibfield  {author} {\bibinfo {author} {\bibfnamefont
  {M.-A.}\ \bibnamefont {Lemonde}}, \bibinfo {author} {\bibfnamefont
  {V.}~\bibnamefont {Peano}}, \bibinfo {author} {\bibfnamefont
  {P.}~\bibnamefont {Rabl}},\ and\ \bibinfo {author} {\bibfnamefont {D.~G.}\
  \bibnamefont {Angelakis}},\ }\bibfield  {title} {\bibinfo {title} {Quantum
  state transfer via acoustic edge states in a \text{2D} optomechanical
  array},\ }\href {https://doi.org/10.1088/1367-2630/ab51f5} {\bibfield
  {journal} {\bibinfo  {journal} {New J. Phys.}\ }\textbf {\bibinfo {volume}
  {21}},\ \bibinfo {pages} {113030} (\bibinfo {year} {2019})}\BibitemShut
  {NoStop}%
\bibitem [{\citenamefont {Li}\ \emph {et~al.}(2020{\natexlab{b}})\citenamefont
  {Li}, \citenamefont {Li},\ and\ \citenamefont
  {Li}}]{PhysRevResearch.2.013121}%
  \BibitemOpen
  \bibfield  {author} {\bibinfo {author} {\bibfnamefont {X.-X.}\ \bibnamefont
  {Li}}, \bibinfo {author} {\bibfnamefont {B.}~\bibnamefont {Li}},\ and\
  \bibinfo {author} {\bibfnamefont {P.-B.}\ \bibnamefont {Li}},\ }\bibfield
  {title} {\bibinfo {title} {Simulation of topological phases with color center
  arrays in phononic crystals},\ }\href
  {https://doi.org/10.1103/PhysRevResearch.2.013121} {\bibfield  {journal}
  {\bibinfo  {journal} {Phys. Rev. Research}\ }\textbf {\bibinfo {volume}
  {2}},\ \bibinfo {pages} {013121} (\bibinfo {year}
  {2020}{\natexlab{b}})}\BibitemShut {NoStop}%
\bibitem [{\citenamefont {Qiao}\ \emph {et~al.}(2020)\citenamefont {Qiao},
  \citenamefont {Li}, \citenamefont {Dong}, \citenamefont {Chen}, \citenamefont
  {Zhou},\ and\ \citenamefont {Li}}]{PhysRevA.101.042313}%
  \BibitemOpen
  \bibfield  {author} {\bibinfo {author} {\bibfnamefont {Y.-F.}\ \bibnamefont
  {Qiao}}, \bibinfo {author} {\bibfnamefont {H.-Z.}\ \bibnamefont {Li}},
  \bibinfo {author} {\bibfnamefont {X.-L.}\ \bibnamefont {Dong}}, \bibinfo
  {author} {\bibfnamefont {J.-Q.}\ \bibnamefont {Chen}}, \bibinfo {author}
  {\bibfnamefont {Y.}~\bibnamefont {Zhou}},\ and\ \bibinfo {author}
  {\bibfnamefont {P.-B.}\ \bibnamefont {Li}},\ }\bibfield  {title} {\bibinfo
  {title} {Phononic-waveguide-assisted steady-state entanglement of
  silicon-vacancy centers},\ }\href
  {https://doi.org/10.1103/PhysRevA.101.042313} {\bibfield  {journal} {\bibinfo
   {journal} {Phys. Rev. A}\ }\textbf {\bibinfo {volume} {101}},\ \bibinfo
  {pages} {042313} (\bibinfo {year} {2020})}\BibitemShut {NoStop}%
\bibitem [{\citenamefont {Safavi-Naeini}\ \emph {et~al.}(2011)\citenamefont
  {Safavi-Naeini}, \citenamefont {Alegre}, \citenamefont {Chan}, \citenamefont
  {Eichenfield}, \citenamefont {Winger}, \citenamefont {Lin}, \citenamefont
  {Hill}, \citenamefont {Chang},\ and\ \citenamefont
  {Painter}}]{Electromagnetically2011Safavi}%
  \BibitemOpen
  \bibfield  {author} {\bibinfo {author} {\bibfnamefont {A.~H.}\ \bibnamefont
  {Safavi-Naeini}}, \bibinfo {author} {\bibfnamefont {T.~P.~M.}\ \bibnamefont
  {Alegre}}, \bibinfo {author} {\bibfnamefont {J.}~\bibnamefont {Chan}},
  \bibinfo {author} {\bibfnamefont {M.}~\bibnamefont {Eichenfield}}, \bibinfo
  {author} {\bibfnamefont {M.}~\bibnamefont {Winger}}, \bibinfo {author}
  {\bibfnamefont {Q.}~\bibnamefont {Lin}}, \bibinfo {author} {\bibfnamefont
  {J.~T.}\ \bibnamefont {Hill}}, \bibinfo {author} {\bibfnamefont {D.~E.}\
  \bibnamefont {Chang}},\ and\ \bibinfo {author} {\bibfnamefont
  {O.}~\bibnamefont {Painter}},\ }\bibfield  {title} {\bibinfo {title}
  {Electromagnetically induced transparency and slow light with
  optomechanics},\ }\href {https://doi.org/10.1038/nature09933} {\bibfield
  {journal} {\bibinfo  {journal} {Nature}\ }\textbf {\bibinfo {volume} {472}},\
  \bibinfo {pages} {69} (\bibinfo {year} {2011})}\BibitemShut {NoStop}%
\bibitem [{\citenamefont {Safavi-Naeini}\ \emph {et~al.}(2014)\citenamefont
  {Safavi-Naeini}, \citenamefont {Hill}, \citenamefont {Meenehan},
  \citenamefont {Chan}, \citenamefont {Gr\"oblacher},\ and\ \citenamefont
  {Painter}}]{PhysRevLett.112.153603}%
  \BibitemOpen
  \bibfield  {author} {\bibinfo {author} {\bibfnamefont {A.~H.}\ \bibnamefont
  {Safavi-Naeini}}, \bibinfo {author} {\bibfnamefont {J.~T.}\ \bibnamefont
  {Hill}}, \bibinfo {author} {\bibfnamefont {S.}~\bibnamefont {Meenehan}},
  \bibinfo {author} {\bibfnamefont {J.}~\bibnamefont {Chan}}, \bibinfo {author}
  {\bibfnamefont {S.}~\bibnamefont {Gr\"oblacher}},\ and\ \bibinfo {author}
  {\bibfnamefont {O.}~\bibnamefont {Painter}},\ }\bibfield  {title} {\bibinfo
  {title} {Two-dimensional phononic-photonic band gap optomechanical crystal
  cavity},\ }\href {https://doi.org/10.1103/PhysRevLett.112.153603} {\bibfield
  {journal} {\bibinfo  {journal} {Phys. Rev. Lett.}\ }\textbf {\bibinfo
  {volume} {112}},\ \bibinfo {pages} {153603} (\bibinfo {year}
  {2014})}\BibitemShut {NoStop}%
\bibitem [{\citenamefont {Khanaliloo}\ \emph {et~al.}(2015)\citenamefont
  {Khanaliloo}, \citenamefont {Jayakumar}, \citenamefont {Hryciw},
  \citenamefont {Lake}, \citenamefont {Kaviani},\ and\ \citenamefont
  {Barclay}}]{PhysRevX.5.041051}%
  \BibitemOpen
  \bibfield  {author} {\bibinfo {author} {\bibfnamefont {B.}~\bibnamefont
  {Khanaliloo}}, \bibinfo {author} {\bibfnamefont {H.}~\bibnamefont
  {Jayakumar}}, \bibinfo {author} {\bibfnamefont {A.~C.}\ \bibnamefont
  {Hryciw}}, \bibinfo {author} {\bibfnamefont {D.~P.}\ \bibnamefont {Lake}},
  \bibinfo {author} {\bibfnamefont {H.}~\bibnamefont {Kaviani}},\ and\ \bibinfo
  {author} {\bibfnamefont {P.~E.}\ \bibnamefont {Barclay}},\ }\bibfield
  {title} {\bibinfo {title} {Single-crystal diamond nanobeam waveguide
  optomechanics},\ }\href {https://doi.org/10.1103/PhysRevX.5.041051}
  {\bibfield  {journal} {\bibinfo  {journal} {Phys. Rev. X}\ }\textbf {\bibinfo
  {volume} {5}},\ \bibinfo {pages} {041051} (\bibinfo {year}
  {2015})}\BibitemShut {NoStop}%
\bibitem [{\citenamefont {Burek}\ \emph {et~al.}(2016)\citenamefont {Burek},
  \citenamefont {Cohen}, \citenamefont {Meenehan}, \citenamefont {El-Sawah},
  \citenamefont {Chia}, \citenamefont {Ruelle}, \citenamefont {Meesala},
  \citenamefont {Rochman}, \citenamefont {Atikian}, \citenamefont {Markham},
  \citenamefont {Twitchen}, \citenamefont {Lukin}, \citenamefont {Painter},\
  and\ \citenamefont {Lon\v{c}ar}}]{Burek:16}%
  \BibitemOpen
  \bibfield  {author} {\bibinfo {author} {\bibfnamefont {M.~J.}\ \bibnamefont
  {Burek}}, \bibinfo {author} {\bibfnamefont {J.~D.}\ \bibnamefont {Cohen}},
  \bibinfo {author} {\bibfnamefont {S.~M.}\ \bibnamefont {Meenehan}}, \bibinfo
  {author} {\bibfnamefont {N.}~\bibnamefont {El-Sawah}}, \bibinfo {author}
  {\bibfnamefont {C.}~\bibnamefont {Chia}}, \bibinfo {author} {\bibfnamefont
  {T.}~\bibnamefont {Ruelle}}, \bibinfo {author} {\bibfnamefont
  {S.}~\bibnamefont {Meesala}}, \bibinfo {author} {\bibfnamefont
  {J.}~\bibnamefont {Rochman}}, \bibinfo {author} {\bibfnamefont {H.~A.}\
  \bibnamefont {Atikian}}, \bibinfo {author} {\bibfnamefont {M.}~\bibnamefont
  {Markham}}, \bibinfo {author} {\bibfnamefont {D.~J.}\ \bibnamefont
  {Twitchen}}, \bibinfo {author} {\bibfnamefont {M.~D.}\ \bibnamefont {Lukin}},
  \bibinfo {author} {\bibfnamefont {O.}~\bibnamefont {Painter}},\ and\ \bibinfo
  {author} {\bibfnamefont {M.}~\bibnamefont {Lon\v{c}ar}},\ }\bibfield  {title}
  {\bibinfo {title} {Diamond optomechanical crystals},\ }\href
  {https://doi.org/10.1364/OPTICA.3.001404} {\bibfield  {journal} {\bibinfo
  {journal} {Optica}\ }\textbf {\bibinfo {volume} {3}},\ \bibinfo {pages}
  {1404} (\bibinfo {year} {2016})}\BibitemShut {NoStop}%
\bibitem [{\citenamefont {Sipahigil}\ \emph {et~al.}(2016)\citenamefont
  {Sipahigil}, \citenamefont {Evans}, \citenamefont {Sukachev}, \citenamefont
  {Burek}, \citenamefont {Borregaard}, \citenamefont {Bhaskar}, \citenamefont
  {Nguyen}, \citenamefont {Pacheco}, \citenamefont {Atikian}, \citenamefont
  {Meuwly}, \citenamefont {Camacho}, \citenamefont {Jelezko}, \citenamefont
  {Bielejec}, \citenamefont {Park}, \citenamefont {Lon{\v c}ar},\ and\
  \citenamefont {Lukin}}]{Sipahigil847}%
  \BibitemOpen
  \bibfield  {author} {\bibinfo {author} {\bibfnamefont {A.}~\bibnamefont
  {Sipahigil}}, \bibinfo {author} {\bibfnamefont {R.~E.}\ \bibnamefont
  {Evans}}, \bibinfo {author} {\bibfnamefont {D.~D.}\ \bibnamefont {Sukachev}},
  \bibinfo {author} {\bibfnamefont {M.~J.}\ \bibnamefont {Burek}}, \bibinfo
  {author} {\bibfnamefont {J.}~\bibnamefont {Borregaard}}, \bibinfo {author}
  {\bibfnamefont {M.~K.}\ \bibnamefont {Bhaskar}}, \bibinfo {author}
  {\bibfnamefont {C.~T.}\ \bibnamefont {Nguyen}}, \bibinfo {author}
  {\bibfnamefont {J.~L.}\ \bibnamefont {Pacheco}}, \bibinfo {author}
  {\bibfnamefont {H.~A.}\ \bibnamefont {Atikian}}, \bibinfo {author}
  {\bibfnamefont {C.}~\bibnamefont {Meuwly}}, \bibinfo {author} {\bibfnamefont
  {R.~M.}\ \bibnamefont {Camacho}}, \bibinfo {author} {\bibfnamefont
  {F.}~\bibnamefont {Jelezko}}, \bibinfo {author} {\bibfnamefont
  {E.}~\bibnamefont {Bielejec}}, \bibinfo {author} {\bibfnamefont
  {H.}~\bibnamefont {Park}}, \bibinfo {author} {\bibfnamefont {M.}~\bibnamefont
  {Lon{\v c}ar}},\ and\ \bibinfo {author} {\bibfnamefont {M.~D.}\ \bibnamefont
  {Lukin}},\ }\bibfield  {title} {\bibinfo {title} {An integrated diamond
  nanophotonics platform for quantum-optical networks},\ }\href
  {https://doi.org/10.1126/science.aah6875} {\bibfield  {journal} {\bibinfo
  {journal} {Science}\ }\textbf {\bibinfo {volume} {354}},\ \bibinfo {pages}
  {847} (\bibinfo {year} {2016})}\BibitemShut {NoStop}%
\bibitem [{\citenamefont {Burek}\ \emph {et~al.}(2017)\citenamefont {Burek},
  \citenamefont {Meuwly}, \citenamefont {Evans}, \citenamefont {Bhaskar},
  \citenamefont {Sipahigil}, \citenamefont {Meesala}, \citenamefont
  {Machielse}, \citenamefont {Sukachev}, \citenamefont {Nguyen}, \citenamefont
  {Pacheco}, \citenamefont {Bielejec}, \citenamefont {Lukin},\ and\
  \citenamefont {Lon\ifmmode~\check{c}\else
  \v{c}\fi{}ar}}]{PhysRevApplied.8.024026}%
  \BibitemOpen
  \bibfield  {author} {\bibinfo {author} {\bibfnamefont {M.~J.}\ \bibnamefont
  {Burek}}, \bibinfo {author} {\bibfnamefont {C.}~\bibnamefont {Meuwly}},
  \bibinfo {author} {\bibfnamefont {R.~E.}\ \bibnamefont {Evans}}, \bibinfo
  {author} {\bibfnamefont {M.~K.}\ \bibnamefont {Bhaskar}}, \bibinfo {author}
  {\bibfnamefont {A.}~\bibnamefont {Sipahigil}}, \bibinfo {author}
  {\bibfnamefont {S.}~\bibnamefont {Meesala}}, \bibinfo {author} {\bibfnamefont
  {B.}~\bibnamefont {Machielse}}, \bibinfo {author} {\bibfnamefont {D.~D.}\
  \bibnamefont {Sukachev}}, \bibinfo {author} {\bibfnamefont {C.~T.}\
  \bibnamefont {Nguyen}}, \bibinfo {author} {\bibfnamefont {J.~L.}\
  \bibnamefont {Pacheco}}, \bibinfo {author} {\bibfnamefont {E.}~\bibnamefont
  {Bielejec}}, \bibinfo {author} {\bibfnamefont {M.~D.}\ \bibnamefont
  {Lukin}},\ and\ \bibinfo {author} {\bibfnamefont {M.}~\bibnamefont
  {Lon\ifmmode~\check{c}\else \v{c}\fi{}ar}},\ }\bibfield  {title} {\bibinfo
  {title} {Fiber-coupled diamond quantum nanophotonic interface},\ }\href
  {https://doi.org/10.1103/PhysRevApplied.8.024026} {\bibfield  {journal}
  {\bibinfo  {journal} {Phys. Rev. Applied}\ }\textbf {\bibinfo {volume} {8}},\
  \bibinfo {pages} {024026} (\bibinfo {year} {2017})}\BibitemShut {NoStop}%
\bibitem [{\citenamefont {Evans}\ \emph {et~al.}(2018)\citenamefont {Evans},
  \citenamefont {Bhaskar}, \citenamefont {Sukachev}, \citenamefont {Nguyen},
  \citenamefont {Sipahigil}, \citenamefont {Burek}, \citenamefont {Machielse},
  \citenamefont {Zhang}, \citenamefont {Zibrov}, \citenamefont {Bielejec},
  \citenamefont {Park}, \citenamefont {Lon{\v c}ar},\ and\ \citenamefont
  {Lukin}}]{Evans662}%
  \BibitemOpen
  \bibfield  {author} {\bibinfo {author} {\bibfnamefont {R.~E.}\ \bibnamefont
  {Evans}}, \bibinfo {author} {\bibfnamefont {M.~K.}\ \bibnamefont {Bhaskar}},
  \bibinfo {author} {\bibfnamefont {D.~D.}\ \bibnamefont {Sukachev}}, \bibinfo
  {author} {\bibfnamefont {C.~T.}\ \bibnamefont {Nguyen}}, \bibinfo {author}
  {\bibfnamefont {A.}~\bibnamefont {Sipahigil}}, \bibinfo {author}
  {\bibfnamefont {M.~J.}\ \bibnamefont {Burek}}, \bibinfo {author}
  {\bibfnamefont {B.}~\bibnamefont {Machielse}}, \bibinfo {author}
  {\bibfnamefont {G.~H.}\ \bibnamefont {Zhang}}, \bibinfo {author}
  {\bibfnamefont {A.~S.}\ \bibnamefont {Zibrov}}, \bibinfo {author}
  {\bibfnamefont {E.}~\bibnamefont {Bielejec}}, \bibinfo {author}
  {\bibfnamefont {H.}~\bibnamefont {Park}}, \bibinfo {author} {\bibfnamefont
  {M.}~\bibnamefont {Lon{\v c}ar}},\ and\ \bibinfo {author} {\bibfnamefont
  {M.~D.}\ \bibnamefont {Lukin}},\ }\bibfield  {title} {\bibinfo {title}
  {Photon-mediated interactions between quantum emitters in a diamond
  nanocavity},\ }\href {https://doi.org/10.1126/science.aau4691} {\bibfield
  {journal} {\bibinfo  {journal} {Science}\ }\textbf {\bibinfo {volume}
  {362}},\ \bibinfo {pages} {662} (\bibinfo {year} {2018})}\BibitemShut
  {NoStop}%
\bibitem [{\citenamefont {Cady}\ \emph {et~al.}(2019)\citenamefont {Cady},
  \citenamefont {Michel}, \citenamefont {Lee}, \citenamefont {Patel},
  \citenamefont {Sarabalis}, \citenamefont {Safavi-Naeini},\ and\ \citenamefont
  {Jayich}}]{Cady_2019}%
  \BibitemOpen
  \bibfield  {author} {\bibinfo {author} {\bibfnamefont {J.~V.}\ \bibnamefont
  {Cady}}, \bibinfo {author} {\bibfnamefont {O.}~\bibnamefont {Michel}},
  \bibinfo {author} {\bibfnamefont {K.~W.}\ \bibnamefont {Lee}}, \bibinfo
  {author} {\bibfnamefont {R.~N.}\ \bibnamefont {Patel}}, \bibinfo {author}
  {\bibfnamefont {C.~J.}\ \bibnamefont {Sarabalis}}, \bibinfo {author}
  {\bibfnamefont {A.~H.}\ \bibnamefont {Safavi-Naeini}},\ and\ \bibinfo
  {author} {\bibfnamefont {A.~C.~B.}\ \bibnamefont {Jayich}},\ }\bibfield
  {title} {\bibinfo {title} {Diamond optomechanical crystals with embedded
  nitrogen-vacancy centers},\ }\href {https://doi.org/10.1088/2058-9565/ab043e}
  {\bibfield  {journal} {\bibinfo  {journal} {Quantum Sci. Technol.}\ }\textbf
  {\bibinfo {volume} {4}},\ \bibinfo {pages} {024009} (\bibinfo {year}
  {2019})}\BibitemShut {NoStop}%
\bibitem [{\citenamefont {Calaj\'o}\ \emph {et~al.}(2019)\citenamefont
  {Calaj\'o}, \citenamefont {Schuetz}, \citenamefont {Pichler}, \citenamefont
  {Lukin}, \citenamefont {Schneeweiss}, \citenamefont {Volz},\ and\
  \citenamefont {Rabl}}]{PhysRevA.99.053852}%
  \BibitemOpen
  \bibfield  {author} {\bibinfo {author} {\bibfnamefont {G.}~\bibnamefont
  {Calaj\'o}}, \bibinfo {author} {\bibfnamefont {M.~J.~A.}\ \bibnamefont
  {Schuetz}}, \bibinfo {author} {\bibfnamefont {H.}~\bibnamefont {Pichler}},
  \bibinfo {author} {\bibfnamefont {M.~D.}\ \bibnamefont {Lukin}}, \bibinfo
  {author} {\bibfnamefont {P.}~\bibnamefont {Schneeweiss}}, \bibinfo {author}
  {\bibfnamefont {J.}~\bibnamefont {Volz}},\ and\ \bibinfo {author}
  {\bibfnamefont {P.}~\bibnamefont {Rabl}},\ }\bibfield  {title} {\bibinfo
  {title} {Quantum acousto-optic control of light-matter interactions in
  nanophotonic networks},\ }\href {https://doi.org/10.1103/PhysRevA.99.053852}
  {\bibfield  {journal} {\bibinfo  {journal} {Phys. Rev. A}\ }\textbf {\bibinfo
  {volume} {99}},\ \bibinfo {pages} {053852} (\bibinfo {year}
  {2019})}\BibitemShut {NoStop}%
\bibitem [{\citenamefont {Nguyen}\ \emph {et~al.}(2019)\citenamefont {Nguyen},
  \citenamefont {Sukachev}, \citenamefont {Bhaskar}, \citenamefont {Machielse},
  \citenamefont {Levonian}, \citenamefont {Knall}, \citenamefont {Stroganov},
  \citenamefont {Riedinger}, \citenamefont {Park}, \citenamefont
  {Lon\ifmmode~\check{c}\else \v{c}\fi{}ar},\ and\ \citenamefont
  {Lukin}}]{PhysRevLett.123.183602}%
  \BibitemOpen
  \bibfield  {author} {\bibinfo {author} {\bibfnamefont {C.~T.}\ \bibnamefont
  {Nguyen}}, \bibinfo {author} {\bibfnamefont {D.~D.}\ \bibnamefont
  {Sukachev}}, \bibinfo {author} {\bibfnamefont {M.~K.}\ \bibnamefont
  {Bhaskar}}, \bibinfo {author} {\bibfnamefont {B.}~\bibnamefont {Machielse}},
  \bibinfo {author} {\bibfnamefont {D.~S.}\ \bibnamefont {Levonian}}, \bibinfo
  {author} {\bibfnamefont {E.~N.}\ \bibnamefont {Knall}}, \bibinfo {author}
  {\bibfnamefont {P.}~\bibnamefont {Stroganov}}, \bibinfo {author}
  {\bibfnamefont {R.}~\bibnamefont {Riedinger}}, \bibinfo {author}
  {\bibfnamefont {H.}~\bibnamefont {Park}}, \bibinfo {author} {\bibfnamefont
  {M.}~\bibnamefont {Lon\ifmmode~\check{c}\else \v{c}\fi{}ar}},\ and\ \bibinfo
  {author} {\bibfnamefont {M.~D.}\ \bibnamefont {Lukin}},\ }\bibfield  {title}
  {\bibinfo {title} {Quantum network nodes based on diamond qubits with an
  efficient nanophotonic interface},\ }\href
  {https://doi.org/10.1103/PhysRevLett.123.183602} {\bibfield  {journal}
  {\bibinfo  {journal} {Phys. Rev. Lett.}\ }\textbf {\bibinfo {volume} {123}},\
  \bibinfo {pages} {183602} (\bibinfo {year} {2019})}\BibitemShut {NoStop}%
\bibitem [{\citenamefont {Li}\ \emph {et~al.}(2012)\citenamefont {Li},
  \citenamefont {Gao}, \citenamefont {Li}, \citenamefont {Ma},\ and\
  \citenamefont {Li}}]{PhysRevA.85.042306}%
  \BibitemOpen
  \bibfield  {author} {\bibinfo {author} {\bibfnamefont {P.-B.}\ \bibnamefont
  {Li}}, \bibinfo {author} {\bibfnamefont {S.-Y.}\ \bibnamefont {Gao}},
  \bibinfo {author} {\bibfnamefont {H.-R.}\ \bibnamefont {Li}}, \bibinfo
  {author} {\bibfnamefont {S.-L.}\ \bibnamefont {Ma}},\ and\ \bibinfo {author}
  {\bibfnamefont {F.-L.}\ \bibnamefont {Li}},\ }\bibfield  {title} {\bibinfo
  {title} {Dissipative preparation of entangled states between two spatially
  separated nitrogen-vacancy centers},\ }\href
  {https://doi.org/10.1103/PhysRevA.85.042306} {\bibfield  {journal} {\bibinfo
  {journal} {Phys. Rev. A}\ }\textbf {\bibinfo {volume} {85}},\ \bibinfo
  {pages} {042306} (\bibinfo {year} {2012})}\BibitemShut {NoStop}%
\bibitem [{\citenamefont {Stannigel}\ \emph {et~al.}(2012)\citenamefont
  {Stannigel}, \citenamefont {Rabl},\ and\ \citenamefont
  {Zoller}}]{Stannigel_2012}%
  \BibitemOpen
  \bibfield  {author} {\bibinfo {author} {\bibfnamefont {K.}~\bibnamefont
  {Stannigel}}, \bibinfo {author} {\bibfnamefont {P.}~\bibnamefont {Rabl}},\
  and\ \bibinfo {author} {\bibfnamefont {P.}~\bibnamefont {Zoller}},\
  }\bibfield  {title} {\bibinfo {title} {Driven-dissipative preparation of
  entangled states in cascaded quantum-optical networks},\ }\href
  {https://doi.org/10.1088/1367-2630/14/6/063014} {\bibfield  {journal}
  {\bibinfo  {journal} {New J. Phys.}\ }\textbf {\bibinfo {volume} {14}},\
  \bibinfo {pages} {063014} (\bibinfo {year} {2012})}\BibitemShut {NoStop}%
\bibitem [{Sup()}]{SupplementalMaterial}%
  \BibitemOpen
  \href@noop {} {}\bibinfo {note} {See Supplemental Material at https://xxx for
  more details, which includes Refs. [78--85].}\BibitemShut {Stop}%
\bibitem [{\citenamefont {Chan}\ \emph {et~al.}(2012)\citenamefont {Chan},
  \citenamefont {Safavi-Naeini}, \citenamefont {Hill}, \citenamefont
  {Meenehan},\ and\ \citenamefont {Painter}}]{doi:10.1063/1.4747726}%
  \BibitemOpen
  \bibfield  {author} {\bibinfo {author} {\bibfnamefont {J.}~\bibnamefont
  {Chan}}, \bibinfo {author} {\bibfnamefont {A.~H.}\ \bibnamefont
  {Safavi-Naeini}}, \bibinfo {author} {\bibfnamefont {J.~T.}\ \bibnamefont
  {Hill}}, \bibinfo {author} {\bibfnamefont {S.}~\bibnamefont {Meenehan}},\
  and\ \bibinfo {author} {\bibfnamefont {O.}~\bibnamefont {Painter}},\
  }\bibfield  {title} {\bibinfo {title} {Optimized optomechanical crystal
  cavity with acoustic radiation shield},\ }\href
  {https://doi.org/10.1063/1.4747726} {\bibfield  {journal} {\bibinfo
  {journal} {Appl. Phys. Lett.}\ }\textbf {\bibinfo {volume} {101}},\ \bibinfo
  {pages} {081115} (\bibinfo {year} {2012})}\BibitemShut {NoStop}%
\bibitem [{\citenamefont {MacCabe}\ \emph {et~al.}(2020)\citenamefont
  {MacCabe}, \citenamefont {Ren}, \citenamefont {Luo}, \citenamefont {Cohen},
  \citenamefont {Zhou}, \citenamefont {Sipahigil}, \citenamefont
  {Mirhosseini},\ and\ \citenamefont {Painter}}]{MacCabe840}%
  \BibitemOpen
  \bibfield  {author} {\bibinfo {author} {\bibfnamefont {G.~S.}\ \bibnamefont
  {MacCabe}}, \bibinfo {author} {\bibfnamefont {H.}~\bibnamefont {Ren}},
  \bibinfo {author} {\bibfnamefont {J.}~\bibnamefont {Luo}}, \bibinfo {author}
  {\bibfnamefont {J.~D.}\ \bibnamefont {Cohen}}, \bibinfo {author}
  {\bibfnamefont {H.}~\bibnamefont {Zhou}}, \bibinfo {author} {\bibfnamefont
  {A.}~\bibnamefont {Sipahigil}}, \bibinfo {author} {\bibfnamefont
  {M.}~\bibnamefont {Mirhosseini}},\ and\ \bibinfo {author} {\bibfnamefont
  {O.}~\bibnamefont {Painter}},\ }\bibfield  {title} {\bibinfo {title}
  {Nano-acoustic resonator with ultralong phonon lifetime},\ }\href
  {https://doi.org/10.1126/science.abc7312} {\bibfield  {journal} {\bibinfo
  {journal} {Science}\ }\textbf {\bibinfo {volume} {370}},\ \bibinfo {pages}
  {840} (\bibinfo {year} {2020})}\BibitemShut {NoStop}%
\bibitem [{\citenamefont {Schmidt}\ \emph {et~al.}()\citenamefont {Schmidt},
  \citenamefont {Peano},\ and\ \citenamefont {Marquardt}}]{opto2013michael}%
  \BibitemOpen
  \bibfield  {author} {\bibinfo {author} {\bibfnamefont {M.}~\bibnamefont
  {Schmidt}}, \bibinfo {author} {\bibfnamefont {V.}~\bibnamefont {Peano}},\
  and\ \bibinfo {author} {\bibfnamefont {F.}~\bibnamefont {Marquardt}},\
  }\href@noop {} {\bibinfo {title} {Optomechanical metamaterials: Dirac
  polaritons, gauge fields, and instabilities}},\ \Eprint
  {https://arxiv.org/abs/1311.7095} {arXiv:1311.7095} \BibitemShut {NoStop}%
\bibitem [{\citenamefont {Sokolov}\ \emph {et~al.}(2017)\citenamefont
  {Sokolov}, \citenamefont {Lian}, \citenamefont {Y\"{u}ce}, \citenamefont
  {Combri\'{e}}, \citenamefont {Rossi},\ and\ \citenamefont
  {Mosk}}]{Sokolov:17}%
  \BibitemOpen
  \bibfield  {author} {\bibinfo {author} {\bibfnamefont {S.}~\bibnamefont
  {Sokolov}}, \bibinfo {author} {\bibfnamefont {J.}~\bibnamefont {Lian}},
  \bibinfo {author} {\bibfnamefont {E.}~\bibnamefont {Y\"{u}ce}}, \bibinfo
  {author} {\bibfnamefont {S.}~\bibnamefont {Combri\'{e}}}, \bibinfo {author}
  {\bibfnamefont {A.~D.}\ \bibnamefont {Rossi}},\ and\ \bibinfo {author}
  {\bibfnamefont {A.~P.}\ \bibnamefont {Mosk}},\ }\bibfield  {title} {\bibinfo
  {title} {Tuning out disorder-induced localization in nanophotonic cavity
  arrays},\ }\href {https://doi.org/10.1364/OE.25.004598} {\bibfield  {journal}
  {\bibinfo  {journal} {Opt. Express}\ }\textbf {\bibinfo {volume} {25}},\
  \bibinfo {pages} {4598} (\bibinfo {year} {2017})}\BibitemShut {NoStop}%
\bibitem [{\citenamefont {Sumetsky}\ and\ \citenamefont
  {Dulashko}(2012)}]{Sumetsky:12}%
  \BibitemOpen
  \bibfield  {author} {\bibinfo {author} {\bibfnamefont {M.}~\bibnamefont
  {Sumetsky}}\ and\ \bibinfo {author} {\bibfnamefont {Y.}~\bibnamefont
  {Dulashko}},\ }\bibfield  {title} {\bibinfo {title} {Snap: Fabrication of
  long coupled microresonator chains with sub-angstrom precision},\ }\href
  {https://doi.org/10.1364/OE.20.027896} {\bibfield  {journal} {\bibinfo
  {journal} {Opt. Express}\ }\textbf {\bibinfo {volume} {20}},\ \bibinfo
  {pages} {27896} (\bibinfo {year} {2012})}\BibitemShut {NoStop}%
\bibitem [{\citenamefont {S\'anchez-Burillo}\ \emph
  {et~al.}(2020{\natexlab{b}})\citenamefont {S\'anchez-Burillo}, \citenamefont
  {Porras},\ and\ \citenamefont {Gonz\'alez-Tudela}}]{PhysRevA.102.013709}%
  \BibitemOpen
  \bibfield  {author} {\bibinfo {author} {\bibfnamefont {E.}~\bibnamefont
  {S\'anchez-Burillo}}, \bibinfo {author} {\bibfnamefont {D.}~\bibnamefont
  {Porras}},\ and\ \bibinfo {author} {\bibfnamefont {A.}~\bibnamefont
  {Gonz\'alez-Tudela}},\ }\bibfield  {title} {\bibinfo {title} {Limits of
  photon-mediated interactions in one-dimensional photonic baths},\ }\href
  {https://doi.org/10.1103/PhysRevA.102.013709} {\bibfield  {journal} {\bibinfo
   {journal} {Phys. Rev. A}\ }\textbf {\bibinfo {volume} {102}},\ \bibinfo
  {pages} {013709} (\bibinfo {year} {2020}{\natexlab{b}})}\BibitemShut
  {NoStop}%
\bibitem [{\citenamefont {Cohen-Tannoudji}\ \emph {et~al.}(1992)\citenamefont
  {Cohen-Tannoudji}, \citenamefont {Dupont-Roc}, \citenamefont {Grynberg},\
  and\ \citenamefont {Thickstun}}]{doi:10.1002/sca.4950140612}%
  \BibitemOpen
  \bibfield  {author} {\bibinfo {author} {\bibfnamefont {C.}~\bibnamefont
  {Cohen-Tannoudji}}, \bibinfo {author} {\bibfnamefont {J.}~\bibnamefont
  {Dupont-Roc}}, \bibinfo {author} {\bibfnamefont {G.}~\bibnamefont
  {Grynberg}},\ and\ \bibinfo {author} {\bibfnamefont {P.}~\bibnamefont
  {Thickstun}},\ }\href {https://doi.org/10.1002/sca.4950140612} {\emph
  {\bibinfo {title} {Atom-photon interactions: Basic processes and
  applications}}}\ (\bibinfo {year} {Wiley Online Library, 1992})\BibitemShut
  {NoStop}%
\bibitem [{\citenamefont {Leonforte}\ \emph {et~al.}(2021)\citenamefont
  {Leonforte}, \citenamefont {Carollo},\ and\ \citenamefont
  {Ciccarello}}]{PhysRevLett.126.063601}%
  \BibitemOpen
  \bibfield  {author} {\bibinfo {author} {\bibfnamefont {L.}~\bibnamefont
  {Leonforte}}, \bibinfo {author} {\bibfnamefont {A.}~\bibnamefont {Carollo}},\
  and\ \bibinfo {author} {\bibfnamefont {F.}~\bibnamefont {Ciccarello}},\
  }\bibfield  {title} {\bibinfo {title} {Vacancy-like dressed states in
  topological waveguide \text{QED}},\ }\href
  {https://doi.org/10.1103/PhysRevLett.126.063601} {\bibfield  {journal}
  {\bibinfo  {journal} {Phys. Rev. Lett.}\ }\textbf {\bibinfo {volume} {126}},\
  \bibinfo {pages} {063601} (\bibinfo {year} {2021})}\BibitemShut {NoStop}%
\bibitem [{\citenamefont {Barik}\ \emph {et~al.}(2018)\citenamefont {Barik},
  \citenamefont {Karasahin}, \citenamefont {Flower}, \citenamefont {Cai},
  \citenamefont {Miyake}, \citenamefont {DeGottardi}, \citenamefont {Hafezi},\
  and\ \citenamefont {Waks}}]{Barik666}%
  \BibitemOpen
  \bibfield  {author} {\bibinfo {author} {\bibfnamefont {S.}~\bibnamefont
  {Barik}}, \bibinfo {author} {\bibfnamefont {A.}~\bibnamefont {Karasahin}},
  \bibinfo {author} {\bibfnamefont {C.}~\bibnamefont {Flower}}, \bibinfo
  {author} {\bibfnamefont {T.}~\bibnamefont {Cai}}, \bibinfo {author}
  {\bibfnamefont {H.}~\bibnamefont {Miyake}}, \bibinfo {author} {\bibfnamefont
  {W.}~\bibnamefont {DeGottardi}}, \bibinfo {author} {\bibfnamefont
  {M.}~\bibnamefont {Hafezi}},\ and\ \bibinfo {author} {\bibfnamefont
  {E.}~\bibnamefont {Waks}},\ }\bibfield  {title} {\bibinfo {title} {A
  topological quantum optics interface},\ }\href
  {https://doi.org/10.1126/science.aaq0327} {\bibfield  {journal} {\bibinfo
  {journal} {Science}\ }\textbf {\bibinfo {volume} {359}},\ \bibinfo {pages}
  {666} (\bibinfo {year} {2018})}\BibitemShut {NoStop}%
\bibitem [{\citenamefont {Mehrabad}\ \emph {et~al.}(2020)\citenamefont
  {Mehrabad}, \citenamefont {Foster}, \citenamefont {Dost}, \citenamefont
  {Clarke}, \citenamefont {Patil}, \citenamefont {Fox}, \citenamefont
  {Skolnick},\ and\ \citenamefont {Wilson}}]{JalaliMehrabad:20}%
  \BibitemOpen
  \bibfield  {author} {\bibinfo {author} {\bibfnamefont {M.~J.}\ \bibnamefont
  {Mehrabad}}, \bibinfo {author} {\bibfnamefont {A.~P.}\ \bibnamefont
  {Foster}}, \bibinfo {author} {\bibfnamefont {R.}~\bibnamefont {Dost}},
  \bibinfo {author} {\bibfnamefont {E.}~\bibnamefont {Clarke}}, \bibinfo
  {author} {\bibfnamefont {P.~K.}\ \bibnamefont {Patil}}, \bibinfo {author}
  {\bibfnamefont {A.~M.}\ \bibnamefont {Fox}}, \bibinfo {author} {\bibfnamefont
  {M.~S.}\ \bibnamefont {Skolnick}},\ and\ \bibinfo {author} {\bibfnamefont
  {L.~R.}\ \bibnamefont {Wilson}},\ }\bibfield  {title} {\bibinfo {title}
  {Chiral topological photonics with an embedded quantum emitter},\ }\href
  {https://doi.org/10.1364/OPTICA.393035} {\bibfield  {journal} {\bibinfo
  {journal} {Optica}\ }\textbf {\bibinfo {volume} {7}},\ \bibinfo {pages}
  {1690} (\bibinfo {year} {2020})}\BibitemShut {NoStop}%
\bibitem [{\citenamefont {Breuer}\ and\ \citenamefont
  {Petruccione}(2002)}]{breuer2002theory}%
  \BibitemOpen
  \bibfield  {author} {\bibinfo {author} {\bibfnamefont {H.-P.}\ \bibnamefont
  {Breuer}}\ and\ \bibinfo {author} {\bibfnamefont {F.}~\bibnamefont
  {Petruccione}},\ }\href@noop {} {\emph {\bibinfo {title} {The theory of open
  quantum systems}}}\ (\bibinfo  {publisher} {Oxford University Press on
  Demand},\ \bibinfo {year} {2002})\BibitemShut {NoStop}%
\bibitem [{\citenamefont {Pichler}\ \emph {et~al.}(2015)\citenamefont
  {Pichler}, \citenamefont {Ramos}, \citenamefont {Daley},\ and\ \citenamefont
  {Zoller}}]{PhysRevA.91.042116}%
  \BibitemOpen
  \bibfield  {author} {\bibinfo {author} {\bibfnamefont {H.}~\bibnamefont
  {Pichler}}, \bibinfo {author} {\bibfnamefont {T.}~\bibnamefont {Ramos}},
  \bibinfo {author} {\bibfnamefont {A.~J.}\ \bibnamefont {Daley}},\ and\
  \bibinfo {author} {\bibfnamefont {P.}~\bibnamefont {Zoller}},\ }\bibfield
  {title} {\bibinfo {title} {Quantum optics of chiral spin networks},\ }\href
  {https://doi.org/10.1103/PhysRevA.91.042116} {\bibfield  {journal} {\bibinfo
  {journal} {Phys. Rev. A}\ }\textbf {\bibinfo {volume} {91}},\ \bibinfo
  {pages} {042116} (\bibinfo {year} {2015})}\BibitemShut {NoStop}%
\bibitem [{\citenamefont {Ramos}\ \emph {et~al.}(2016)\citenamefont {Ramos},
  \citenamefont {Vermersch}, \citenamefont {Hauke}, \citenamefont {Pichler},\
  and\ \citenamefont {Zoller}}]{PhysRevA.93.062104}%
  \BibitemOpen
  \bibfield  {author} {\bibinfo {author} {\bibfnamefont {T.}~\bibnamefont
  {Ramos}}, \bibinfo {author} {\bibfnamefont {B.}~\bibnamefont {Vermersch}},
  \bibinfo {author} {\bibfnamefont {P.}~\bibnamefont {Hauke}}, \bibinfo
  {author} {\bibfnamefont {H.}~\bibnamefont {Pichler}},\ and\ \bibinfo {author}
  {\bibfnamefont {P.}~\bibnamefont {Zoller}},\ }\bibfield  {title} {\bibinfo
  {title} {Non-markovian dynamics in chiral quantum networks with spins and
  photons},\ }\href {https://doi.org/10.1103/PhysRevA.93.062104} {\bibfield
  {journal} {\bibinfo  {journal} {Phys. Rev. A}\ }\textbf {\bibinfo {volume}
  {93}},\ \bibinfo {pages} {062104} (\bibinfo {year} {2016})}\BibitemShut
  {NoStop}%
\bibitem [{\citenamefont {Zhou}\ \emph {et~al.}(2008)\citenamefont {Zhou},
  \citenamefont {Dong}, \citenamefont {Liu}, \citenamefont {Sun},\ and\
  \citenamefont {Nori}}]{PhysRevA.78.063827}%
  \BibitemOpen
  \bibfield  {author} {\bibinfo {author} {\bibfnamefont {L.}~\bibnamefont
  {Zhou}}, \bibinfo {author} {\bibfnamefont {H.}~\bibnamefont {Dong}}, \bibinfo
  {author} {\bibfnamefont {Y.-x.}\ \bibnamefont {Liu}}, \bibinfo {author}
  {\bibfnamefont {C.~P.}\ \bibnamefont {Sun}},\ and\ \bibinfo {author}
  {\bibfnamefont {F.}~\bibnamefont {Nori}},\ }\bibfield  {title} {\bibinfo
  {title} {Quantum supercavity with atomic mirrors},\ }\href
  {https://doi.org/10.1103/PhysRevA.78.063827} {\bibfield  {journal} {\bibinfo
  {journal} {Phys. Rev. A}\ }\textbf {\bibinfo {volume} {78}},\ \bibinfo
  {pages} {063827} (\bibinfo {year} {2008})}\BibitemShut {NoStop}%
\bibitem [{\citenamefont {Liao}\ \emph {et~al.}(2010)\citenamefont {Liao},
  \citenamefont {Gong}, \citenamefont {Zhou}, \citenamefont {Liu},
  \citenamefont {Sun},\ and\ \citenamefont {Nori}}]{PhysRevA.81.042304}%
  \BibitemOpen
  \bibfield  {author} {\bibinfo {author} {\bibfnamefont {J.-Q.}\ \bibnamefont
  {Liao}}, \bibinfo {author} {\bibfnamefont {Z.~R.}\ \bibnamefont {Gong}},
  \bibinfo {author} {\bibfnamefont {L.}~\bibnamefont {Zhou}}, \bibinfo {author}
  {\bibfnamefont {Y.-x.}\ \bibnamefont {Liu}}, \bibinfo {author} {\bibfnamefont
  {C.~P.}\ \bibnamefont {Sun}},\ and\ \bibinfo {author} {\bibfnamefont
  {F.}~\bibnamefont {Nori}},\ }\bibfield  {title} {\bibinfo {title}
  {Controlling the transport of single photons by tuning the frequency of
  either one or two cavities in an array of coupled cavities},\ }\href
  {https://doi.org/10.1103/PhysRevA.81.042304} {\bibfield  {journal} {\bibinfo
  {journal} {Phys. Rev. A}\ }\textbf {\bibinfo {volume} {81}},\ \bibinfo
  {pages} {042304} (\bibinfo {year} {2010})}\BibitemShut {NoStop}%
\bibitem [{\citenamefont {Calaj\'o}\ \emph {et~al.}(2016)\citenamefont
  {Calaj\'o}, \citenamefont {Ciccarello}, \citenamefont {Chang},\ and\
  \citenamefont {Rabl}}]{PhysRevA.93.033833}%
  \BibitemOpen
  \bibfield  {author} {\bibinfo {author} {\bibfnamefont {G.}~\bibnamefont
  {Calaj\'o}}, \bibinfo {author} {\bibfnamefont {F.}~\bibnamefont
  {Ciccarello}}, \bibinfo {author} {\bibfnamefont {D.}~\bibnamefont {Chang}},\
  and\ \bibinfo {author} {\bibfnamefont {P.}~\bibnamefont {Rabl}},\ }\bibfield
  {title} {\bibinfo {title} {Atom-field dressed states in slow-light waveguide
  \text{QED}},\ }\href {https://doi.org/10.1103/PhysRevA.93.033833} {\bibfield
  {journal} {\bibinfo  {journal} {Phys. Rev. A}\ }\textbf {\bibinfo {volume}
  {93}},\ \bibinfo {pages} {033833} (\bibinfo {year} {2016})}\BibitemShut
  {NoStop}%
\bibitem [{\citenamefont {Nori}\ \emph {et~al.}(1995)\citenamefont {Nori},
  \citenamefont {Merlin}, \citenamefont {Haas}, \citenamefont {Sandvik},\ and\
  \citenamefont {Dagotto}}]{PhysRevLett.75.553}%
  \BibitemOpen
  \bibfield  {author} {\bibinfo {author} {\bibfnamefont {F.}~\bibnamefont
  {Nori}}, \bibinfo {author} {\bibfnamefont {R.}~\bibnamefont {Merlin}},
  \bibinfo {author} {\bibfnamefont {S.}~\bibnamefont {Haas}}, \bibinfo {author}
  {\bibfnamefont {A.~W.}\ \bibnamefont {Sandvik}},\ and\ \bibinfo {author}
  {\bibfnamefont {E.}~\bibnamefont {Dagotto}},\ }\bibfield  {title} {\bibinfo
  {title} {Magnetic raman scattering in two-dimensional spin-1/2 heisenberg
  antiferromagnets: Spectral shape anomaly and magnetostrictive effects},\
  }\href {https://doi.org/10.1103/PhysRevLett.75.553} {\bibfield  {journal}
  {\bibinfo  {journal} {Phys. Rev. Lett.}\ }\textbf {\bibinfo {volume} {75}},\
  \bibinfo {pages} {553} (\bibinfo {year} {1995})}\BibitemShut {NoStop}%
\bibitem [{\citenamefont {Peropadre}\ \emph {et~al.}(2013)\citenamefont
  {Peropadre}, \citenamefont {Zueco}, \citenamefont {Wulschner}, \citenamefont
  {Deppe}, \citenamefont {Marx}, \citenamefont {Gross},\ and\ \citenamefont
  {Garc\'{\i}a-Ripoll}}]{PhysRevB.87.134504}%
  \BibitemOpen
  \bibfield  {author} {\bibinfo {author} {\bibfnamefont {B.}~\bibnamefont
  {Peropadre}}, \bibinfo {author} {\bibfnamefont {D.}~\bibnamefont {Zueco}},
  \bibinfo {author} {\bibfnamefont {F.}~\bibnamefont {Wulschner}}, \bibinfo
  {author} {\bibfnamefont {F.}~\bibnamefont {Deppe}}, \bibinfo {author}
  {\bibfnamefont {A.}~\bibnamefont {Marx}}, \bibinfo {author} {\bibfnamefont
  {R.}~\bibnamefont {Gross}},\ and\ \bibinfo {author} {\bibfnamefont {J.~J.}\
  \bibnamefont {Garc\'{\i}a-Ripoll}},\ }\bibfield  {title} {\bibinfo {title}
  {Tunable coupling engineering between superconducting resonators: From
  sidebands to effective gauge fields},\ }\href
  {https://doi.org/10.1103/PhysRevB.87.134504} {\bibfield  {journal} {\bibinfo
  {journal} {Phys. Rev. B}\ }\textbf {\bibinfo {volume} {87}},\ \bibinfo
  {pages} {134504} (\bibinfo {year} {2013})}\BibitemShut {NoStop}%
\bibitem [{\citenamefont {Roushan}\ \emph {et~al.}(2017)\citenamefont
  {Roushan}, \citenamefont {Neill}, \citenamefont {Megrant}, \citenamefont
  {Chen}, \citenamefont {Babbush}, \citenamefont {Barends}, \citenamefont
  {Campbell}, \citenamefont {Chen}, \citenamefont {Chiaro}, \citenamefont
  {Dunsworth}, \citenamefont {Fowler}, \citenamefont {Jeffrey}, \citenamefont
  {Kelly}, \citenamefont {Lucero}, \citenamefont {Mutus}, \citenamefont
  {O'alley}, \citenamefont {Neeley}, \citenamefont {Quintana}, \citenamefont
  {Sank}, \citenamefont {Vainsencher}, \citenamefont {Wenner}, \citenamefont
  {White}, \citenamefont {Kapit}, \citenamefont {Neven},\ and\ \citenamefont
  {Martinis}}]{Chiral2017Roushan}%
  \BibitemOpen
  \bibfield  {author} {\bibinfo {author} {\bibfnamefont {P.}~\bibnamefont
  {Roushan}}, \bibinfo {author} {\bibfnamefont {C.}~\bibnamefont {Neill}},
  \bibinfo {author} {\bibfnamefont {A.}~\bibnamefont {Megrant}}, \bibinfo
  {author} {\bibfnamefont {Y.}~\bibnamefont {Chen}}, \bibinfo {author}
  {\bibfnamefont {R.}~\bibnamefont {Babbush}}, \bibinfo {author} {\bibfnamefont
  {R.}~\bibnamefont {Barends}}, \bibinfo {author} {\bibfnamefont
  {B.}~\bibnamefont {Campbell}}, \bibinfo {author} {\bibfnamefont
  {Z.}~\bibnamefont {Chen}}, \bibinfo {author} {\bibfnamefont {B.}~\bibnamefont
  {Chiaro}}, \bibinfo {author} {\bibfnamefont {A.}~\bibnamefont {Dunsworth}},
  \bibinfo {author} {\bibfnamefont {A.}~\bibnamefont {Fowler}}, \bibinfo
  {author} {\bibfnamefont {E.}~\bibnamefont {Jeffrey}}, \bibinfo {author}
  {\bibfnamefont {J.}~\bibnamefont {Kelly}}, \bibinfo {author} {\bibfnamefont
  {E.}~\bibnamefont {Lucero}}, \bibinfo {author} {\bibfnamefont
  {J.}~\bibnamefont {Mutus}}, \bibinfo {author} {\bibfnamefont {P.~J.~J.}\
  \bibnamefont {O'alley}}, \bibinfo {author} {\bibfnamefont
  {M.}~\bibnamefont {Neeley}}, \bibinfo {author} {\bibfnamefont
  {C.}~\bibnamefont {Quintana}}, \bibinfo {author} {\bibfnamefont
  {D.}~\bibnamefont {Sank}}, \bibinfo {author} {\bibfnamefont {A.}~\bibnamefont
  {Vainsencher}}, \bibinfo {author} {\bibfnamefont {J.}~\bibnamefont {Wenner}},
  \bibinfo {author} {\bibfnamefont {T.}~\bibnamefont {White}}, \bibinfo
  {author} {\bibfnamefont {E.}~\bibnamefont {Kapit}}, \bibinfo {author}
  {\bibfnamefont {H.}~\bibnamefont {Neven}},\ and\ \bibinfo {author}
  {\bibfnamefont {J.}~\bibnamefont {Martinis}},\ }\bibfield  {title} {\bibinfo
  {title} {Chiral ground-state currents of interacting photons in a synthetic
  magnetic field},\ }\href {https://doi.org/10.1038/nphys3930} {\bibfield
  {journal} {\bibinfo  {journal} {Nat. Phys.}\ }\textbf {\bibinfo {volume}
  {13}},\ \bibinfo {pages} {146} (\bibinfo {year} {2017})}\BibitemShut
  {NoStop}%
\bibitem [{\citenamefont {Mirhosseini}\ \emph {et~al.}(2018)\citenamefont
  {Mirhosseini}, \citenamefont {Kim}, \citenamefont {Ferreira}, \citenamefont
  {Kalaee}, \citenamefont {Sipahigil}, \citenamefont {Keller},\ and\
  \citenamefont {Painter}}]{Superconducting2018Mirhosseini}%
  \BibitemOpen
  \bibfield  {author} {\bibinfo {author} {\bibfnamefont {M.}~\bibnamefont
  {Mirhosseini}}, \bibinfo {author} {\bibfnamefont {E.}~\bibnamefont {Kim}},
  \bibinfo {author} {\bibfnamefont {V.~S.}\ \bibnamefont {Ferreira}}, \bibinfo
  {author} {\bibfnamefont {M.}~\bibnamefont {Kalaee}}, \bibinfo {author}
  {\bibfnamefont {A.}~\bibnamefont {Sipahigil}}, \bibinfo {author}
  {\bibfnamefont {A.~J.}\ \bibnamefont {Keller}},\ and\ \bibinfo {author}
  {\bibfnamefont {O.}~\bibnamefont {Painter}},\ }\bibfield  {title} {\bibinfo
  {title} {Superconducting metamaterials for waveguide quantum
  electrodynamics},\ }\href {https://doi.org/10.1038/s41467-018-06142-z}
  {\bibfield  {journal} {\bibinfo  {journal} {Nat. Commun.}\ }\textbf {\bibinfo
  {volume} {9}},\ \bibinfo {pages} {3706} (\bibinfo {year} {2018})}\BibitemShut
  {NoStop}%
\bibitem [{\citenamefont {Fang}\ \emph {et~al.}(2012)\citenamefont {Fang},
  \citenamefont {Yu},\ and\ \citenamefont {Fan}}]{Realizing2012Fang}%
  \BibitemOpen
  \bibfield  {author} {\bibinfo {author} {\bibfnamefont {K.}~\bibnamefont
  {Fang}}, \bibinfo {author} {\bibfnamefont {Z.}~\bibnamefont {Yu}},\ and\
  \bibinfo {author} {\bibfnamefont {S.}~\bibnamefont {Fan}},\ }\bibfield
  {title} {\bibinfo {title} {Realizing effective magnetic field for photons by
  controlling the phase of dynamic modulation},\ }\href
  {https://doi.org/10.1038/nphoton.2012.236} {\bibfield  {journal} {\bibinfo
  {journal} {Nat. Photonics}\ }\textbf {\bibinfo {volume} {6}},\ \bibinfo
  {pages} {782} (\bibinfo {year} {2012})}\BibitemShut {NoStop}%
\bibitem [{\citenamefont {Majumdar}\ \emph {et~al.}(2012)\citenamefont
  {Majumdar}, \citenamefont {Rundquist}, \citenamefont {Bajcsy}, \citenamefont
  {Dasika}, \citenamefont {Bank},\ and\ \citenamefont {Vu\ifmmode
  \check{c}\else \v{c}\fi{}kovi\ifmmode~\acute{c}\else
  \'{c}\fi{}}}]{PhysRevB.86.195312}%
  \BibitemOpen
  \bibfield  {author} {\bibinfo {author} {\bibfnamefont {A.}~\bibnamefont
  {Majumdar}}, \bibinfo {author} {\bibfnamefont {A.}~\bibnamefont {Rundquist}},
  \bibinfo {author} {\bibfnamefont {M.}~\bibnamefont {Bajcsy}}, \bibinfo
  {author} {\bibfnamefont {V.~D.}\ \bibnamefont {Dasika}}, \bibinfo {author}
  {\bibfnamefont {S.~R.}\ \bibnamefont {Bank}},\ and\ \bibinfo {author}
  {\bibfnamefont {J.}~\bibnamefont {Vu\ifmmode \check{c}\else
  \v{c}\fi{}kovi\ifmmode~\acute{c}\else \'{c}\fi{}}},\ }\bibfield  {title}
  {\bibinfo {title} {Design and analysis of photonic crystal coupled cavity
  arrays for quantum simulation},\ }\href
  {https://doi.org/10.1103/PhysRevB.86.195312} {\bibfield  {journal} {\bibinfo
  {journal} {Phys. Rev. B}\ }\textbf {\bibinfo {volume} {86}},\ \bibinfo
  {pages} {195312} (\bibinfo {year} {2012})}\BibitemShut {NoStop}%
\bibitem [{\citenamefont {Lodahl}\ \emph {et~al.}(2015)\citenamefont {Lodahl},
  \citenamefont {Mahmoodian},\ and\ \citenamefont
  {Stobbe}}]{RevModPhys.87.347}%
  \BibitemOpen
  \bibfield  {author} {\bibinfo {author} {\bibfnamefont {P.}~\bibnamefont
  {Lodahl}}, \bibinfo {author} {\bibfnamefont {S.}~\bibnamefont {Mahmoodian}},\
  and\ \bibinfo {author} {\bibfnamefont {S.}~\bibnamefont {Stobbe}},\
  }\bibfield  {title} {\bibinfo {title} {Interfacing single photons and single
  quantum dots with photonic nanostructures},\ }\href
  {https://doi.org/10.1103/RevModPhys.87.347} {\bibfield  {journal} {\bibinfo
  {journal} {Rev. Mod. Phys.}\ }\textbf {\bibinfo {volume} {87}},\ \bibinfo
  {pages} {347} (\bibinfo {year} {2015})}\BibitemShut {NoStop}%
\bibitem [{\citenamefont {Fang}\ \emph {et~al.}(2017)\citenamefont {Fang},
  \citenamefont {Luo}, \citenamefont {Metelmann}, \citenamefont {Matheny},
  \citenamefont {Marquardt}, \citenamefont {Clerk},\ and\ \citenamefont
  {Painter}}]{Generalized2017Fang}%
  \BibitemOpen
  \bibfield  {author} {\bibinfo {author} {\bibfnamefont {K.}~\bibnamefont
  {Fang}}, \bibinfo {author} {\bibfnamefont {J.}~\bibnamefont {Luo}}, \bibinfo
  {author} {\bibfnamefont {A.}~\bibnamefont {Metelmann}}, \bibinfo {author}
  {\bibfnamefont {M.~H.}\ \bibnamefont {Matheny}}, \bibinfo {author}
  {\bibfnamefont {F.}~\bibnamefont {Marquardt}}, \bibinfo {author}
  {\bibfnamefont {A.~A.}\ \bibnamefont {Clerk}},\ and\ \bibinfo {author}
  {\bibfnamefont {O.}~\bibnamefont {Painter}},\ }\bibfield  {title} {\bibinfo
  {title} {Generalized non-reciprocity in an optomechanical circuit via
  synthetic magnetism and reservoir engineering},\ }\href
  {https://doi.org/10.1038/nphys4009} {\bibfield  {journal} {\bibinfo
  {journal} {Nat. Phys.}\ }\textbf {\bibinfo {volume} {13}},\ \bibinfo {pages}
  {465} (\bibinfo {year} {2017})}\BibitemShut {NoStop}%
\bibitem [{\citenamefont {Mirhosseini}\ \emph {et~al.}(2020)\citenamefont
  {Mirhosseini}, \citenamefont {Sipahigil}, \citenamefont {Kalaee},\ and\
  \citenamefont {Painter}}]{Mirhosseini2020Mirhosseini}%
  \BibitemOpen
  \bibfield  {author} {\bibinfo {author} {\bibfnamefont {M.}~\bibnamefont
  {Mirhosseini}}, \bibinfo {author} {\bibfnamefont {A.}~\bibnamefont
  {Sipahigil}}, \bibinfo {author} {\bibfnamefont {M.}~\bibnamefont {Kalaee}},\
  and\ \bibinfo {author} {\bibfnamefont {O.}~\bibnamefont {Painter}},\
  }\bibfield  {title} {\bibinfo {title} {Superconducting qubit to optical
  photon transduction},\ }\href {https://doi.org/10.1038/s41586-020-3038-6}
  {\bibfield  {journal} {\bibinfo  {journal} {Nature}\ }\textbf {\bibinfo
  {volume} {588}},\ \bibinfo {pages} {599} (\bibinfo {year}
  {2020})}\BibitemShut {NoStop}%
\bibitem [{\citenamefont {Mittal}\ \emph {et~al.}(2016)\citenamefont {Mittal},
  \citenamefont {Ganeshan}, \citenamefont {Fan}, \citenamefont {Vaezi},\ and\
  \citenamefont {Hafezi}}]{Measurement2016Mittal}%
  \BibitemOpen
  \bibfield  {author} {\bibinfo {author} {\bibfnamefont {S.}~\bibnamefont
  {Mittal}}, \bibinfo {author} {\bibfnamefont {S.}~\bibnamefont {Ganeshan}},
  \bibinfo {author} {\bibfnamefont {J.}~\bibnamefont {Fan}}, \bibinfo {author}
  {\bibfnamefont {A.}~\bibnamefont {Vaezi}},\ and\ \bibinfo {author}
  {\bibfnamefont {M.}~\bibnamefont {Hafezi}},\ }\bibfield  {title} {\bibinfo
  {title} {Measurement of topological invariants in a \text{2D} photonic
  system},\ }\href {https://doi.org/10.1038/nphoton.2016.10} {\bibfield
  {journal} {\bibinfo  {journal} {Nat. Photonics}\ }\textbf {\bibinfo {volume}
  {10}},\ \bibinfo {pages} {180} (\bibinfo {year} {2016})}\BibitemShut
  {NoStop}%
\bibitem [{\citenamefont {Johansson}\ \emph {et~al.}(2012)\citenamefont
  {Johansson}, \citenamefont {Nation},\ and\ \citenamefont
  {Nori}}]{JOHANSSON20121760}%
  \BibitemOpen
  \bibfield  {author} {\bibinfo {author} {\bibfnamefont {J.}~\bibnamefont
  {Johansson}}, \bibinfo {author} {\bibfnamefont {P.}~\bibnamefont {Nation}},\
  and\ \bibinfo {author} {\bibfnamefont {F.}~\bibnamefont {Nori}},\ }\bibfield
  {title} {\bibinfo {title} {Qutip: An open-source python framework for the
  dynamics of open quantum systems},\ }\href
  {https://doi.org/https://doi.org/10.1016/j.cpc.2012.02.021} {\bibfield
  {journal} {\bibinfo  {journal} {Comput. Phys. Commun.}\ }\textbf {\bibinfo
  {volume} {183}},\ \bibinfo {pages} {1760} (\bibinfo {year}
  {2012})}\BibitemShut {NoStop}%
\bibitem [{\citenamefont {Johansson}\ \emph {et~al.}(2013)\citenamefont
  {Johansson}, \citenamefont {Nation},\ and\ \citenamefont
  {Nori}}]{JOHANSSON20131234}%
  \BibitemOpen
  \bibfield  {author} {\bibinfo {author} {\bibfnamefont {J.}~\bibnamefont
  {Johansson}}, \bibinfo {author} {\bibfnamefont {P.}~\bibnamefont {Nation}},\
  and\ \bibinfo {author} {\bibfnamefont {F.}~\bibnamefont {Nori}},\ }\bibfield
  {title} {\bibinfo {title} {Qutip 2: A python framework for the dynamics of
  open quantum systems},\ }\href
  {https://doi.org/https://doi.org/10.1016/j.cpc.2012.11.019} {\bibfield
  {journal} {\bibinfo  {journal} {Comput. Phys. Commun.}\ }\textbf {\bibinfo
  {volume} {184}},\ \bibinfo {pages} {1234} (\bibinfo {year}
  {2013})}\BibitemShut {NoStop}%
\end{thebibliography}

\begin{thebibliography}{14}%
\makeatletter
\providecommand \@ifxundefined [1]{%
 \@ifx{#1\undefined}
}%
\providecommand \@ifnum [1]{%
 \ifnum #1\expandafter \@firstoftwo
 \else \expandafter \@secondoftwo
 \fi
}%
\providecommand \@ifx [1]{%
 \ifx #1\expandafter \@firstoftwo
 \else \expandafter \@secondoftwo
 \fi
}%
\providecommand \natexlab [1]{#1}%
\providecommand \enquote  [1]{``#1''}%
\providecommand \bibnamefont  [1]{#1}%
\providecommand \bibfnamefont [1]{#1}%
\providecommand \citenamefont [1]{#1}%
\providecommand \href@noop [0]{\@secondoftwo}%
\providecommand \href [0]{\begingroup \@sanitize@url \@href}%
\providecommand \@href[1]{\@@startlink{#1}\@@href}%
\providecommand \@@href[1]{\endgroup#1\@@endlink}%
\providecommand \@sanitize@url [0]{\catcode `\\12\catcode `\$12\catcode
  `\&12\catcode `\#12\catcode `\^12\catcode `\_12\catcode `\%12\relax}%
\providecommand \@@startlink[1]{}%
\providecommand \@@endlink[0]{}%
\providecommand \url  [0]{\begingroup\@sanitize@url \@url }%
\providecommand \@url [1]{\endgroup\@href {#1}{\urlprefix }}%
\providecommand \urlprefix  [0]{URL }%
\providecommand \Eprint [0]{\href }%
\providecommand \doibase [0]{https://doi.org/}%
\providecommand \selectlanguage [0]{\@gobble}%
\providecommand \bibinfo  [0]{\@secondoftwo}%
\providecommand \bibfield  [0]{\@secondoftwo}%
\providecommand \translation [1]{[#1]}%
\providecommand \BibitemOpen [0]{}%
\providecommand \bibitemStop [0]{}%
\providecommand \bibitemNoStop [0]{.\EOS\space}%
\providecommand \EOS [0]{\spacefactor3000\relax}%
\providecommand \BibitemShut  [1]{\csname bibitem#1\endcsname}%
\let\auto@bib@innerbib\@empty
\bibitem [{\citenamefont {Chan}\ \emph {et~al.}(2012)\citenamefont {Chan},
  \citenamefont {Safavi-Naeini}, \citenamefont {Hill}, \citenamefont
  {Meenehan},\ and\ \citenamefont {Painter}}]{doi:10.1063/1.4747726}%
  \BibitemOpen
  \bibfield  {author} {\bibinfo {author} {\bibfnamefont {J.}~\bibnamefont
  {Chan}}, \bibinfo {author} {\bibfnamefont {A.~H.}\ \bibnamefont
  {Safavi-Naeini}}, \bibinfo {author} {\bibfnamefont {J.~T.}\ \bibnamefont
  {Hill}}, \bibinfo {author} {\bibfnamefont {S.}~\bibnamefont {Meenehan}},\
  and\ \bibinfo {author} {\bibfnamefont {O.}~\bibnamefont {Painter}},\
  }\bibfield  {title} {\bibinfo {title} {Optimized optomechanical crystal
  cavity with acoustic radiation shield},\ }\href
  {https://doi.org/10.1063/1.4747726} {\bibfield  {journal} {\bibinfo
  {journal} {Appl. Phys. Lett.}\ }\textbf {\bibinfo {volume} {101}},\ \bibinfo
  {pages} {081115} (\bibinfo {year} {2012})}\BibitemShut {NoStop}%
\bibitem [{\citenamefont {MacCabe}\ \emph {et~al.}(2020)\citenamefont
  {MacCabe}, \citenamefont {Ren}, \citenamefont {Luo}, \citenamefont {Cohen},
  \citenamefont {Zhou}, \citenamefont {Sipahigil}, \citenamefont
  {Mirhosseini},\ and\ \citenamefont {Painter}}]{MacCabe840}%
  \BibitemOpen
  \bibfield  {author} {\bibinfo {author} {\bibfnamefont {G.~S.}\ \bibnamefont
  {MacCabe}}, \bibinfo {author} {\bibfnamefont {H.}~\bibnamefont {Ren}},
  \bibinfo {author} {\bibfnamefont {J.}~\bibnamefont {Luo}}, \bibinfo {author}
  {\bibfnamefont {J.~D.}\ \bibnamefont {Cohen}}, \bibinfo {author}
  {\bibfnamefont {H.}~\bibnamefont {Zhou}}, \bibinfo {author} {\bibfnamefont
  {A.}~\bibnamefont {Sipahigil}}, \bibinfo {author} {\bibfnamefont
  {M.}~\bibnamefont {Mirhosseini}},\ and\ \bibinfo {author} {\bibfnamefont
  {O.}~\bibnamefont {Painter}},\ }\bibfield  {title} {\bibinfo {title}
  {Nano-acoustic resonator with ultralong phonon lifetime},\ }\href
  {https://doi.org/10.1126/science.abc7312} {\bibfield  {journal} {\bibinfo
  {journal} {Science}\ }\textbf {\bibinfo {volume} {370}},\ \bibinfo {pages}
  {840} (\bibinfo {year} {2020})}\BibitemShut {NoStop}%
\bibitem [{\citenamefont {Seif}\ \emph {et~al.}(2018)\citenamefont {Seif},
  \citenamefont {DeGottardi}, \citenamefont {Esfarjani},\ and\ \citenamefont
  {Hafezi}}]{Thermal2018Seif}%
  \BibitemOpen
  \bibfield  {author} {\bibinfo {author} {\bibfnamefont {A.}~\bibnamefont
  {Seif}}, \bibinfo {author} {\bibfnamefont {W.}~\bibnamefont {DeGottardi}},
  \bibinfo {author} {\bibfnamefont {K.}~\bibnamefont {Esfarjani}},\ and\
  \bibinfo {author} {\bibfnamefont {M.}~\bibnamefont {Hafezi}},\ }\bibfield
  {title} {\bibinfo {title} {Thermal management and non-reciprocal control of
  phonon flow via optomechanics},\ }\href
  {https://doi.org/10.1038/s41467-018-03624-y} {\bibfield  {journal} {\bibinfo
  {journal} {Nat. Commun.}\ }\textbf {\bibinfo {volume} {9}},\ \bibinfo {pages}
  {1207} (\bibinfo {year} {2018})}\BibitemShut {NoStop}%
\bibitem [{\citenamefont {Schmidt}\ \emph {et~al.}()\citenamefont {Schmidt},
  \citenamefont {Peano},\ and\ \citenamefont {Marquardt}}]{opto2013michael}%
  \BibitemOpen
  \bibfield  {author} {\bibinfo {author} {\bibfnamefont {M.}~\bibnamefont
  {Schmidt}}, \bibinfo {author} {\bibfnamefont {V.}~\bibnamefont {Peano}},\
  and\ \bibinfo {author} {\bibfnamefont {F.}~\bibnamefont {Marquardt}},\
  }\href@noop {} {\bibinfo {title} {Optomechanical metamaterials: Dirac
  polaritons, gauge fields, and instabilities}},\ \Eprint
  {https://arxiv.org/abs/1311.7095} {arXiv:1311.7095} \BibitemShut {NoStop}%
\bibitem [{\citenamefont {Lemonde}\ \emph {et~al.}(2019)\citenamefont
  {Lemonde}, \citenamefont {Peano}, \citenamefont {Rabl},\ and\ \citenamefont
  {Angelakis}}]{Lemonde_2019}%
  \BibitemOpen
  \bibfield  {author} {\bibinfo {author} {\bibfnamefont {M.-A.}\ \bibnamefont
  {Lemonde}}, \bibinfo {author} {\bibfnamefont {V.}~\bibnamefont {Peano}},
  \bibinfo {author} {\bibfnamefont {P.}~\bibnamefont {Rabl}},\ and\ \bibinfo
  {author} {\bibfnamefont {D.~G.}\ \bibnamefont {Angelakis}},\ }\bibfield
  {title} {\bibinfo {title} {Quantum state transfer via acoustic edge states in
  a \text{2D} optomechanical array},\ }\href
  {https://doi.org/10.1088/1367-2630/ab51f5} {\bibfield  {journal} {\bibinfo
  {journal} {New J. Phys.}\ }\textbf {\bibinfo {volume} {21}},\ \bibinfo
  {pages} {113030} (\bibinfo {year} {2019})}\BibitemShut {NoStop}%
\bibitem [{\citenamefont {Sokolov}\ \emph {et~al.}(2017)\citenamefont
  {Sokolov}, \citenamefont {Lian}, \citenamefont {Y\"{u}ce}, \citenamefont
  {Combri\'{e}}, \citenamefont {Rossi},\ and\ \citenamefont
  {Mosk}}]{Sokolov:17}%
  \BibitemOpen
  \bibfield  {author} {\bibinfo {author} {\bibfnamefont {S.}~\bibnamefont
  {Sokolov}}, \bibinfo {author} {\bibfnamefont {J.}~\bibnamefont {Lian}},
  \bibinfo {author} {\bibfnamefont {E.}~\bibnamefont {Y\"{u}ce}}, \bibinfo
  {author} {\bibfnamefont {S.}~\bibnamefont {Combri\'{e}}}, \bibinfo {author}
  {\bibfnamefont {A.~D.}\ \bibnamefont {Rossi}},\ and\ \bibinfo {author}
  {\bibfnamefont {A.~P.}\ \bibnamefont {Mosk}},\ }\bibfield  {title} {\bibinfo
  {title} {Tuning out disorder-induced localization in nanophotonic cavity
  arrays},\ }\href {https://doi.org/10.1364/OE.25.004598} {\bibfield  {journal}
  {\bibinfo  {journal} {Opt. Express}\ }\textbf {\bibinfo {volume} {25}},\
  \bibinfo {pages} {4598} (\bibinfo {year} {2017})}\BibitemShut {NoStop}%
\bibitem [{\citenamefont {Sumetsky}\ and\ \citenamefont
  {Dulashko}(2012)}]{Sumetsky:12}%
  \BibitemOpen
  \bibfield  {author} {\bibinfo {author} {\bibfnamefont {M.}~\bibnamefont
  {Sumetsky}}\ and\ \bibinfo {author} {\bibfnamefont {Y.}~\bibnamefont
  {Dulashko}},\ }\bibfield  {title} {\bibinfo {title} {Snap: Fabrication of
  long coupled microresonator chains with sub-angstrom precision},\ }\href
  {https://doi.org/10.1364/OE.20.027896} {\bibfield  {journal} {\bibinfo
  {journal} {Opt. Express}\ }\textbf {\bibinfo {volume} {20}},\ \bibinfo
  {pages} {27896} (\bibinfo {year} {2012})}\BibitemShut {NoStop}%
\bibitem [{\citenamefont {Lemonde}\ \emph {et~al.}(2018)\citenamefont
  {Lemonde}, \citenamefont {Meesala}, \citenamefont {Sipahigil}, \citenamefont
  {Schuetz}, \citenamefont {Lukin}, \citenamefont {Loncar},\ and\ \citenamefont
  {Rabl}}]{PhysRevLett.120.213603}%
  \BibitemOpen
  \bibfield  {author} {\bibinfo {author} {\bibfnamefont {M.-A.}\ \bibnamefont
  {Lemonde}}, \bibinfo {author} {\bibfnamefont {S.}~\bibnamefont {Meesala}},
  \bibinfo {author} {\bibfnamefont {A.}~\bibnamefont {Sipahigil}}, \bibinfo
  {author} {\bibfnamefont {M.~J.~A.}\ \bibnamefont {Schuetz}}, \bibinfo
  {author} {\bibfnamefont {M.~D.}\ \bibnamefont {Lukin}}, \bibinfo {author}
  {\bibfnamefont {M.}~\bibnamefont {Loncar}},\ and\ \bibinfo {author}
  {\bibfnamefont {P.}~\bibnamefont {Rabl}},\ }\bibfield  {title} {\bibinfo
  {title} {Phonon networks with silicon-vacancy centers in diamond
  waveguides},\ }\href {https://doi.org/10.1103/PhysRevLett.120.213603}
  {\bibfield  {journal} {\bibinfo  {journal} {Phys. Rev. Lett.}\ }\textbf
  {\bibinfo {volume} {120}},\ \bibinfo {pages} {213603} (\bibinfo {year}
  {2018})}\BibitemShut {NoStop}%
\bibitem [{\citenamefont {Breuer}\ and\ \citenamefont
  {Petruccione}(2002)}]{breuer2002theory}%
  \BibitemOpen
  \bibfield  {author} {\bibinfo {author} {\bibfnamefont {H.-P.}\ \bibnamefont
  {Breuer}}\ and\ \bibinfo {author} {\bibfnamefont {F.}~\bibnamefont
  {Petruccione}},\ }\href@noop {} {\emph {\bibinfo {title} {The theory of open
  quantum systems}}}\ (\bibinfo  {publisher} {Oxford University Press},\
  \bibinfo {year} {2002})\BibitemShut {NoStop}%
\bibitem [{\citenamefont {S\'anchez-Burillo}\ \emph {et~al.}(2020)\citenamefont
  {S\'anchez-Burillo}, \citenamefont {Porras},\ and\ \citenamefont
  {Gonz\'alez-Tudela}}]{PhysRevA.102.013709}%
  \BibitemOpen
  \bibfield  {author} {\bibinfo {author} {\bibfnamefont {E.}~\bibnamefont
  {S\'anchez-Burillo}}, \bibinfo {author} {\bibfnamefont {D.}~\bibnamefont
  {Porras}},\ and\ \bibinfo {author} {\bibfnamefont {A.}~\bibnamefont
  {Gonz\'alez-Tudela}},\ }\bibfield  {title} {\bibinfo {title} {Limits of
  photon-mediated interactions in one-dimensional photonic baths},\ }\href
  {https://doi.org/10.1103/PhysRevA.102.013709} {\bibfield  {journal} {\bibinfo
   {journal} {Phys. Rev. A}\ }\textbf {\bibinfo {volume} {102}},\ \bibinfo
  {pages} {013709} (\bibinfo {year} {2020})}\BibitemShut {NoStop}%
\bibitem [{\citenamefont {Pichler}\ \emph {et~al.}(2015)\citenamefont
  {Pichler}, \citenamefont {Ramos}, \citenamefont {Daley},\ and\ \citenamefont
  {Zoller}}]{PhysRevA.91.042116}%
  \BibitemOpen
  \bibfield  {author} {\bibinfo {author} {\bibfnamefont {H.}~\bibnamefont
  {Pichler}}, \bibinfo {author} {\bibfnamefont {T.}~\bibnamefont {Ramos}},
  \bibinfo {author} {\bibfnamefont {A.~J.}\ \bibnamefont {Daley}},\ and\
  \bibinfo {author} {\bibfnamefont {P.}~\bibnamefont {Zoller}},\ }\bibfield
  {title} {\bibinfo {title} {Quantum optics of chiral spin networks},\ }\href
  {https://doi.org/10.1103/PhysRevA.91.042116} {\bibfield  {journal} {\bibinfo
  {journal} {Phys. Rev. A}\ }\textbf {\bibinfo {volume} {91}},\ \bibinfo
  {pages} {042116} (\bibinfo {year} {2015})}\BibitemShut {NoStop}%
\bibitem [{\citenamefont {Stannigel}\ \emph {et~al.}(2012)\citenamefont
  {Stannigel}, \citenamefont {Rabl},\ and\ \citenamefont
  {Zoller}}]{Stannigel_2012}%
  \BibitemOpen
  \bibfield  {author} {\bibinfo {author} {\bibfnamefont {K.}~\bibnamefont
  {Stannigel}}, \bibinfo {author} {\bibfnamefont {P.}~\bibnamefont {Rabl}},\
  and\ \bibinfo {author} {\bibfnamefont {P.}~\bibnamefont {Zoller}},\
  }\bibfield  {title} {\bibinfo {title} {Driven-dissipative preparation of
  entangled states in cascaded quantum-optical networks},\ }\href
  {https://doi.org/10.1088/1367-2630/14/6/063014} {\bibfield  {journal}
  {\bibinfo  {journal} {New J. Phys.}\ }\textbf {\bibinfo {volume} {14}},\
  \bibinfo {pages} {063014} (\bibinfo {year} {2012})}\BibitemShut {NoStop}%
\bibitem [{\citenamefont {Cohen-Tannoudji}\ \emph {et~al.}(1992)\citenamefont
  {Cohen-Tannoudji}, \citenamefont {Dupont-Roc}, \citenamefont {Grynberg},\
  and\ \citenamefont {Thickstun}}]{doi:10.1002/sca.4950140612}%
  \BibitemOpen
  \bibfield  {author} {\bibinfo {author} {\bibfnamefont {C.}~\bibnamefont
  {Cohen-Tannoudji}}, \bibinfo {author} {\bibfnamefont {J.}~\bibnamefont
  {Dupont-Roc}}, \bibinfo {author} {\bibfnamefont {G.}~\bibnamefont
  {Grynberg}},\ and\ \bibinfo {author} {\bibfnamefont {P.}~\bibnamefont
  {Thickstun}},\ }\href {https://doi.org/10.1002/sca.4950140612} {\emph
  {\bibinfo {title} {Atom-photon interactions: Basic processes and
  applications}}}\ (\bibinfo {year} {Wiley Online Library, 1992})\BibitemShut
  {NoStop}%
\bibitem [{\citenamefont {Leonforte}\ \emph {et~al.}(2021)\citenamefont
  {Leonforte}, \citenamefont {Carollo},\ and\ \citenamefont
  {Ciccarello}}]{PhysRevLett.126.063601}%
  \BibitemOpen
  \bibfield  {author} {\bibinfo {author} {\bibfnamefont {L.}~\bibnamefont
  {Leonforte}}, \bibinfo {author} {\bibfnamefont {A.}~\bibnamefont {Carollo}},\
  and\ \bibinfo {author} {\bibfnamefont {F.}~\bibnamefont {Ciccarello}},\
  }\bibfield  {title} {\bibinfo {title} {Vacancy-like dressed states in
  topological waveguide \text{QED}},\ }\href
  {https://doi.org/10.1103/PhysRevLett.126.063601} {\bibfield  {journal}
  {\bibinfo  {journal} {Phys. Rev. Lett.}\ }\textbf {\bibinfo {volume} {126}},\
  \bibinfo {pages} {063601} (\bibinfo {year} {2021})}\BibitemShut {NoStop}%
\end{thebibliography}
%

\end{document}